\documentclass[paper,10pt]{iopart}
\usepackage{setspace}

\usepackage{setstack}
\usepackage{amssymb}
\usepackage{graphicx}
\usepackage{subfigure}

\newcommand{\norm}[1]{| #1 |}

\newcommand{\no}[1]{#1_{0}}

\usepackage{iopams}

\usepackage{subfigure}
\usepackage{bm}
\usepackage{psfig}
\usepackage{float}
\usepackage{multirow}
\usepackage{psfrag}

\usepackage{psfragx}
\usepackage{graphicx}
\usepackage{cite}
\usepackage{caption}
\usepackage{color}
\usepackage{longtable}

\makeatletter

\newcommand{\Rmnum}[1]{\expandafter\@slowromancap\romannumeral #1@}
\makeatother

\newcommand{\figref}[1]{Fig.~\ref{#1}}

\newcommand{\eqref}[1]{(\ref{#1})}

\def \x{\mathbf{x}}

\def \y{\mathbf{y}}
\def \z{\mathbf{z}}
\def \v{\mathbf{v}}

\def \0{\mathbf{0}}

\DeclareMathAlphabet{\bi}{OML}{cmm}{b}{it}
\DeclareMathAlphabet{\bcal}{OMS}{cmsy}{b}{n}
\DeclareMathAlphabet{\brmn}{OT1}{cmr}{bx}{n}

\begin{document}

\title[Synthetic Aperture Imaging of Moving Targets using Ultra-Narrowband CW]{Bistatic Synthetic Aperture Radar Imaging of Moving Targets using Ultra-Narrowband Continuous Waveforms}

\author{Ling Wang and Birsen Yaz{\i}c{\i}}

\address{Department of Electrical, Computer and Systems Engineering,
Rensselaer Polytechnic Institute, Troy, NY 12180} 
\ead{yazici@ecse.rpi.edu}

\begin{abstract}
We consider a synthetic aperture radar (SAR) system that uses ultra-narrowband continuous waveforms (CW) as an illumination source. Such a system has many practical advantages, such as the use of relatively simple, low-cost and low-power transmitters, and in some cases, using the transmitters of opportunity, such as TV, radio stations. Additionally, ultra-narrowband CW signals are suitable for motion estimation due to their ability to acquire high resolution Doppler information.

In this paper, we present a novel synthetic aperture imaging method for moving targets using a bi-static SAR system transmitting ultra-narrowband continuous waveforms. Our method exploits the high Doppler resolution provided by ultra-narrowband CW signals to image both the scene reflectivity and to determine the velocity of multiple moving targets. Starting from the first principle, we develop a novel forward model based on the temporal Doppler induced by the movement of antennas and moving targets. We form the reflectivity image of the scene and estimate the motion parameters using a filtered-backprojection technique combined with a contrast optimization method. Analysis of the point spread function of our image formation method shows that reflectivity images are focused when the motion parameters are estimated correctly. We present analysis of the velocity resolution and the resolution of reconstructed reflectivity images. We analyze the error between the correct and reconstructed position of targets due to errors in velocity estimation. 
Extensive numerical simulations demonstrate the performance of our method and validate the theoretical results.
\end{abstract}

\section{Introduction}
\subsection{Motivations}
Conventional synthetic aperture radar (SAR) is designed for stationary target imaging \cite{Jakowatz96,CGM}. Moving targets are typically smeared or defocused in SAR images \cite{Raney}. Many different approaches have been suggested to address the moving target imaging problem for conventional SAR systems 
\cite{Yang93,Barbarossa92_1,Barbarossa92_2,Soumekh02,Kirscht03,Stuff04,Zhou07,Borcea12,
Werness90,Perry99,Munson00,Munson02,Zhu11,Martorella11,Jakowatz98,Jao01,MGZ05,Hack11,MC12}. 
Both the imaging of static scenes and moving targets in conventional SAR rely on the high range resolution provided by wideband transmitted waveforms.
Such waveforms are ideal in localizing the targets, but poor in determining their motion parameters.

In this paper, we consider a SAR system that uses ultra-narrowband continuous waveforms (CW) as an illumination source.
Unlike the high range resolution waveforms used by conventional SAR systems, ultra-narrowband CW signals have high Doppler resolution which can be used to determine the velocity of moving targets with high resolution. CW radar systems also have the advantage of using relatively simple and low cost transmitters and receivers which can be made small and lightweight \cite{skolnik_2nd,Griffiths1988,Griffiths1998,Meta2007}. Additionally, a SAR system that uses ultra-narrowband CW signals may not need a dedicated transmitter. Ambient radio frequency signals, such as those provided by radio and television stations, etc., can be used as illumination sources.
In \cite{LB2010}, 
we presented a synthetic aperture imaging method of stationary scenes using ultra-narrowband CW signals.
(See also the introduction of [27] for a survey of stationary target imaging using Doppler only measurements.)
In this paper, we present a new and novel method for synthetic aperture imaging of both stationary scenes and multiple moving targets using such waveforms. 
Our method exploits the high Doppler resolution provided by such waveforms to form high resolution images of both the stationary scatters and to determine the velocity of moving targets. To the best of our knowledge, our method is the first in the literature that addresses the synthetic aperture imaging of moving targets using ultra-narrowband continuous waveforms.

\subsection{Related Work}
Conventional wideband SAR moving target imaging techniques can be roughly categorized into two classes depending on the assumptions made on the motion parameters.

The first class of techniques either assume \emph{a priori} knowledge of
target motion parameters or estimate this information prior to image reconstruction \cite{Yang93,Barbarossa92_1,Barbarossa92_2,Soumekh02,Kirscht03,Stuff04,Zhou07,Borcea12}. These motion parameters, which include relative velocity of targets with respect to antennas, Doppler shift, Doppler rate etc. are then used to reconstruct ``focused'' reflectivity images.
However, in practice a priori knowledge of motion parameters is either unavailable or difficult to determine. Therefore a great deal of effort has been devoted to develop techniques that do not require a priori knowledge of unknown motion parameters for image formation \cite{Werness90,Perry99,Munson00,Munson02,Zhu11,Martorella11,Jakowatz98,Jao01,MGZ05,Hack11}. 
In this class of techniques, the estimation of motion parameters and image formation process are performed jointly. Our approach falls into this class of methods where we couple the estimation of multiple target velocities with the reconstruction of scene reflectivity.

Techniques for the joint estimation of motion parameters and SAR image formation are based on variety of approaches. These include adaptation of inverse synthetic aperture imaging type methods \cite{Werness90}, \cite{Martorella11}, \cite{Martorella2005_2}; autofocus type methods \cite{Jakowatz98}, \cite{Zhu11}; the keystone transform \cite{Perry99}; time-frequency transform based imaging methods \cite{Munson00,Munson02}; and generalized likelihood ratio type of ideas where the reflectivity images are formed for a range of hypothesized motion parameters from which the unknown motion parameters are estimated while simultaneously forming focused reflectivity images \cite{Jakowatz98,Jao01,MGZ05,Hack11}.


\subsection{Overview and Advantages of Our Work}
We note that all the work in SAR imaging of moving targets has been developed for the conventional wideband SAR \cite{Yang93,Barbarossa92_1,Barbarossa92_2,Soumekh02,Kirscht03,Stuff04,Zhou07,Borcea12,Werness90,Perry99,
Munson00,Munson02,Zhu11,Martorella11,Jakowatz98,Jao01,MGZ05,Hack11,MC12}.
Our work differs significantly from the existing work in SAR imaging of moving targets. Conventional SAR moving target imaging methods ignore the ``temporal'' Doppler since wideband waveforms have poor Doppler resolution. Furthermore, they rely on start-stop approximation \cite{Jakowatz96,CGM}. We instead begin with the wave equation and derive a novel forward model that includes temporal Doppler parameters induced by the movement of the antennas and moving targets.
Next, we develop a novel filtered-backprojection type method combined with image contrast optimization to reconstruct the scene reflectivity and to determine the velocity of moving targets.

Similar to \cite{MGZ05,Hack11,Jakowatz98,Jao01}, we adopt a generalized likelihood ratio type approach and form a set of reflectivity images for a range of hypothesized velocities for each scatterer. Our imaging method exploits high Doppler resolution of the transmitted waveforms in that we form reflectivity images by filtering and backprojecting the preprocessed received signal onto the position-space iso-Doppler contours defined in this paper. The scatterers that lie on the position-space iso-Doppler contours can be determined with high resolution due to high resolution Doppler measurements. We show that when the hypothesized velocity is equal to the correct velocity of a scatterer at a given location, the singularities of the scene are reconstructed at the correct location and orientation. We design the filter so that the the singularities of the scene are reconstructed at the correct strength whenever the hypothesized velocity is equal to the true velocity of a scatterer. This filter depends not only the antenna beam patterns, geometric spreading factors etc., but also the hypothesized target velocity. We next use the contrast of the reflectivity images to determine the velocity of moving targets. We present the point spread function (PSF) analysis and the resolution analysis of our method.
The PSF analysis shows that our reflectivity image reconstruction method uses temporal Doppler and Doppler-rate in forming a high resolution image. 
We analyze the resolution of the reconstructed reflectivity images and the resolution of achievable velocity estimation. Our analysis identifies several factors related to the imaging geometry and the transmitted waveforms that effect the resolution of reflectivity images and velocity field. We analyze the error between the correct and reconstructed positions of the scatterers due to error in the hypothesized velocity.
We derive an analytic formula that predicts the positioning errors/smearing caused by moving targets in reflectivity images reconstructed under the stationary scene assumption.
Specifically, we show that small errors in the velocity estimation results in small positioning errors in the reconstructed reflectivity images. 
We present extensive numerical simulations to demonstrate the performance of our method and to validate the theoretical findings.

In addition to the advantages provided by the ultra-narrowband CW signals, our moving target imaging method also has the following advantages as compared to the existing SAR moving target imaging methods:
(1) Unlike \cite{Soumekh02,Kirscht03,Stuff04,Zhou07,Borcea12,Werness90,Perry99,Zhu11,
Martorella11,Jakowatz98,Jao01,MGZ05}, our method can reconstruct the images of multiple moving targets regardless of the target speed, the direction of target velocity and target location; and determine the two-dimensional velocity of ground moving targets. Furthermore, our method can reconstruct high-resolution images of stationary and moving targets simultaneously.
(2) Unlike \cite{Yang93,Barbarossa92_1,Barbarossa92_2,Soumekh02,Kirscht03,Stuff04,Zhou07,Borcea12},
our imaging method does not require \emph{a  priori} knowledge of the target motion parameters. Furthermore it does not require a priori knowledge of the number of moving targets present in the scene.
(3) Our method focuses moving targets at the correct locations in the reconstructed reflectivity images. The localization and repositioning techniques of moving targets used in most conventional SAR or ground moving target indicator methods are not needed \cite{Zhu11,Jakowatz98,Jao01,MGZ05}. 
(4) We use a linear model for the target motion. However, our method can be easily extended to accommodate arbitrary target motions, such as nonlinear, accelerating targets. 
(5) It can be used for arbitrary imaging geometries including arbitrary flight trajectories and non-flat topography. Furthermore, our image formation method is analytic which can be implemented computationally efficiently \cite{Demanet10,Demanet09,Demanet07}. 

\subsection{Organization of the Paper}
The remainder of the paper is organized as follows:
In Section \ref{sec:measModel}, we present our moving target, incident and scattered field models, and the received signal model from a moving scene.
In Section \ref{sec:ForwardModel}, we develop a novel forward model that maps the reflectivity and velocity field of a moving scene to a correlated received signal. 
In Section \ref{sec:imaging}, we develop an FBP-type image formation method to reconstruct the reflectivity of the scene and a contrast-maximization based velocity estimation method. In Section \ref{sec:ResoAna} we analyze the resolution of the reconstructed reflectivity images and the velocity resolution. In Section \ref{sec:PosError} we present the error in the position of scatterers due to error in the hypothesized velocity.
In Section \ref{sec:simulation}, we present numerical simulations. Section \ref{sec:conclusion} concludes our paper.
\addtocounter{table}{0}
\begin{table}[!h!t!p]
\linespread{1.1}
\centering
\caption{Table of Notations}
{\small
  \begin{tabular}{p{1in}p{3.7in}}
  \hline\hline
  Symbol & Designation \\ 
  \hline 
  $\omega_0 (f_0)$ & (Angular) carrier frequency of the ultra--narrowband waveform\\
  $\brmn x = (\bi x,\psi(\bi x))$ & Earth's surface\\
  $V(\brmn x)$ & 3D Reflectivity function\\
  $\rho(\bi x)$ & Surface reflectivity\\
  $\bGamma(\bi x,t)$ & Location of the moving target at time $t$ located at $\x$ at $t=0$\\
  $\v_\x$ & Velocity of the moving target located at $\x$ at time $t=0$\\
  $\widehat{\brmn x}$ & Unit vector in the direction of $\x \in \mathbb{R}^3$\\
  $\bgamma_T(t), \dot{\bgamma}_T(t)$ & Flight trajectory and velocity of the transmitter\\
  $\bgamma_R(t), \dot{\bgamma}_R(t)$ & Flight trajectory and velocity of the receiver\\
  $r(t)$ & Received signal along the receiver trajectory $\bgamma_R(t)$ due to a transmitter
  traversing the trajectory $\bgamma_T(t)$\\
  $q(\bi x,\bi v)$ & Reflectivity function of the moving target that takes into account the target movement\\
  $p(t), \tilde{p}(t)$ & Transmitted waveform and its complex amplitude\\
  $s$ & Temporal translation variable\\
  $\mu$ & Temporal scaling factor\\
  $\phi(t)$ & Temporal windowing function\\
  $d(s,\mu)$ & Windowed, scaled-and-translated correlations of the received signal and the transmitted waveform\\
  $\mathcal{F}$ & Forward modeling operator\\
  $\varphi(t,\bi x,\bi v,s,\mu)$ & Phase of the operator $\mathcal{F}$\\
  $A(t,\bi x,\bi v,s,\mu)$ & Amplitude of the operator $\mathcal{F}$\\
  $\mathrm{supp}(A)$ & Support of $A$\\
  $f_{d}(s,\x,\v)$ & Bistatic Doppler frequency with respect to a moving target\\
  $F(s,\mu)$ & Four-dimensional bistatic iso-Doppler manifold\\
  $F_{\bi v_0}(s,\mu)$ &  Two-dimensional position-space bistatic iso-Doppler contours\\
  $F_{\bi x_0}(s,\mu)$ &  Two-dimensional velocity-space bistatic iso-Doppler contours\\
  $\dot{f}_{d}(s,\x,\v)$ & Bistatic Doppler-rate with respect to a moving target\\
  $\dot{F}(s,C)$ & Four-dimensional Bistatic iso-Doppler-rate manifold\\
  $\dot{F}_{\bi v_0}(s,C)$ & Two-dimensional position-space bistatic iso-Doppler-rate contours\\
  $\dot{F}_{\bi x_0}(s,C)$ & Two-dimensional velocity-space bistatic iso-Doppler-rate contours\\
  $\mathcal{K}_{\bi v_h}$ & Filtered-backprojection reflectivity imaging operator for a hypothesized velocity $\v_h$\\
  $L^{\bi v_\x}_{\bi v_h}(\bi z,\bi x)$ & Point spread function of $\mathcal{K}_{\bi v_h}$\\
  $Q_{\bi v_h}(\bi z,t,s)$ & Reconstruction filter of the reflectivity imaging operator $\mathcal{K}_{\bi v_h}$\\
  ${\rho}_{\bi v_h}(\bi z)$ & Reconstructed reflectivity image for a hypothesized velocity\\
  $\Omega_{\bi v_\x,\bi z}$ & Data collection manifold at $\x=(\bi x,\bpsi(\bi x))$ for $\v_h=\v_0$\\
  $L_\phi$ & Length of the support of the temporal windowing function $\phi(t)$\\
  $I(\bi v_h)$ & Contrast-image \\
  $\mathcal{M}$ & Sample mean over the spatial coordinates \\
  $\boldsymbol{\varsigma}$ & Fourier vector associated with the velocity \\
  \hline\hline 
  \end{tabular}}
\end{table} 
\section{Model for Moving Targets, Incident Field, Scattered Field and Received Signal Models}\label{sec:measModel}
We use the following notational conventions throughout the paper. The bold Roman, bold
italic and Roman lower-case letters are used to denote variables in $\mathbb{R}^3$, $\mathbb{R}^2$
and $\mathbb{R}$, respectively, i.e., $\brmn x = (\bi x,x)\in\mathbb{R}^3$, with $\bi
x\in\mathbb{R}^2$ and $x\in\mathbb{R}$. The calligraphic letters ($\mathcal{F}, \mathcal{K}$ etc.) are used to denote operators. Table I lists the notations used throughout the paper.

Let the earth's surface be denoted by $\brmn x = (\bi x, \psi(\bi x))\in\mathbb{R}^3$,
where $\bi x\in \mathbb{R}^2$ and $\psi:\mathbb{R}^2\rightarrow \mathbb{R}$ is a known function for the ground topography. Furthermore, we assume that
the scattering takes place in a thin region near the surface. Thus,
the reflectivity function has the form
\begin{equation}\label{eq:vapp}
    V(\brmn x) = \rho(\bi x) \delta(x - \psi(\bi x)).
\end{equation}

\subsection{Model for a Moving Target}\label{sec:MovtgModel}
Let $\brmn z = \bGamma(\brmn x, t)$ denote the location of a moving target at time $t$, where $\brmn x$ denotes the location of the target at some reference time, say $t = 0$. We assume that for each $t \in [0, T]$, the function $\bGamma(\cdot, t):\mathbb{R}^3 \rightarrow \mathbb{R}^3$ is a diffeomorphism. Physically, this means that two distinct scatterers cannot move into the same location. Furthermore, we assume that for each  $\brmn x \in\mathbb{R}^3$,  $\bGamma(\brmn x, \cdot):\mathbb{R} \rightarrow \mathbb{R}$ is differentiable.

Let the inverse $\bGamma^{-1}(\cdot, t)$, of the function $\bGamma(\cdot, t)$ be $\balpha(\cdot, t)$, i.e., $\brmn x = \balpha(\brmn z, t)=\bGamma^{-1}(\brmn z, t)$. We assume that the refractive indices of the scatterers are preserved over time, however, the scatterer at $\brmn x$ moves along the trajectory $\brmn z = \bGamma(\brmn x, t)$. Thus, $V(\brmn x)$ at time $t=0$ translates as $V(\balpha(\brmn z, t))$ at time $t$. 

Let ${\brmn v}_{\brmn x}(t)$ denote the velocity of the target at time $t$, located at $\x$ when $t=0$, i.e.,
\begin{eqnarray}
    {\brmn v}_{\brmn x}(t)&=&\dot{\bGamma}(\brmn x, t)
    \label{eq:v_def}
    \\
    &=&[\dot{\Gamma}_1(\brmn x, t),\dot{\Gamma}_2(\brmn x, t),\dot{\Gamma}_3(\brmn x, t)]
    \nonumber
\end{eqnarray}
where $\bGamma(\brmn x, t) = [\Gamma_{1}(\brmn x, t), \ \Gamma_{2}(\brmn x, t), \ \Gamma_{3}(\brmn x, t)]^T$ and $\dot{\bGamma}(\brmn x, \cdot)$ denotes the derivative of $\bGamma(\brmn x, \cdot)$ with respect to $t$. 
We define
\begin{equation}\label{eq:2dV}
    {\bi v}_{\brmn x}(t)= [\dot{\Gamma}_{1}({\brmn x}, t), \ \dot{\Gamma}_{2}({\brmn x},t)].
\end{equation}
For ground moving targets, since
\begin{equation}
\Gamma_{3}(\brmn x, t)=\psi((\Gamma_{1}(\brmn x, t), \ \Gamma_{2}(\brmn x, t))),
\end{equation}
we write
\begin{eqnarray}
    \dot{\bGamma}_{3}({\brmn x}, t)
    &=& \nabla_{\bi x}\psi(\bi x) \cdot [\dot{\Gamma}_{1}({\brmn x}, t), \ \dot{\Gamma}_{2}({\brmn x},t)]
    \nonumber
    \\
    &=& \nabla_{\bi x}\psi(\bi x) \cdot {\bi v}_{\brmn x}(t)
\end{eqnarray}
where $\nabla_{\bi x}\psi(\bi x)$ is the gradient of $\psi(\bi x)$ with respect to $\bi x$. Thus,
\begin{equation}\label{eq:velocity3D}
    {\brmn v}_{\brmn x}(t) = [{\bi v}_{\brmn x}(t), \ \nabla_{\bi x}\psi(\bi x) \cdot {\bi v}_{\brmn x}(t)]\,.
\end{equation}


%

\subsection{Model for the Incident Field}
For a transmitter with isotropic antenna located at $\z$ transmitting a waveform $s(t)$, the propagation of electromagnetic waves in a medium can be described using the scalar wave equation \cite{colton98,ghoshroy02},
\begin{equation}\label{eq:wave_eq}
    [\nabla^2-\frac{1}{c^2}\partial_t^2]E(t,\y)=\delta(\y-\z)s(t)
\end{equation}
where $c$ is the speed of electromagnetic waves in the medium and $E(t,\x)$ is the electric field. Note that this model can be extended to include realistic antenna models in a straightforward manner.

The propagation medium is characterized by the \emph{Green's function}, which satisfies
\begin{equation}
    [\nabla^2-\frac{1}{c^2_0}\partial_t^2]g(t,\y)=-\delta(\y)\delta(t).
\end{equation}
In free-space, the Green's function is given by
\begin{equation}\label{eq:fsgreens}
    g(\y,t)=\frac{\delta(t-\norm{\y}/c_0)}{4\pi\norm{\y}}
\end{equation}
where $c_0$ is the speed of light in vacuum.

Let $\bgamma_T(t)$ be the trajectory of the transmitter and $p(t)$ be the transmitted waveform.
The incident field $E^{\mathrm{in}}$ satisfies the scalar wave equation in (\ref{eq:wave_eq}) where $c$ is replaced by $c_0$ and $\z$ is replaced by $\bgamma_T(t)$:
\begin{equation}
    [\nabla^2-\frac{1}{c^2_0}\partial_t^2]E^{\mathrm{in}}(t,\z)=\delta(\z-\bgamma_T(t))p(t).
\end{equation}

Thus, using (\ref{eq:fsgreens}), we have
\begin{eqnarray}
   E^{\mathrm{in}}(\brmn z, t) = -\int \frac{\delta(t-t'-|\brmn z - \bgamma_T(t')|/c_0)}{4\pi|\brmn z - \bgamma_T(t')|} p(t') \rmd t'.
   \label{eq:IncField}
\end{eqnarray}

\subsection{Models for the Scattered Field and the Received Signal}
Let $E^{\mathrm{sc}}(\brmn y, t)$ denote the scattered field at $\y$ due to the transmitter located at $\bgamma_T(t)$ transmitting waveform $p(t)$. Then, using (\ref{eq:wave_eq}) and under the Born approximation and the assumption of isotropic receiving antenna, we have
\begin{eqnarray}
     E^{\mathrm{sc}}(\brmn y, t) &=& -\int \frac{\delta(t-t'-|\brmn y  - \brmn z|/c_0)}{4\pi|\brmn y  - \brmn z|} \int V(\balpha(\brmn z, t'))
     \nonumber
     \\
     &&  \times \left(-\int \frac{\delta(t'-t''-|\brmn z - \bgamma_T(t'')|/c_0)}{4\pi|\brmn z - \bgamma_T(t'')|}\ddot{p}(t'') \rmd t''\right) \rmd t'\rmd \brmn z.
   \label{eq:ScatField}
\end{eqnarray}

Let $\bgamma_R(t)$ denote the trajectory of the receiver and $r(t)$ denote the received signal at the receiver. Then, we have
\begin{eqnarray}
     r(t) &=& E^{\mathrm{sc}}(\bgamma_R(t),t)
     \nonumber
     \\
     &=& \int \frac{\delta(t-t'-|\bgamma_R(t)  - \brmn z|/c_0)}{4\pi|\bgamma_R(t)  - \brmn z|} V(\balpha(\brmn z, t'))
     \nonumber
     \\
     &&  \times \frac{\delta(t'-t''-|\brmn z - \bgamma_T(t'')|/c_0)}{4\pi|\brmn z - \bgamma_T(t'')|}\ddot{p}(t'') \rmd t'' \rmd t'\rmd \brmn z.
   \label{eq:ScatField1}
\end{eqnarray}

Assuming that the waveform is transmitted starting at time $s$, for a short duration of $t''\in [0, T]$ \footnote{For a typical wideband chirp  pulse, this time interval is in the order of $10^{-6} \mathrm{s}$, while for an ultranarrowband CW signal, it is in the order of $10^{-3} \,\mathrm{s}$ or longer.}, the wave goes out at $s+t'',\,t''\in [0, T]$ from the transmitter, reaches the target at $t'+s$ and arrives at the receiving antenna at $t+s$. Note that $t'',t',t$ are relative time variables within the interval that starts at time $s$. Thus, for this short time interval, using (\ref{eq:ScatField1}), we have
\begin{eqnarray}
     \hspace{-2cm}r(t+s)
     &=& \int \frac{\delta(t-t'-|\bgamma_R(t+s)  - \brmn z|/c_0)}{4\pi|\bgamma_R(t+s)  - \brmn z|} V(\balpha(\brmn z, t'+s))
     \nonumber
     \\
     \hspace{-2cm}&&  \times \frac{\delta(t'-t''-|\brmn z - \bgamma_T(t''+s)|/c_0)}{4\pi|\brmn z - \bgamma_T(t''+s)|}\ddot{p}(t''+s) \rmd t'' \rmd t'\rmd \brmn z.
    \label{eq:ScatField2}
\end{eqnarray}

In \eqref{eq:ScatField2}, we make the following change of variables
\begin{equation}
    \brmn z \rightarrow \brmn x = \balpha(\brmn z, t'+s)=\bGamma^{-1}(\brmn z,t'+s)
\end{equation}
and obtain
\begin{eqnarray}
     \hspace{-2cm}r(t+s)
     &=& \int \frac{\delta(t-t'-|\bgamma_R(t+s)  - \bGamma(\x,t'+s)|/c_0)}{4\pi|\bgamma_R(t+s)  - \bGamma(\x,t'+s)|} V(\brmn x)|\nabla_{\brmn x}\bGamma(\brmn x, t'+s))|
     \nonumber
     \\
     \hspace{-2cm}&&  \times \frac{\delta(t'-t''-|\bGamma(\x,t'+s) - \bgamma_T(t''+s)|/c_0)}{4\pi|\bGamma(\x,t'+s) - \bgamma_T(t''+s)|}\ddot{p}(t''+s) \rmd t'' \rmd t'\rmd \brmn x
    \label{eq:ScatField3}
\end{eqnarray}
where $|\nabla_{\brmn x}\bGamma(\brmn x, t'+s))|$ 
is the determinant of the Jacobian that comes from the change of variables.

We make the assumption that the scatterers are moving linearly and therefore 
\begin{equation}
    \bGamma(\brmn x, t) \approx \brmn x + \v_{\brmn x} t
\end{equation}
where the velocity $\v_{\brmn x}$ is now time independent.
Furthermore, we assume that $|\nabla_\x \bGamma(\x, t)| \approx 1$ since radar scenes are not very compressible. Thus, (\ref{eq:ScatField3}) becomes
\begin{eqnarray}
     \hspace{-2cm}r(t+s)
     &=& \int \frac{\delta(t-t'-|\bgamma_R(t+s)  - (\x + \v_{\x}(t'+s))|/c_0)}{4\pi|\bgamma_R(t+s)  - (\x + \v_{\x}(t'+s))|} V(\brmn x)
     \nonumber
     \\
     \hspace{-2cm}&&  \times \frac{\delta(t'-t''-|\x + \v_{\x}(t'+s) - \bgamma_T(t''+s)|/c_0)}{4\pi|\x + \v_{\x}(t'+s) - \bgamma_T(t''+s)|}\ddot{p}(t''+s) \rmd t'' \rmd t'\rmd \brmn x\,.
    \label{eq:rec_2}
\end{eqnarray}
Note that in (\ref{eq:rec_2})  $\x = [\bi x, \psi(\bi x)]$ and $\v_{\x} = [\bi v_{\x}, \nabla_{\bi x}\psi(\bi x) \cdot \bi v_{\x}]$.

We now define
\begin{eqnarray}\label{eq:Q_def}
    q(\bi x, \bi v) &=&\rho(\bi x)\delta(\bi v-\bi v_{\x})\\
    &\approx&\rho(\bi x)\varphi(\bi v,\bi v_{\x})
\end{eqnarray}
as the \emph{phase-space reflectivity function} 
of the moving scene where $\varphi$ is a differentiable function of $\bi v$ that approximates $\delta(\bi v-\bi v_{\x})$ in the limit. Using (\ref{eq:Q_def}) and (\ref{eq:vapp}), we rewrite (\ref{eq:rec_2}) as follows:
\begin{eqnarray}
     \hspace{-2cm}r(t+s)
     &=& \int \frac{\delta(t-t'-|\bgamma_R(t+s)  - (\x + \v(t'+s))|/c_0)}{4\pi|\bgamma_R(t+s)  - (\x + \v(t'+s))|} q(\bi x, \bi v)
     \nonumber
     \\
     \hspace{-2cm}&&  \times \frac{\delta(t'-t''-|\x + \v(t'+s) - \bgamma_T(t''+s)|/c_0)}{4\pi|\x + \v(t'+s) - \bgamma_T(t''+s)|}\ddot{p}(t''+s) \rmd t'' \rmd t'\rmd \bi x\rmd \bi v\,
    \label{eq:rec_n}
\end{eqnarray}
where $\v = [\bi v, \ \nabla_{\bi x}\psi(\bi x) \cdot \bi v]$.

We next make some approximations to evaluate $t',t''$ integrals in (\ref{eq:rec_n}). First, we make the Taylor series expansions in $ \bgamma_R$ and $ \bgamma_T$ around $t,t''=0$,
\begin{eqnarray}
    \bgamma_R(t+s) &\approx& \bgamma_R(s)+\dot{\bgamma}_R(s)t+\cdots\,,
    \label{eq:bgammaR_app}
    \\
    \bgamma_T(t''+s) &\approx& \bgamma_T(s)+\dot{\bgamma}_T(s)t''+\cdots\,.
    \label{eq:bgammaT_app}
\end{eqnarray}

Next, under the assumptions that
\begin{eqnarray}
    |\x + \v(t'+s) -\bgamma_T(s)|\gg |\v t'|, \ \, |\dot{\bgamma}_T(s)t''|
    \nonumber
    \\
    |\bgamma_R(s)-(\x + \v(t'+s))|\gg |\v t'|, \ \, |\dot{\bgamma}_R(s)t|,
    \label{eq:Vel_assump}
\end{eqnarray}
we approximate
\begin{eqnarray}
    \hspace{-2cm}|\x + \v(t'+s)-\bgamma_T(t''+s)|
    \approx&
    |\x+\v s-\bgamma_T(s)|
    \nonumber
    \\
    & +\widehat{(\x+\v s)-\bgamma_T(s)}\cdot[\v t'-\dot{\bgamma}_T(s)t'']\,,
    \label{eq:RTz_app}
    \\
    \hspace{-2cm}|\bgamma_R(t+s)-(\x + \v(t'+s))|
    \approx&
    |\bgamma_R(s)-(\x+\v s)|
    \nonumber
    \\
    &+\widehat{\bgamma_R(s)-(\x+\v s)}\cdot [\dot{\bgamma}_R(s)t-\v t']\,.
    \label{eq:RRz_app_2}
\end{eqnarray}



Thus, substituting (\ref{eq:RTz_app}) and (\ref{eq:RRz_app_2}) into (\ref{eq:rec_n}) and carrying out  $t''$ and $t'$ integrations, we obtain
\begin{equation}
    \hspace{-1cm} r(t+s)=\int
    \frac{\ddot{p}(\alpha t-\tau+s) q(\bi x,\bi v)}
    {(4\pi)^2 |\bgamma_R(s)-(\x+\v s)| |(\x+\v s)-\bgamma_T(s)|} d\bi x \,d\bi v
    \label{eq:rec_4}
\end{equation}
where the time dilation $\alpha$ is given by
\begin{eqnarray}
    \hspace{-1cm}\alpha &=& \frac{1-\widehat{\bgamma_R(s)-(\x+\v s)} \cdot \dot{\bgamma}_R(s)/c_0}
    {1+\widehat{\bgamma_T(s)-(\x+\v s)} \cdot \dot{\bgamma}_T(s)/c_0}
    \cdot
    \frac{1+\widehat{\bgamma_T(s)-(\x+\v s)} \cdot \v/c_0}
    {1-\widehat{\bgamma_R(s)-(\x+\v s)} \cdot \v/c_0}
    \label{eq:alpha_2}
\end{eqnarray}
and the time delay $\tau$ is given by
\begin{eqnarray}
    \tau &\approx&
    [|\bgamma_T(s)-(\x+\v s)|+|\bgamma_R(s)-(\x+\v s)|]/c_0
    \nonumber
    \\
    &&-[(\widehat{\bgamma_T(s)-(\x+\v s)}+\widehat{\bgamma_R(s)-(\x+\v s)})\cdot \v\,s]/c_0\,.
    \label{eq:tau_app_2}
\end{eqnarray}

We see that the time dilation term $\alpha$ in (\ref{eq:alpha_2}) is the product of two terms. The first term is the Doppler scale factor due to the movement of the transmitting and receiving antennas. The second term is the Doppler scale factor due to the movement of targets.
Similarly, the delay term $\tau$ in (\ref{eq:tau_app_2}) is composed of two terms. The first term represents the \emph{bistatic range} for a target located at $\x+\v s$, while the second term describes the range variation due to the movement of targets.

Note that conventional wideband SAR image formation methods assume that the radar scene is stationary. Therefore, the Doppler scale factor due to the movement of targets is ignored and set to 1. Furthermore, these methods rely on the
``start-stop'' approximation \cite{Jakowatz96,CGM} where the movement of the antennas within each pulse
propagation is neglected. Therefore, the Doppler scale factor induced by the movement of antennas
is also ignored and set to 1. As a result, wideband SAR imaging methods, including the ones developed for moving target imaging \cite{Yang93,Barbarossa92_1,Barbarossa92_2,Soumekh02,Kirscht03,Stuff04,Zhou07,Borcea12,
Werness90,Perry99,Munson00,Munson02,Zhu11,Martorella11,Jakowatz98,Jao01,MGZ05,Hack11,MC12}, ignore the time dilation term $\alpha$ in (\ref{eq:rec_4}) and set it equal to 1 since wideband signals cannot provide high resolution Doppler measurements.


\subsection{Received Signal Model}
For a narrowband waveform, we have
\begin{equation}\label{eq:p_wave}
    p(t)=\rme^{\rmi \omega_0 t}\tilde p(t)
\end{equation}
where $\omega_0$ denotes the carrier frequency and $\tilde{p}(t)$ is the complex envelope of $p$, which is slow varying as a function of $t$ as compared to $\rme^{\rmi \omega_0 t}$.

Substituting (\ref{eq:p_wave}) into (\ref{eq:rec_4}), we obtain
\begin{eqnarray}
    \hspace{-1cm}r(t+s)=-\omega_0^2 \int
    \frac{\tilde{p}(\alpha t-\tau+s)
    \rme^{\rmi \omega_0 (\alpha t-\tau+s)}
    q(\bi x,\bi v)}
    {(4\pi)^2 G_{TR}(s,\bi x,\bi v)} d\bi x d\bi v
    \label{eq:rec_5}
\end{eqnarray}
where $\alpha$ and $\tau$ are as in (\ref{eq:alpha_2}) and (\ref{eq:tau_app_2}) and $G_{TR}$ is the product of the geometrical spreading factors given by
\begin{equation}\label{eq:G_TR}
    G_{TR}(s,\bi x,\bi v)=|\bgamma_R(s)-(\x+\v s)| |\x+\v s-\bgamma_T(s)|\,.
\end{equation}

Note that $\tilde{p}$ is a slow-varying function of time. Therefore, we approximate $\tilde{p}(\alpha t) \approx \tilde{p}(t)$ in the rest of our discussion.
Furthermore, since the speed of the antennas and the scatterers are much less than the speed of light, we approximate (\ref{eq:alpha_2}) as
$\alpha \approx 1+\beta$
where
\begin{eqnarray}
    \hspace{-2cm} \beta =  [\widehat{\bgamma_T(s)-(\x+\v s)} \cdot (\v-\dot{\bgamma}_T(s))
    +\widehat{\bgamma_R(s)-(\x+\v s)} \cdot (\v-\dot{\bgamma}_R(s))]/c_0\,.
\end{eqnarray}
Note that $f_0\beta$, where $f_0=\omega_0/2\pi$, represents the total Doppler frequency induced by the relative radial motion of the antennas and the target. We refer to $-f_0\beta$ as the \emph{bistatic Doppler frequency for moving targets} and denote it with $f_d(s,\bi x,\bi v)$, i.e.,
\begin{eqnarray}\label{eq:fd}
    \fl f_d(s,\bi x,\bi v)=&\frac{f_0}{c_0}[\widehat{\bgamma_T(s)-(\x+\v s)} \cdot (\dot{\bgamma}_T(s)-\v)
    +\widehat{\bgamma_R(s)-(\x+\v s)} \cdot (\dot{\bgamma}_R(s)-\v)].
\end{eqnarray}
Note that in (\ref{eq:G_TR}) and (\ref{eq:fd}), $\x=[\bi x,\psi(\bi x)]$ and $\v=[\bi v,\nabla_{\bi x}\psi(\bi x)\cdot \bi v]$.

\section{Forward Model for Moving Target Imaging}\label{sec:ForwardModel}
In this section, we derive a forward model by correlating the windowed and translated received signal with the scaled or frequency-shifted transmitted waveform, which is a mapping from the four-dimensional position and velocity space to the data space that depends on two variables, translation and  scaling factor. 
We use the forward model to reconstruct the moving targets in two-dimensional position space and to estimate their two-dimensional velocities.

We define the correlation of the received signal given in (\ref{eq:rec_5}) with a scaled or frequency-shifted version of the transmitted signal over a finite time window as follows:
\begin{equation}\label{eq:d_ori}
    d(s,\mu)=\int r(t+s) p^*(\mu t)\phi(t)dt
\end{equation}
for some $s \in \mathbb{R}$ and $\mu \in \mathbb{R}^+$, where $\phi(t)$, $t\in [0,T_{\phi}]$ is a smooth windowing function with a finite support.

Substituting (\ref{eq:rec_5}) into (\ref{eq:d_ori}), we obtain
\begin{eqnarray}
    \hspace{-0.5cm}d(s,\mu)&=&  \int
    \frac{\rme^{\rmi \omega_0(\alpha-\mu)t}\rme^{\rmi \omega_0(s-\tau)}}
    {(4\pi)^2 G_{TR}(s,\bi x,\bi v)}
    \omega_0^4 \tilde{p}(t-\tau+s)
    \tilde{p}^*(t)q(\bi x,\bi v)d\bi x d\bi v dt\,.
    \label{eq:rec_6}
\end{eqnarray}
Note that since  $\tilde{p}(t)$ is a slow-varying function of $t$, we use $\tilde{p}(\mu t)\approx \tilde{p}(t)$ in (\ref{eq:rec_6}).

We define the forward modeling operator, $\mathcal{F}$, as follows: 
\begin{eqnarray}
    d(s,\mu) &  \approx &\mathcal{F}[q](s,\mu)
    \nonumber
    \\
    &:=& \int \rme^{-\rmi \phi(t,\bi x,\bi v,s,\mu)} A(t,\bi x,\bi v,s,\mu)
    q(\bi x,\bi v)d\bi x d\bi v dt
    \label{eq:d_ori2}
\end{eqnarray}
where
\begin{equation}\label{eq:phi}
    \phi(t,\bi x,\bi v,s,\mu)=2\pi f_0 t[(\mu-1)+f_d(s,\bi x,\bi v)/f_0]\,,
\end{equation}
\begin{equation}
    A(t,\bi x,\bi v,s,\mu)=
    \frac{\tilde{p}(t-\tau+s)\tilde{p}^*(t)\rme^{\rmi \omega_0(s-\tau)}\omega_0^4 }
    {(4\pi)^2 G_{TR}(s,\bi x,\bi v)}\,
\end{equation}
and the bistatic Doppler frequency $f_d$ is as defined in (\ref{eq:fd}).

We assume that for some $m_A$, $A$ satisfies the inequality
\begin{eqnarray}
    \hspace{-2cm}&\sup_{(t,\mu,s,\bi x,\bi v)\in \mathcal{U}}
    \left| \partial_t^{\alpha_t} \partial_\mu^{\alpha_\mu}
    \partial_{s}^{\beta_s}
    \partial_{x_1}^{\epsilon_1} \partial_{x_2}^{\epsilon_2}
    \partial_{v_{1}}^{\varepsilon_1} \partial_{v_{2}}^{\varepsilon_2}
    A(t,\bi x,\bi v,s,\mu) \right|
    \leq C_A (1+ t^2)^{(m_A -|\alpha_t|)/2}
    \label{eq:symbol_A12}
\end{eqnarray}
where $\mathcal{U}$ is any compact subset of $\mathbb{R} \times \mathbb{R}^+ \times
\mathbb{R} \times \mathbb{R}^2 \times \mathbb{R}^2$, and the constant $C_A$ depends on
$\mathcal{U},\alpha_{t,\mu},\beta_s$, $\epsilon_{1,2}$, $\varepsilon_{1,2}$.  This assumption is needed in order to make various
stationary phase calculations hold.

\subsection{Leading Order Contributions of the Forward Model}\label{sec:Crit_forward}
Under the assumption (\ref{eq:symbol_A12}), (\ref{eq:d_ori2}) defines $\mathcal{F}$ as a \emph{Fourier integral operator} whose
leading-order contribution comes from the intersection of the illuminated ground topography, the velocity field whose third component lies on the tangent plane of the ground topography and $(\x,\v) \in \mathbb{R}^3 \times \mathbb{R}^3$ that have the same bistatic Doppler frequency.

We denote the four-dimensional manifold formed by this intersection as
\begin{equation}\label{eq:isoDop_4d}
     F(s,\mu)=\{(\bi x,\bi v):f_d(s,\bi x,\bi v)=f_0(1-\mu),\,(\bi x, \bi v) \in \mathrm{supp}(A) \}
\end{equation}
and refer to $F(s,\mu)$ as the \emph{bistatic iso-Doppler manifold}.

In order to visualize the four-dimensional bistatic iso-Doppler manifold for moving targets, we consider the cross-sections of the bistatic iso-Doppler manifold for a constant velocity and a constant position. We define
\begin{equation}\label{eq:isoDop_2dPS}
    F_{\bi v_0}(s,\mu)=\{\bi x: f_d(s,\x,\v_0)=f_0(1-\mu),\, \,(\bi x, \bi v_0) \in \mathrm{supp}(A) \}
\end{equation}
and
\begin{eqnarray}
    F_{\bi x_0}(s,\mu)=\{\bi v: f_d(s,\x_0,\v)=f_0(1-\mu),\, \,(\bi x_0, \bi v) \in \mathrm{supp}(A) \}\,.
    \nonumber
    \\
    \label{eq:isoDop_2dVS}
\end{eqnarray}

\begin{figure}[t!]
    \centering
    \subfigure[]
    {
    \label{fig:bothlinearTra_fixV}
    \includegraphics[width=1.57in]{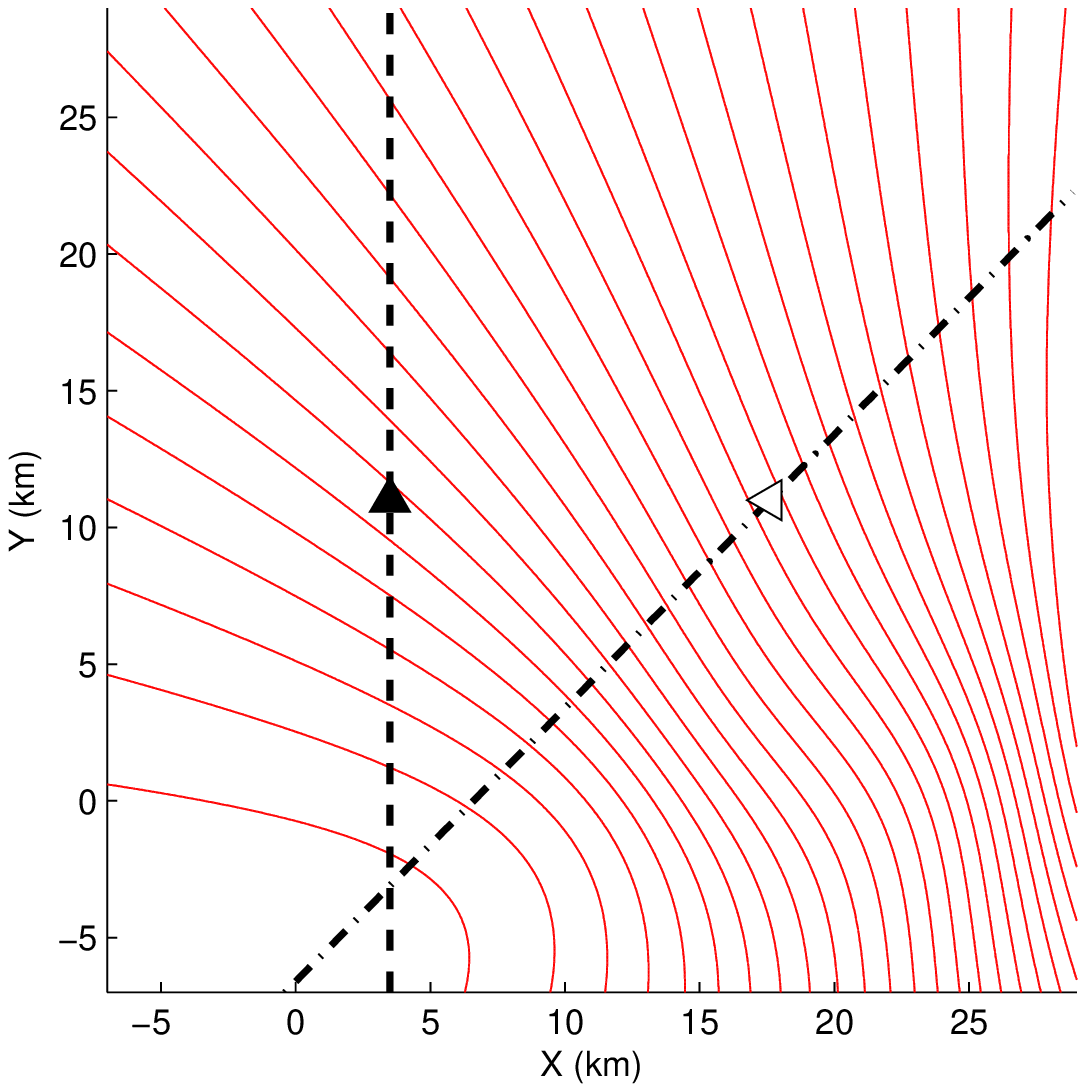}
    }
    \subfigure[]
    {
    \label{fig:linparabTra_fixV}
    \includegraphics[width=1.57in]{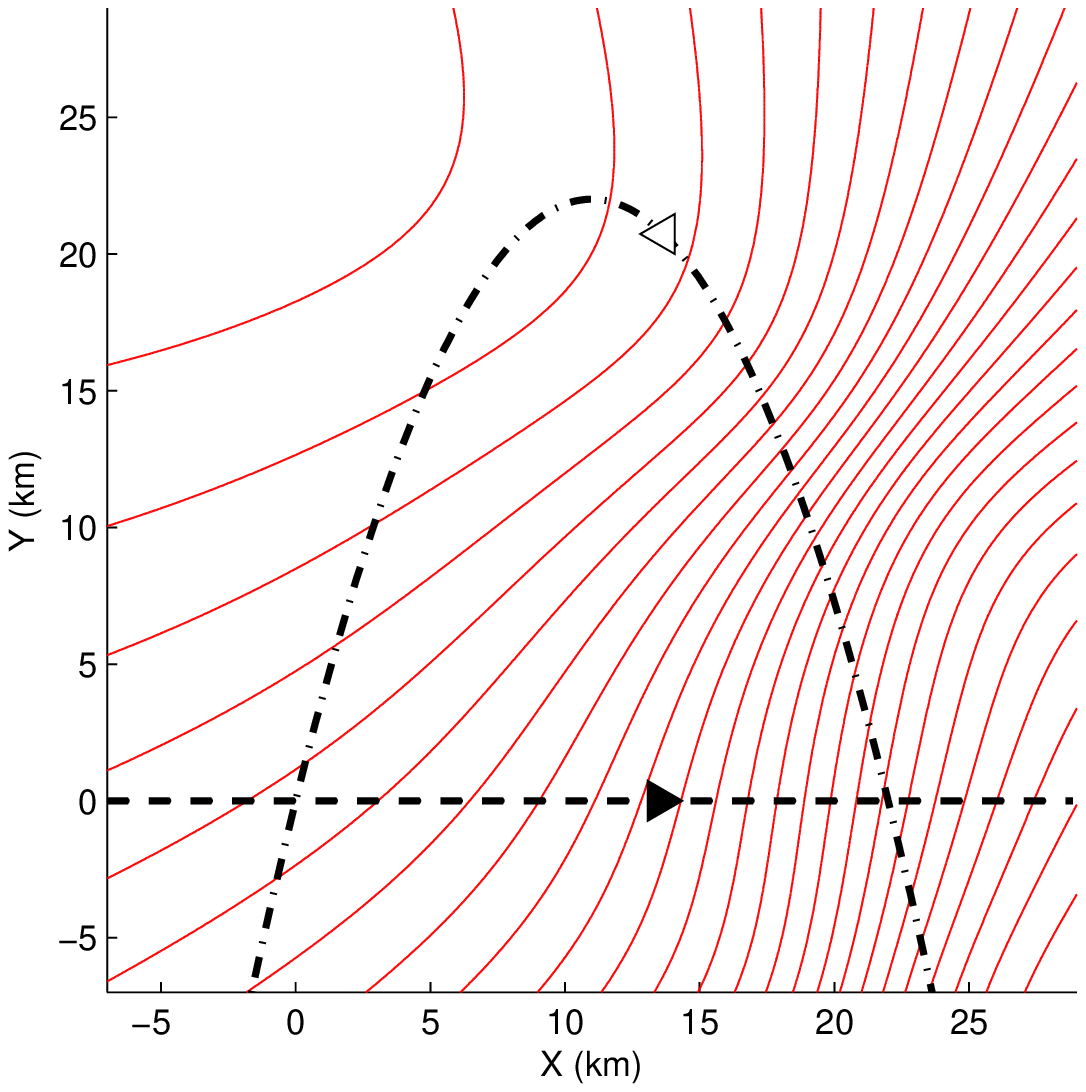}
    }
    \subfigure[]
    {
    \label{fig:bothcircularTra_fixV}
    \includegraphics[width=1.57in]{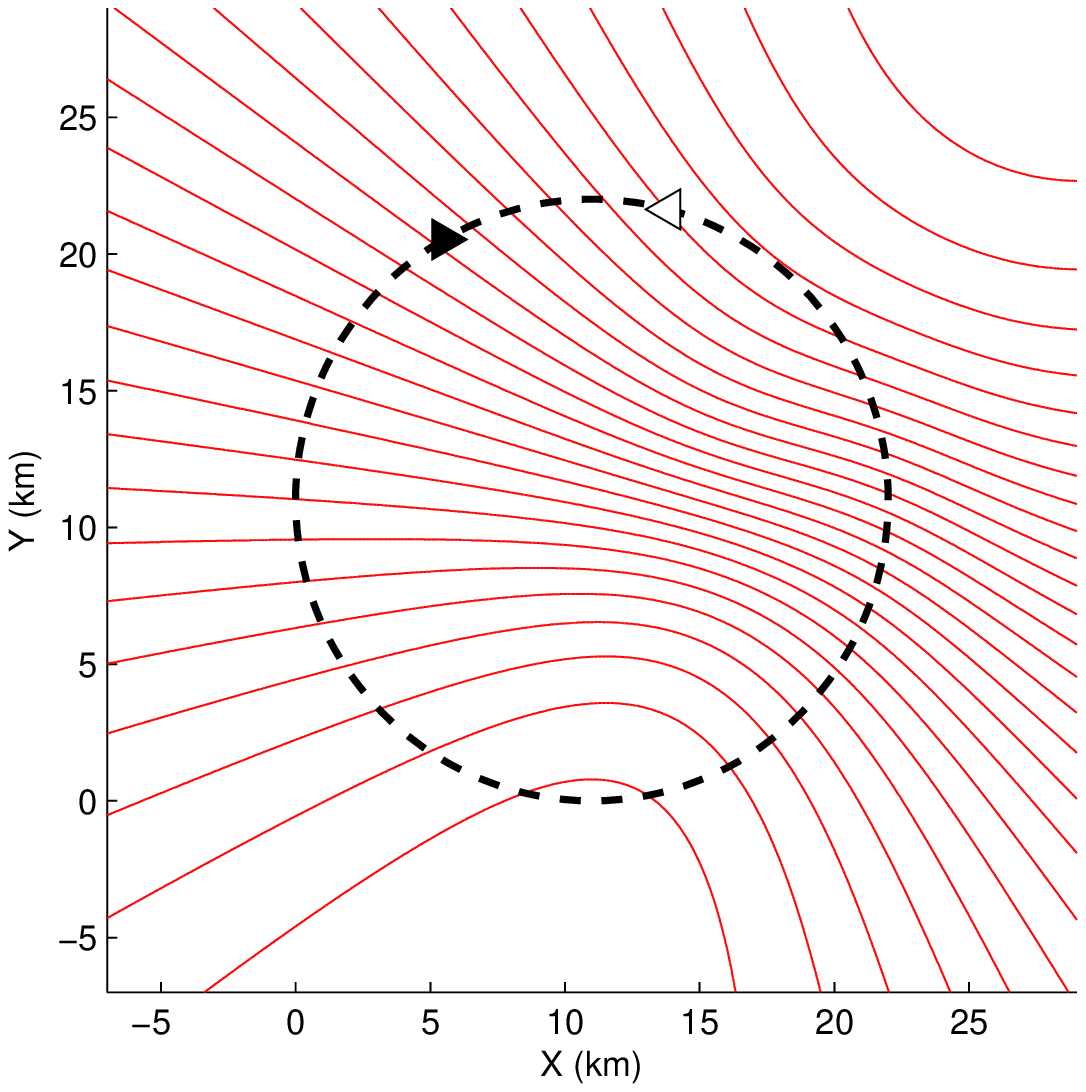}
    }
    \caption{Position-space bistatic iso-Doppler contours determined for a certain $s$ and a fixed $\v_0=[-150,\,150,\,0]\mathrm{m/s}$ for three different transmitter and receiver flight trajectories indicated by the dashed and dash-dot lines, respectively. The black and white triangles denote the corresponding positions of the transmitter and receiver.
    Note that each red curve corresponds to a distinct value of $\mu$.
    \label{fig:fd_contour_PS}}
\end{figure}

\begin{figure}[]
    \centering
    \subfigure[]
    {
    \label{fig:bothlinearTra_fixV}
    \includegraphics[width=1.6in]{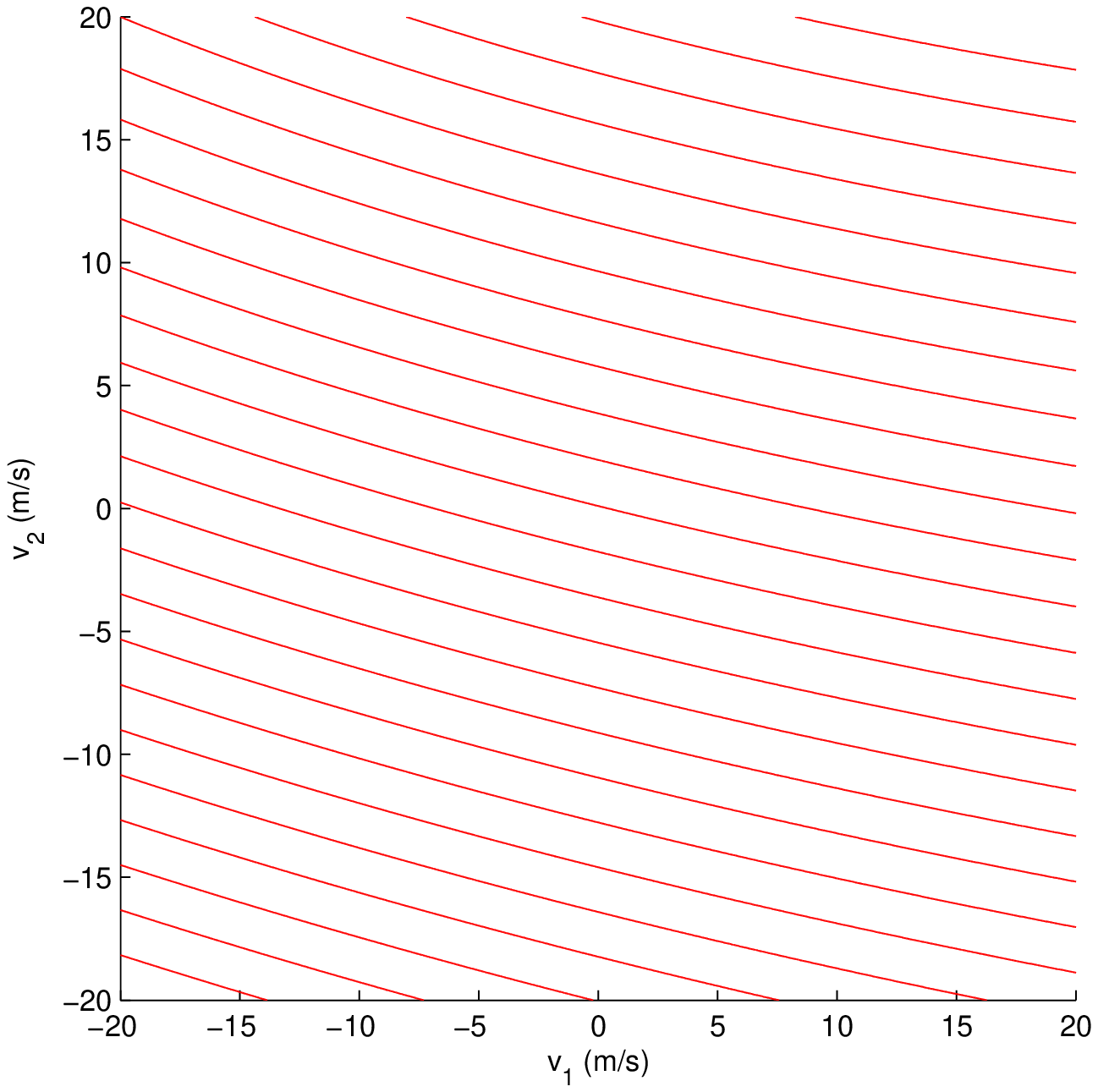}
    }
    \subfigure[]
    {
    \label{fig:linparabTra_fixP}
    \includegraphics[width=1.6in]{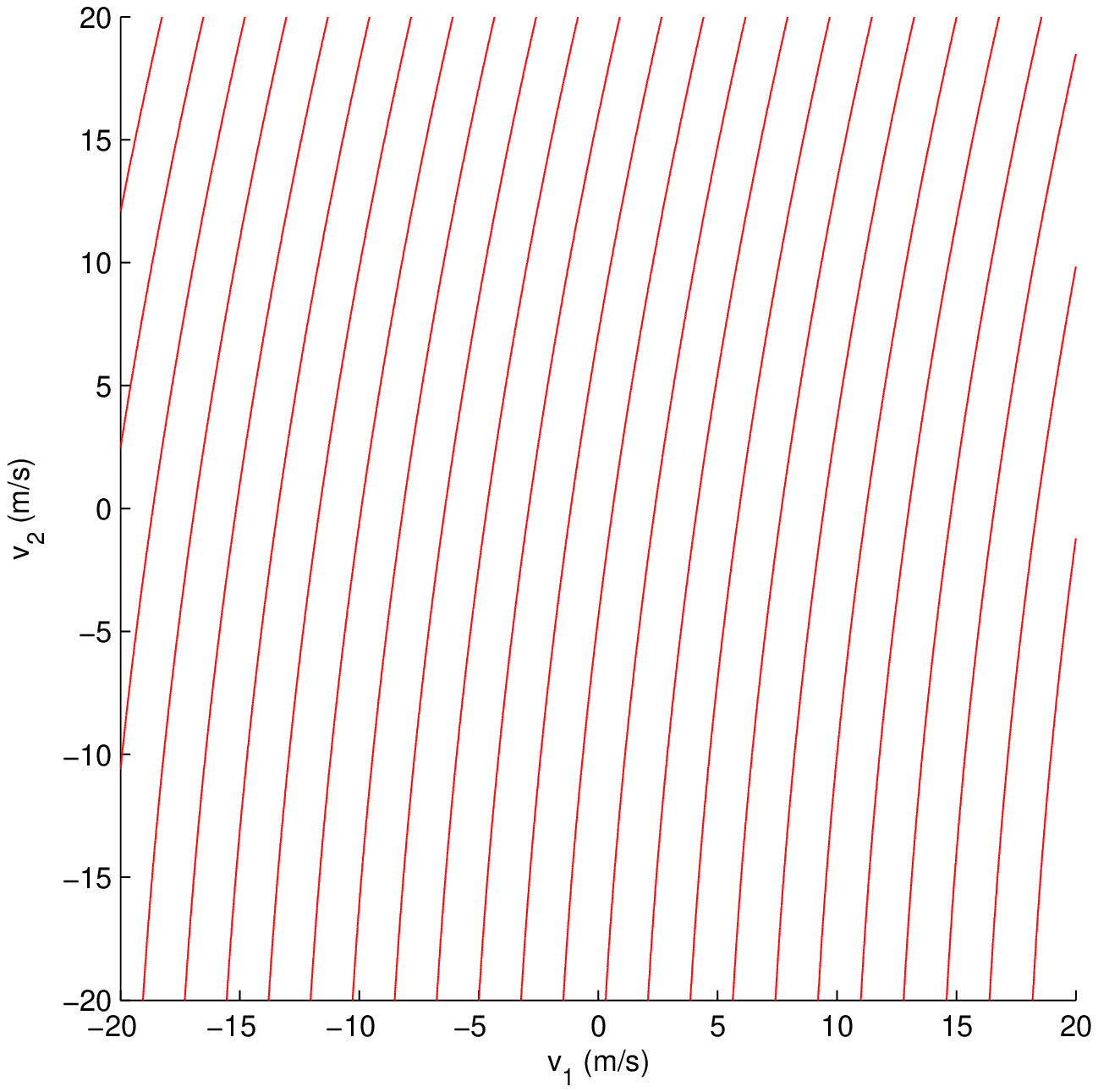}
    }
    \subfigure[]
    {
    \label{fig:bothcircularTra_fixP}
    \includegraphics[width=1.6in]{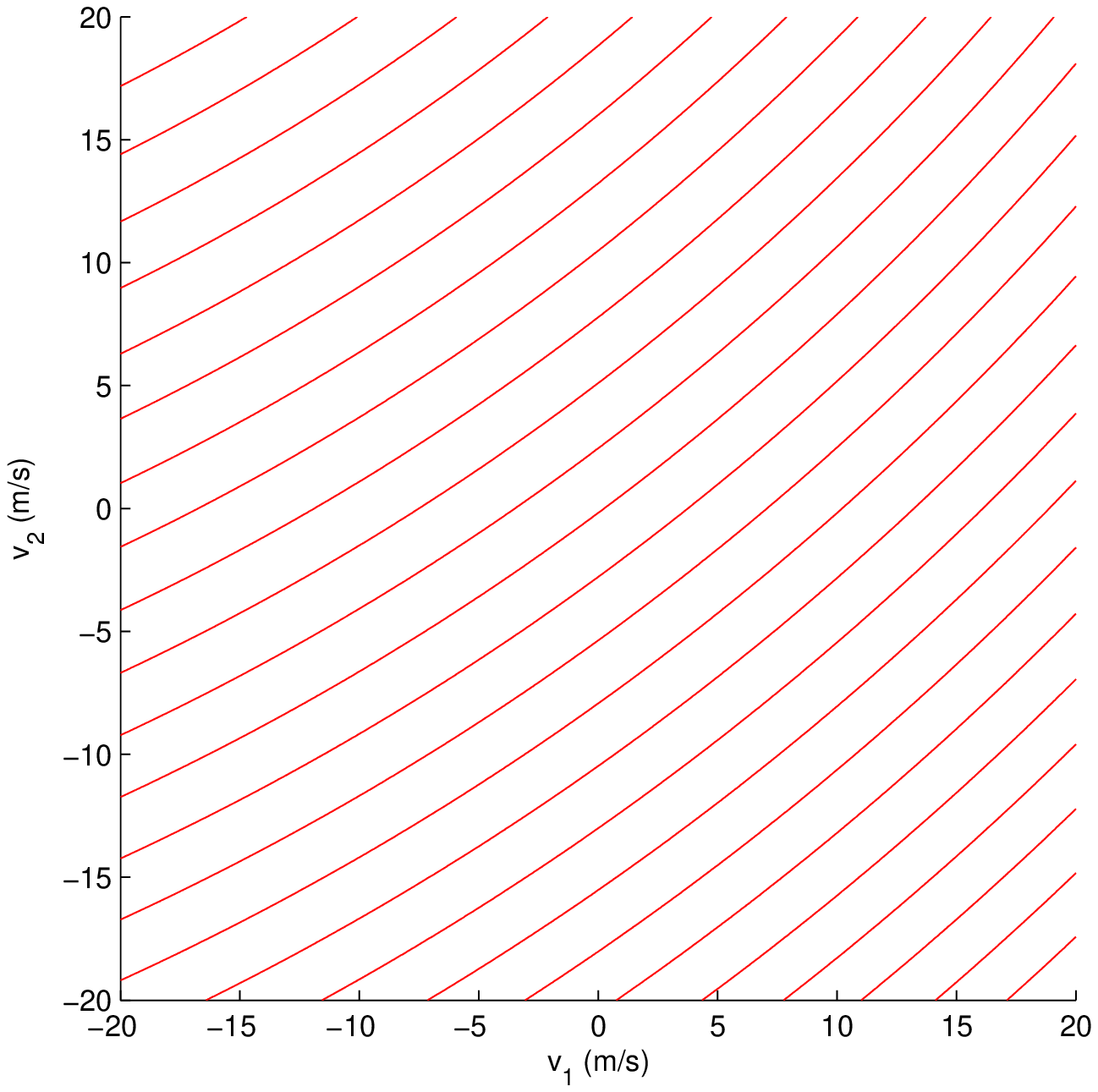}
    }
    \caption{Velocity-space bistatic iso-Doppler contours determined for a certain $s$ and a fixed $\x_0=[5,\,10,\,0]\mathrm{km}$ for three different transmitter and receiver flight trajectories shown in \figref{fig:fd_contour_PS}.
    Note that each red curve corresponds to a distinct value of $\mu$.
    \label{fig:fd_contour_VS}}
\end{figure}

\figref{fig:fd_contour_PS} and \figref{fig:fd_contour_VS} show the position-space and velocity-space bistatic iso-Doppler contours for three different flight trajectories over a flat topography:
(a) The transmitter and receiver are both traversing straight linear flight trajectories. $\bgamma_T(s)=[3.5,s,6.5]\mathrm{km}$ and $\bgamma_R(s)=[(s-7),s,6.5]\mathrm{km}$ where $s=vt$ with speed $v=261\,\mathrm{m}/\mathrm{s}$.
(b) The transmitter is traversing a straight linear flight trajectory, $\bgamma_R(s)=[s,0,6.5]\mathrm{km}$ and the receiver is traversing a parabolic flight trajectory, $\bgamma_T(s)=[s, (s-11)^2*22/121,6.5]\mathrm{km}$ where $s=vt$ with speed $v=261\,\mathrm{m}/\mathrm{s}$.
(c) The transmitter and receiver are both traversing a circular flight trajectory. $\bgamma_T(s)=\bgamma_C(s)$ and $\bgamma_R(s)=\bgamma_C(s-\pi/4)$ where $\bgamma_C(s)=[11+11\cos s,11+11\sin s,6.5]\mathrm{km}$ with $s=\frac{v}{R}t$ where speed $v=261\,\mathrm{m}/\mathrm{s}$ and radius $R=11\mathrm{km}$.


\section{Image Formation}\label{sec:imaging}
A natural choice to form phase-space reflectivity images would be to use a filtered-back projection (FBP) type imaging operator that filters and backprojects the data onto the four-dimensional bistatic iso-Doppler manifolds introduced in Section \ref{sec:ForwardModel}.
Ideally, we wish to reconstruct a phase-space reflectivity image so that the point spread function of the imaging operator is an approximate Dirac-delta function in both position and velocity spaces. 
However, since the data is two-dimensional and the phase-space reflectivity is four-dimensional, it may not be possible to obtain such a point spread function by backprojecting onto the four-dimensional bistatic iso-Doppler manifolds. 

Therefore, we assume that the velocity is constant, say $\bi v_h$, and reconstruct a set of two-dimensional reflectivity images in position space only for a range of hypothesized velocities. We refer to each image as the $\bi v_h$\emph{-reflectivity image} and form it by an FBP-type imaging operator, where we filter and backproject the data onto the \emph{position-space} bistatic iso-Doppler contours, i.e., the cross sections of the bistatic iso-Doppler mainfolds for a range of hypothesized velocities. We show that whenever the hypothesized velocity is equal to the correct velocity for a scatterer, the scatterer can be reconstructed at the correct location in the position space. We design the FBP filter to ensure that the reconstructed reflectivity for a scatterer has the correct strength whenever the hypothesized velocity is equal to the true velocity of the scatterer. From this set of images, we estimate the velocity of the scatterers using a figure of merit that measures the degree to which the images are focused. The reflectivity images corresponding to the estimated velocities provide focused images of the moving scatterers present in the scene. 

Below we introduce the FBP operator in forming the $\bi v_h$-reflectivity images, analyze its point spread function, and next present the design of the FBP filter. Finally, we describe how to determine the velocity of moving targets.

\subsection{$\bi v_{h}$-Reflectivity Image Formation}\label{sec:ReflectFBP}
We form the $\bi v_{h}$-reflectivity image $q_{\bi v_h}(\bi z)$ for a fixed hypothesized velocity $\v_h = [\bi v_h, \nabla_{\bi z}\psi(\bi z) \cdot \bi v_h]$ by filtering and backprojecting the data onto the position-space iso-Doppler contour $F_{\bi v_h}(s,\mu)$:
\begin{eqnarray}
    q_{\bi v_h}(\bi z)& :=\mathcal{K}_{\bi v_h}[d](\bi z)
    \nonumber
    \\
    & =\int \rme^{\rmi \phi_{\bi v_h}(t,\bi z,s,\mu)}Q_{\bi v_h}(\bi z,t,s)d(s,\mu)dtdsd\mu,
    \label{eq:rho_fbpoperator}
\end{eqnarray}
where $\mathcal{K}_{\bi v_{h}}$ is the filtered-backprojection operator for the fixed velocity $\bi v_h$,
 \begin{equation}\label{Phase_Kv}
    \phi_{\bi v_h}(t,\bi z,s,\mu) = \phi(t,\bi z, \bi v_h, s,\mu)
\end{equation}
and $Q_{\bi v_h}$ is the filter to be determined below. Note that $\bi v_h$ is a fixed parameter for $\phi_{\bi v_h}$ and $Q_{\bi v_h}$.

We assume that for some $m_{Q_{\bi v_h}}$, $Q_{\bi v_h}$ satisfies the inequality
\begin{eqnarray}
    \hspace{-2cm}&\sup_{(t,s,\bi z)\in \mathcal{U}}
    \left| \partial_t^{\alpha_t}
    \partial_{s}^{\beta_s}
    \partial_{z_1}^{\epsilon_1} \partial_{z_2}^{\epsilon_2}
    Q_{\bi v_h}(\bi z,t,s) \right|
    \leq C_{Q_{\bi v_h}} (1+ t^2)^{(m_{Q_{\bi v_h}} -|\alpha_t|)/2}
    \label{eq:symbol_Q12}
\end{eqnarray}
where $\mathcal{U}$ is any compact subset of $\mathbb{R} \times \mathbb{R}^+\times \mathbb{R}^2$, and the constant $C_{Q_{\bi v_h}}$ depends on
$\mathcal{U},\alpha_{t},\beta_s$, $\epsilon_{1,2}$. Under the assumption (\ref{eq:symbol_Q12}),  (\ref{eq:rho_fbpoperator}) defines $\mathcal{K}_{\bi v_{h}}$ as a Fourier integral operator.



\subsection{PSF Analysis}

Substituting (\ref{eq:d_ori2}) into (\ref{eq:rho_fbpoperator}), we rewrite (\ref{eq:rho_fbpoperator}) as
\begin{eqnarray}
    q_{\bi v_h}(\bi z)& :=\mathcal{K}_{\bi v_h}\mathcal{F}[q](\bi z,\bi v_h)
    \nonumber
    \\
    & =\int L^{\bi v_{\x}}_{\bi v_h}(\bi z,\bi x) q(\bi x,\bi v_{\x}) d \bi x
    \label{eq:rho_fbp_psf}
\end{eqnarray}
where $L^{\bi v_{\x}}_{\bi v_h}(\bi z,\bi x)$ is the \emph{Point Spread Function} (PSF) of the two-dimensional reflectivity imaging operator for the hypothesized velocity $\bi v_h$ with respect to the true velocity $\bi v_{\x}$ given by
\begin{eqnarray}
    L^{\bi v_{\x}}_{\bi v_h}(\bi z,\bi x) & =\int \rme^{\rmi [\phi_{\bi v_h}(t,\bi z,s,\mu)-\phi(t,\bi x,\bi v_{\x},s,\mu)]}
    \nonumber
    \\
    & \hspace{0.5cm} \times Q_{\bi v_h}(\bi z,t,s)A(t',\bi x,\bi v_{\x},s,\mu)
    dt ds d\mu dt'\,.
    \label{eq:psf}
\end{eqnarray}

We define
\begin{eqnarray}
    & \hspace{-1cm} \Phi_k=\phi_{\bi v_h}(t,\bi z,s,\mu)-\phi(t,\bi x,\bi v_{\x},s,\mu)
    \nonumber
    \\
    & \hspace{-0.55cm} =2\pi t[(\mu-1)f_0+f_d(s,\bi z,\bi v_h)]-2\pi t'[(\mu-1)f_0+f_d(s,\bi x,\bi v_{\x})]\,.
\end{eqnarray}

Applying the stationary phase theorem to approximate the $t'$ and $\mu$ integrations in (\ref{eq:psf}) \footnote[1]{The determinant of the Hessian of $\Phi_k$ is $(2\pi)^2f_0^2$. Thus, the stationary points are non-degenerate.}, we obtain
\begin{eqnarray}
    & \partial_{t'}\Phi_k  =-2\pi [(\mu-1)f_0+f_d(s,\bi x,\bi v_{\x})]=0
    \nonumber
    \\
    & \hspace{-0.5cm} \Longrightarrow  \mu =1-\frac{f_d(s,\bi x,\bi v_{\x})}{f_0}\,,
    \\
    & \partial_{\mu}\Phi_k  =2\pi(t-t')f_0=0
    \nonumber
    \\
    & \hspace{-0.5cm} \Longrightarrow t=t'\,.
\end{eqnarray}
Substituting the results back into (\ref{eq:psf}), we get the kernel of the image fidelity operator $\mathcal{K}_{\bi v_h}\mathcal{F}$:
\begin{eqnarray}
    & \hspace{-1cm}L^{\bi v_{\x}}_{\bi v_h}(\bi z,\bi x)
    \approx \int \rme^{\rmi 2 \pi t[f_d(s,\bi z,\bi v_h)-f_d(s,\bi x,\bi v_{\x})]}
    \nonumber
    \\
    & \hspace{1cm}
    \times Q_{\bi v_h}(\bi z,t,s)A(t,\bi x,\bi v_{\x},s,1-f_d(s,\bi x,\bi v_{\x})/f_0)
    dt\,d s\,.
    \label{eq:psf2}
\end{eqnarray}
To simplify our notation, we let
\begin{equation}\label{eq:A}
    A(t,\bi x,\bi v_{\x},s)=A(t,\bi x,\bi v_{\x},s,1-f_d(s,\bi x,\bi v_{\x})/f_0)\,.
\end{equation}

The main contribution to $L^{\bi v_{\x}}_{\bi v_h}$ comes from the critical points of the phase of $\mathcal{K}_{\bi v_h}\mathcal{F}$ that satisfy the conditions\cite{T}:
\begin{eqnarray}
    \partial_t(2 \pi t[f_d(s,\bi z,\bi v_h)-f_d(s,\bi x,\bi v_{\x})])=0
    \nonumber
    \\
    \Longrightarrow f_d(s,\bi z,\bi v_h)=f_d(s,\bi x,\bi v_{\x})\,,
    \\
    \partial_s(2 \pi t[f_d(s,\bi x,\bi v_{\x})-f_d(s,\bi z,\bi v_h)])=0
    \nonumber
    \\
    \Longrightarrow \dot{f}_d(s,\bi z,\bi v_h)=\dot{f}_d(s,\bi x,\bi v_{\x})
\end{eqnarray}
where $\dot{f}_d(s,\bi x,\bi v_{\x})$ denotes the first-order derivative of $f_d(s,\bi x,\bi v_{\x})$ with respect to time $s$, i.e., $\dot{f}_d(s,\bi x,\bi v_{\x})=\partial f_d(s,\bi x,\bi v_{\x})/\partial s$. We refer to $\dot{f}_d(s,\bi x,\bi v_{\x})$ as the \emph{bistatic Doppler-rate}.

Using (\ref{eq:fd}), we obtain
\begin{eqnarray}
    \fl \dot{f}_d(s,\bi x,\bi v_{\x})&=\frac{f_0}{c_0}
    \left[\frac{1}{|\bgamma_T(s)-(\x+\v_{\x} s)|}|(\dot{\bgamma}_{T}(s)-\v_{\x})_\perp|^2
    +\widehat{(\bgamma_T(s)-(\x+\v_{\x} s))}\cdot \ddot{\bgamma}_T(s)\right.
    \nonumber
    \\
    \fl & \left.
    +\frac{1}{|\bgamma_R(s)-(\x+\v_{\x} s)|}|(\dot{\bgamma}_{R}(s)-\v_{\x})_\perp|^2
    +\widehat{(\bgamma_R(s)-(\x+\v_{\x} s))}\cdot \ddot{\bgamma}_R(s)\right]
    \label{eq:Doprate2}
\end{eqnarray}
where
\begin{eqnarray}\label{eq:gamaperp}
    \fl (\dot{\bgamma}_{T,R}(s)-\v_{\x})_\perp & =(\dot{\bgamma}_{T,R}(s)-\v_{\x})-
    \nonumber
    \\
    & \hspace{-1cm} (\widehat{\bgamma_{T,R}(s)-(\x+\v_{\x} s)}) [\widehat{(\bgamma_{T,R}(s)-(\x+\v_{\x} s))}\cdot (\dot{\bgamma}_{T,R}(s)-\v_{\x})]
\end{eqnarray}
denotes the projection of the relative velocity
$\dot{\bgamma}_{T,R}(s)-\v_{\x}$ onto the plane whose normal vector is along
$\widehat{\bgamma_{T,R}(s)-(\x+\v_{\x} s)}$.
Note that in (\ref{eq:Doprate2}) $\x=[\bi x,\psi(\bi x)]$ and $\v_{\x}=[\bi v_{\x},\nabla_{\bi x}\psi(\bi x)\cdot \bi v_{\x}]$.

In (\ref{eq:Doprate2}), the summation of the first two terms in the square bracket corresponds to the
relative radial acceleration between the transmitter and the target located at $\x+\v_{\x} s$ at time $s$, 
while the summation of the last two terms in the square bracket corresponds to the
relative radial acceleration between the receiver and the target located at $\x+\v_{\x} s$ at time $s$.
For the derivation of (\ref{eq:Doprate2}), see \ref{app:Appendix1}.

We refer to the locus of the points formed by the intersection of the
illuminated surface, $[\bi x,\psi(\bi x)]$, the velocity field, $[\bi v, \nabla_{\bi x}\psi(\bi x)\cdot \bi v]$, and the set $\{(\x,\v) \in \mathbb{R}^3 \times \mathbb{R}^3: \dot{f}_d(s,\z,\v)=C\}$, for some constant $C$, as the \emph{bistatic iso-Doppler-rate manifold} and denote it by
\begin{equation}
    \dot{F}(s, C)=\{(\bi x,\bi v): \dot{f}_d(s,\bi x,\bi v)=C,\, (\bi x, \bi v) \in \mathrm{supp}(A) \}\,.
\end{equation}

We consider the cross-sections of the bistatic iso-Doppler-rate manifold for a constant velocity and a constant position and define
\begin{equation}\label{eq:isoDoprate_2dPS}
    \dot{F}_{\bi v_0}(s,C)=\{\bi x: \dot{f}_d(s,\bi x,\bi v_0)=C ,\, (\bi x, \bi v_0) \in \mathrm{supp}(A) \}
\end{equation}
and
\begin{equation}\label{eq:isoDoprate_2dVS}
    \dot{F}_{\bi x_0}(s,C)=\{\bi v: \dot{f}_d(s,\x_0,\bi v)=C ,\,  (\bi x_0, \bi v) \in \mathrm{supp}(A) \}\,.
\end{equation}
(\ref{eq:isoDoprate_2dPS}) specifies an iso-Doppler-rate contour in the two-dimensional position space. We refer to this contour as the \emph{position-space bistatic iso-Doppler-rate contour for moving targets}. Similarly,  (\ref{eq:isoDoprate_2dVS}) specifies an iso-Doppler-rate contour in the two-dimensional velocity space. We refer to this contour as the \emph{velocity-space bistatic iso-Doppler-rate contour for moving targets}.

\begin{figure}[t!]
    \centering
    \subfigure[]
    {
    \label{fig:bothlinearTraR_fixV}
    \includegraphics[width=1.57in]{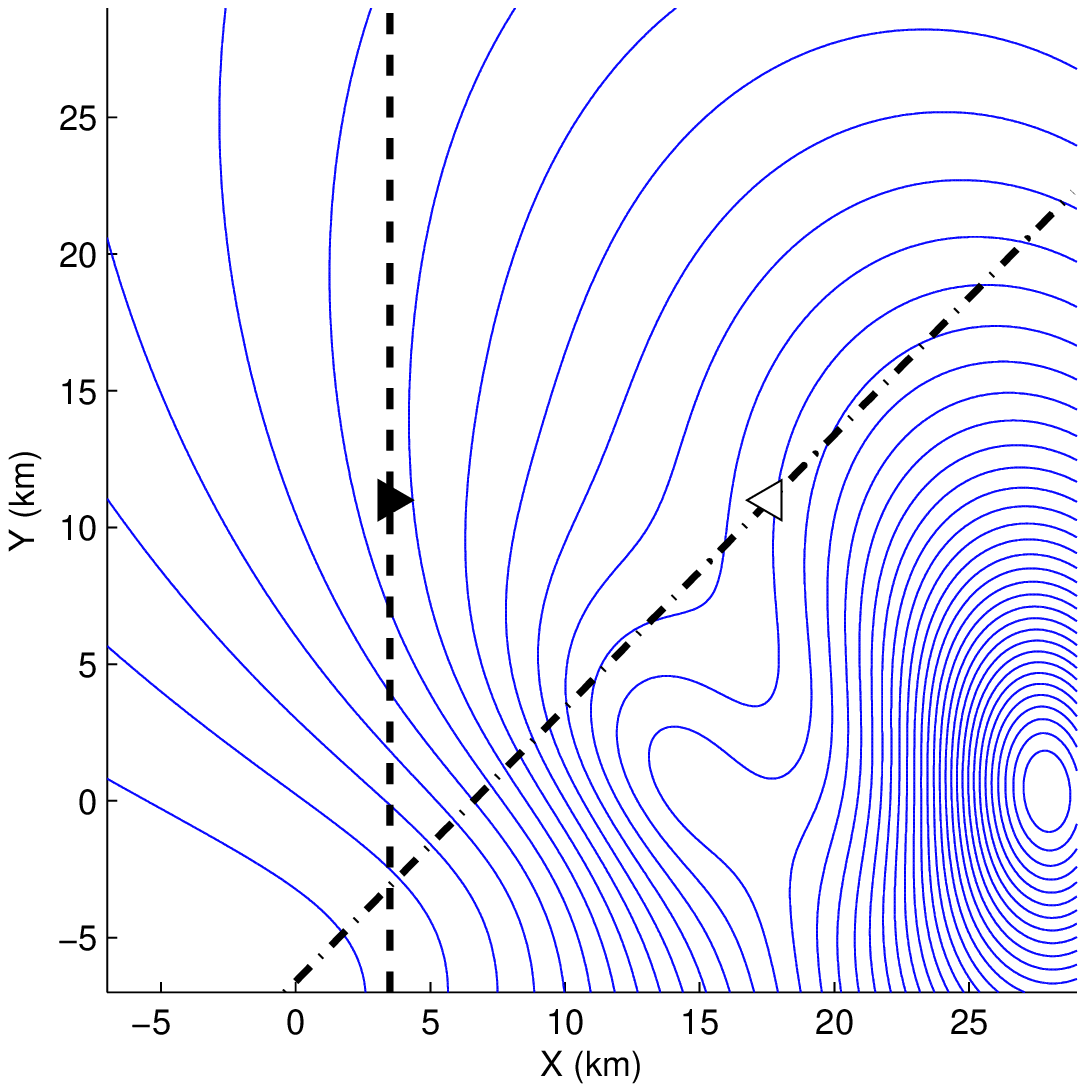}
    }
    \subfigure[]
    {
    \label{fig:linparabTraR_fixV}
    \includegraphics[width=1.57in]{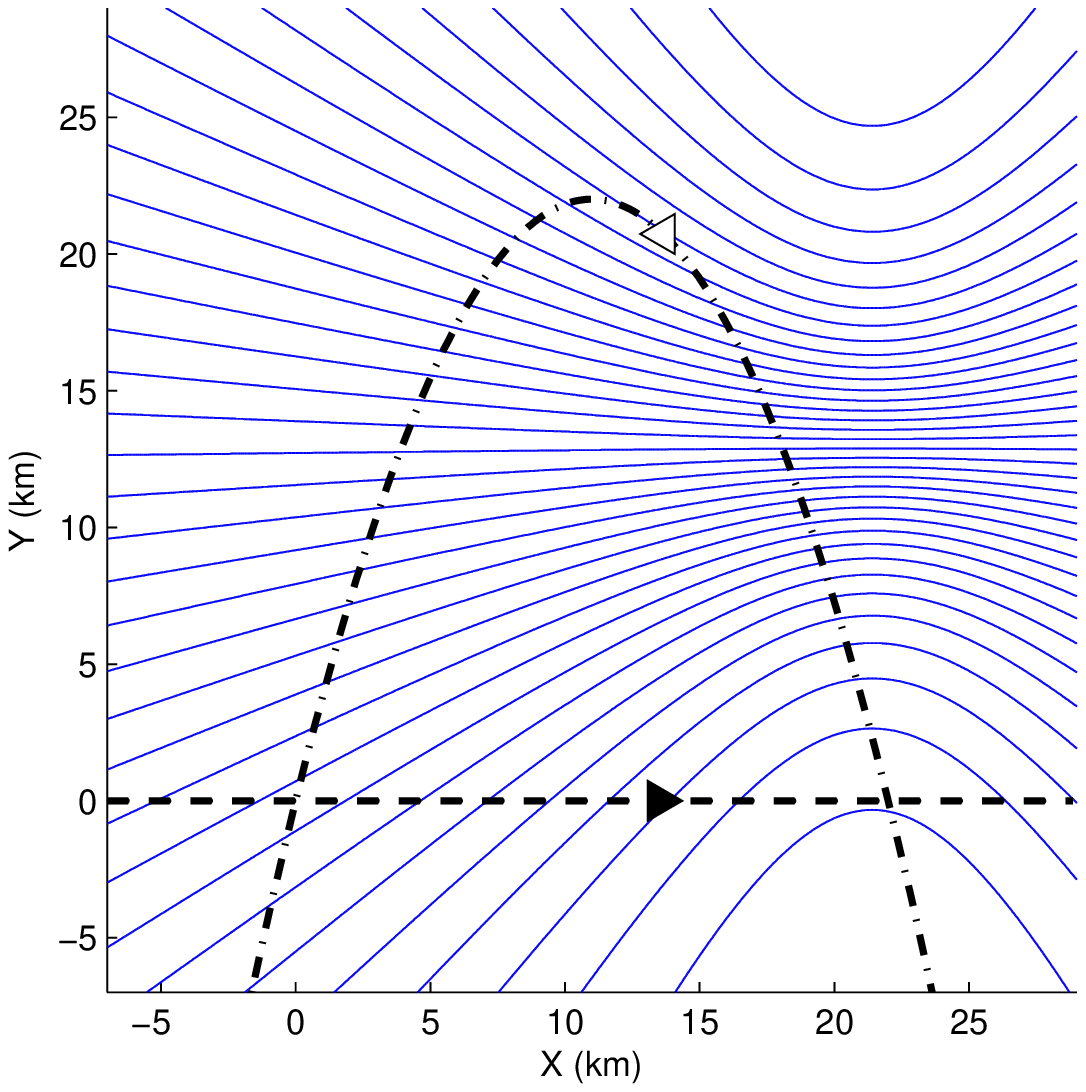}
    }
    \subfigure[]
    {
    \label{fig:bothcircularTraR_fixV}
    \includegraphics[width=1.57in]{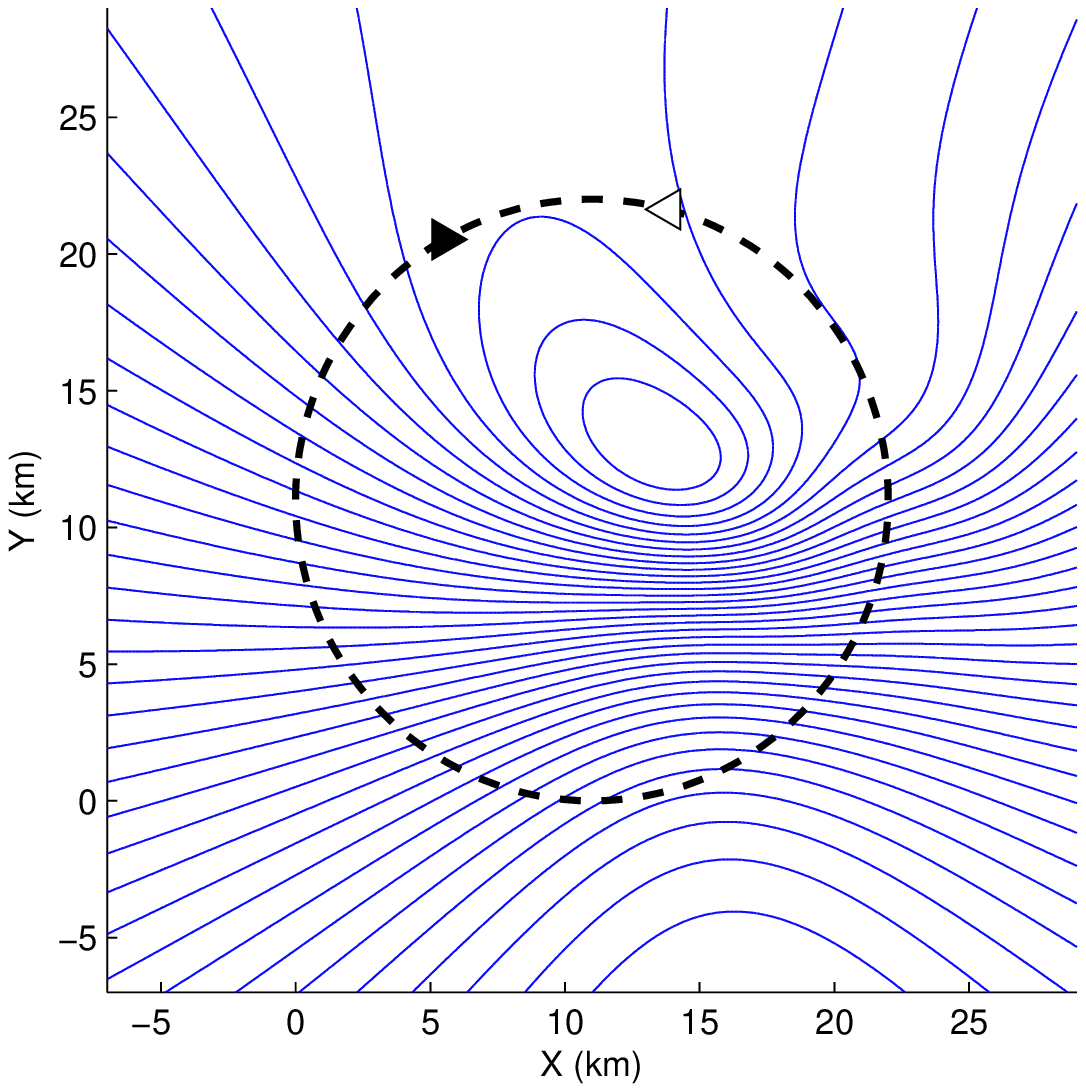}
    }
    \caption{Position-space bistatic iso-Doppler-rate contours determined for a certain $s$ and a fixed $\v_0=[-150,\,150,\,0]\mathrm{m/s}$ for three different transmitter and receiver flight trajectories as described in \ref{sec:Crit_forward}. The black and white triangles denote the corresponding positions of the transmitter and receiver.
    Note that each blue curve corresponds to a distinct value of $C$.
    \label{fig:fdrate_contour_PS}}
\end{figure}

\begin{figure}[]
    \centering
    \subfigure[]
    {
    \label{fig:bothlinearTraR_fixP}
    \includegraphics[width=1.6in]{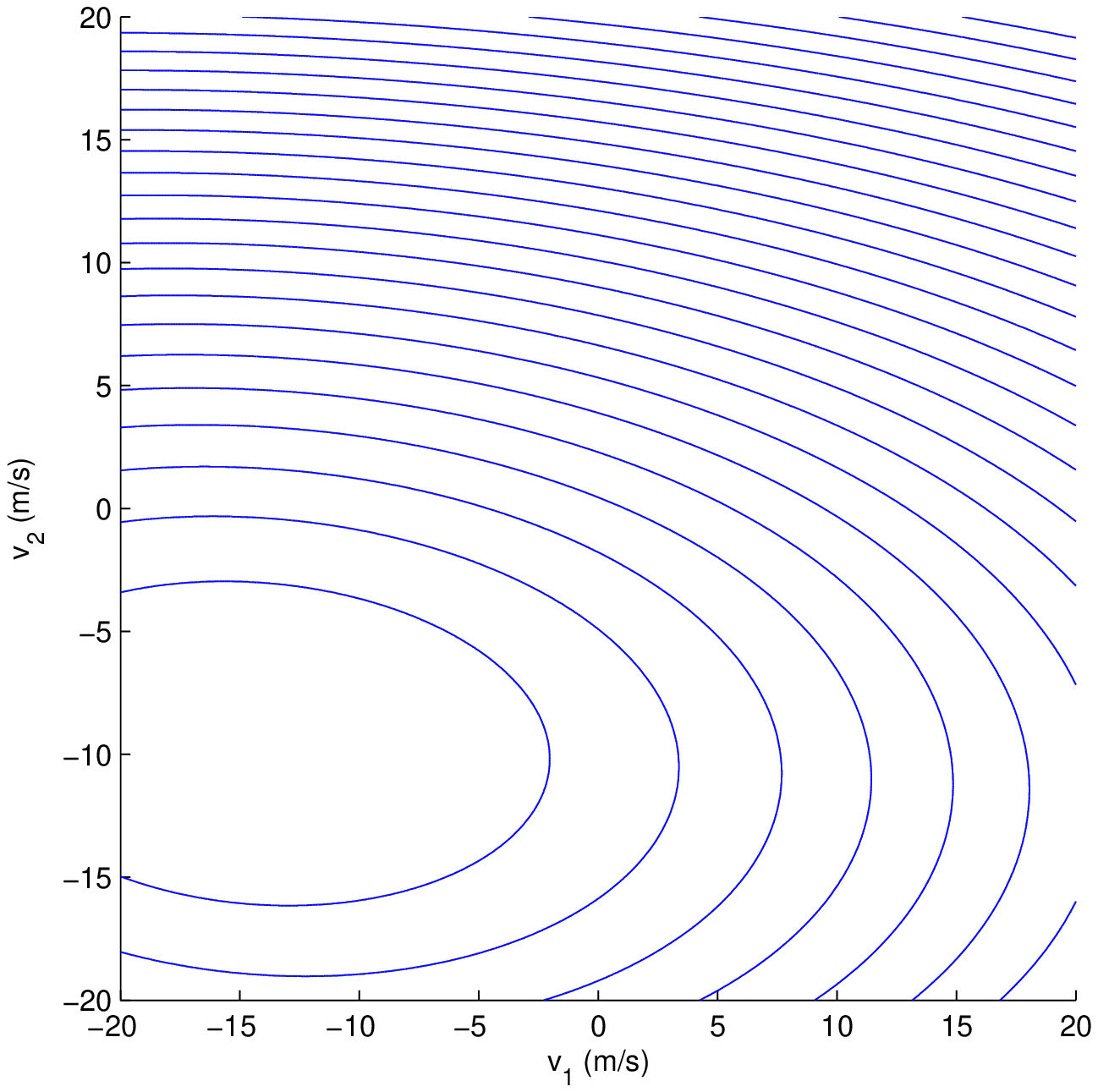}
    }
    \subfigure[]
    {
    \label{fig:linparabTraR_fixP}
    \includegraphics[width=1.6in]{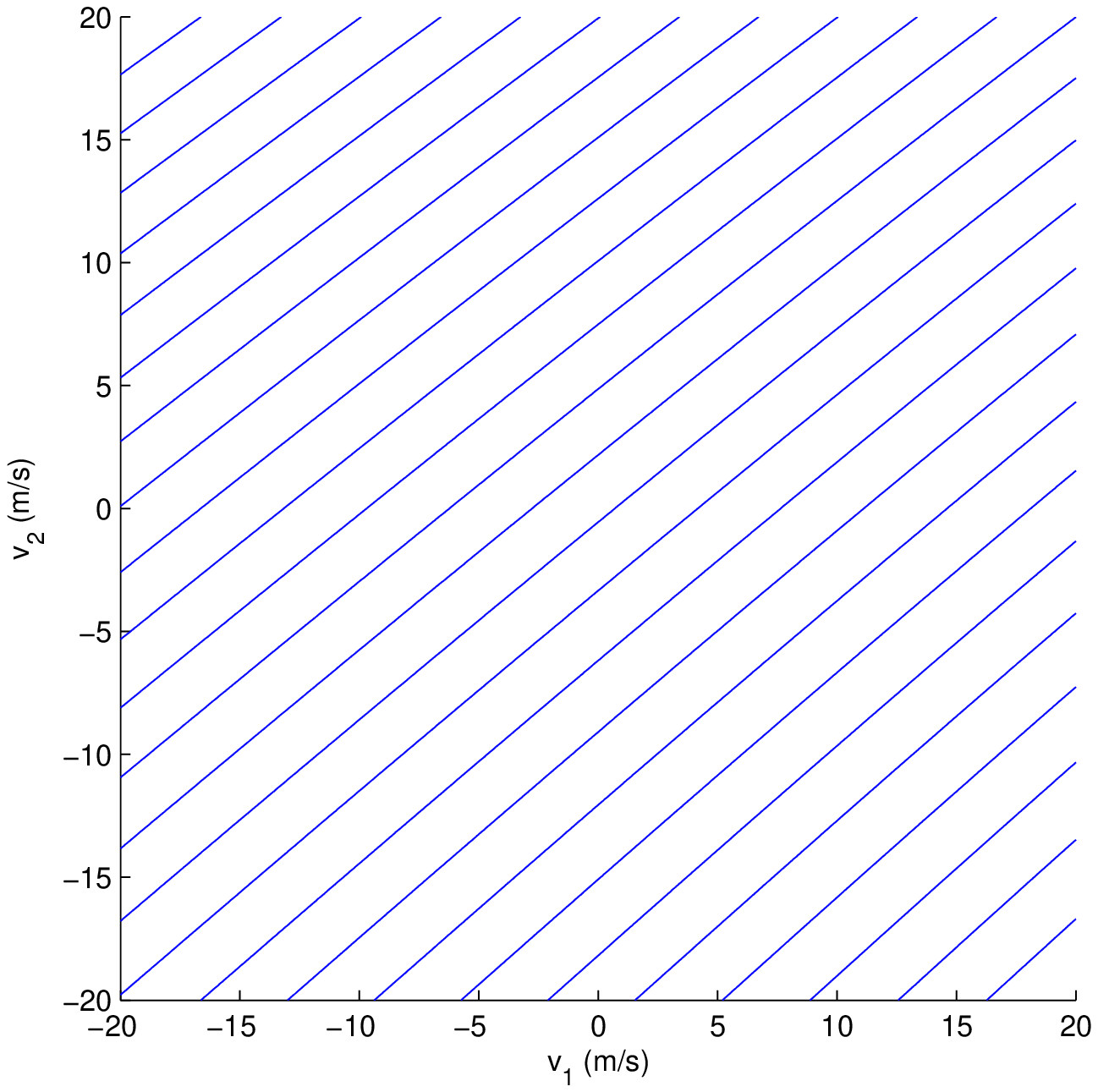}
    }
    \subfigure[]
    {
    \label{fig:bothcircularTraR_fixP}
    \includegraphics[width=1.6in]{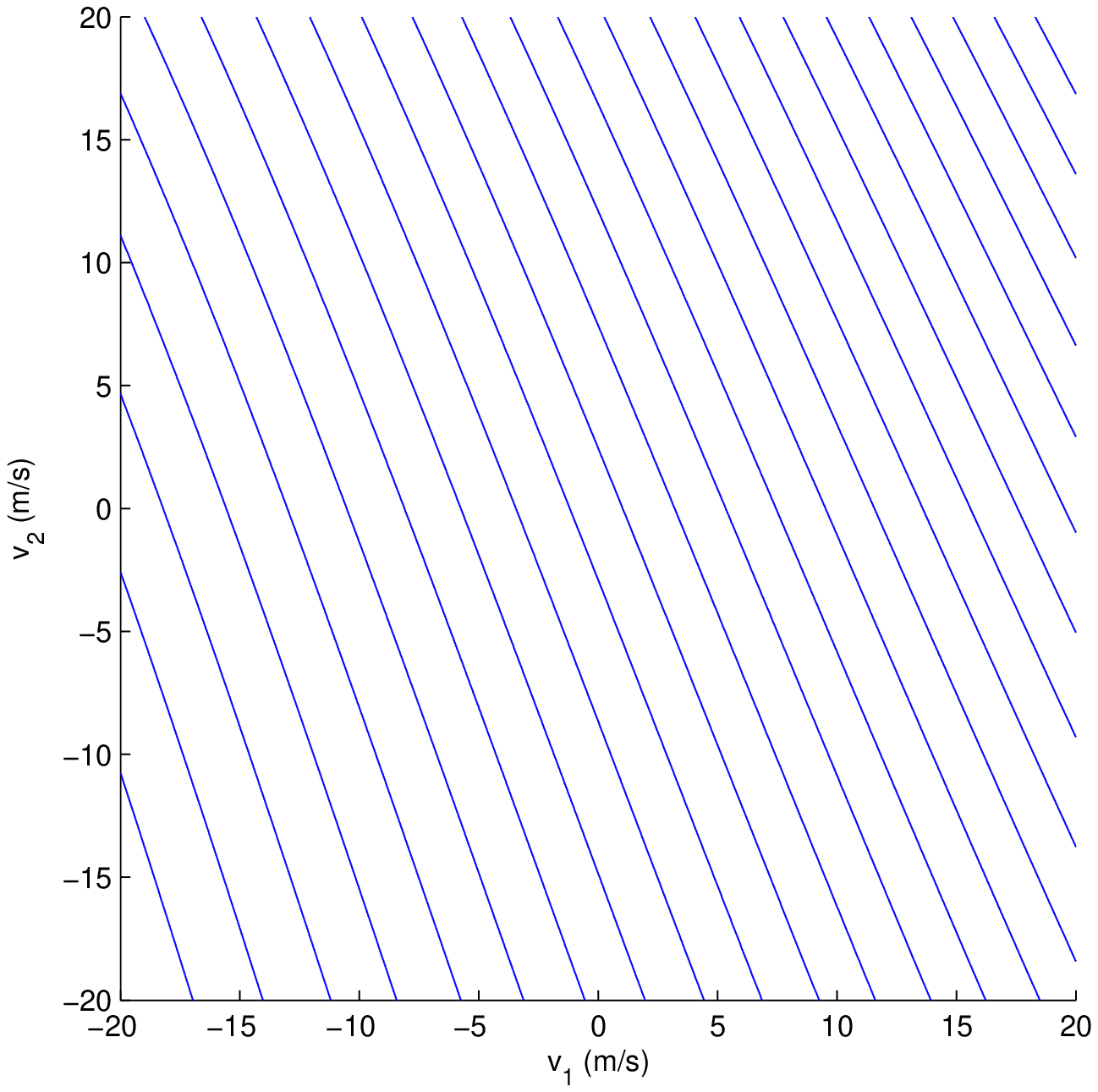}
    }
    \caption{Velocity-space bistatic iso-Doppler-rate contours determined for a certain $s$ and a fixed $\x_0=[5,\,10,\,0]\mathrm{km}$ for three different transmitter and receiver flight trajectories as described in \ref{sec:Crit_forward}. Note that each blue curve corresponds to a distinct value of $C$.
    \label{fig:fdrate_contour_VS}}
\end{figure}

\figref{fig:fdrate_contour_PS} and \figref{fig:fdrate_contour_VS} show the position-space bistatic iso-Doppler-rate contours and velocity-space bistatic iso-Doppler-rate contours for three different flight trajectories over a flat topography that are described in Section  \ref{sec:Crit_forward}.

The critical points of the phase of $\mathcal{K}_{\bi v_h}\mathcal{F}$ that contribute to the reflectivity image formation are those points that lie at the intersection of the position-space bistatic iso-Doppler contours, $F_{\bi v_h}(s,\mu)$ and position-space bistatic iso-Doppler-rate contours, $\dot{F}_{\bi v_h}(s,C)$. For the correct velocity, i.e., $\bi v_h=\bi v_{\x}$, this intersection contributes to the reconstruction of the true target \footnote[2]{We assume that the flight trajectory and the illumination patterns are chosen such that the intersection has a single element avoiding any right-left type of ambiguities.}.
Note that when $\bi v_h \neq \bi v_{\x}$, the points lying at the aforementioned intersection may lead to the artifacts in the reconstructed reflectivity image. 

\subsection{Determination of the FBP Filter}
We determine $Q_{\bi v_h}$ so that the PSF of the two-dimensional reflectivity imaging operator, $L^{\bi v_{\x}}_{\bi v_h}(\bi z,\bi x)$ 
is as close as possible to the Dirac-delta function, $\delta(\bi z-\bi x)$ for $\bi v_h=\bi v_{\x}$, i.e., whenever the reflectivity at $\bi z$ is reconstructed for the correct target velocity $\bi v_{\x}$. 
We assume that at the correct target velocity, the flight trajectory and the illumination pattern are chosen such that the only contribution to $L^{\bi v_{\x}}_{\bi v_{h}}(\bi z,\bi x)$ comes from those points $\bi z=\bi x$.

Thus, we linearize $f_d(s,\bi z,\bi v_{h})$ around $\bi z=\bi x $ for $\bi v_h=\bi v_\x$ and approximate
\begin{eqnarray}
    f_d(s,\bi z,\bi v_{h})-f_d(s,\bi x,\bi v_{h}) \approx (\bi z-\bi x)\cdot \nabla_{\bi z}f_d(s,\bi z,\bi v_h) 
    \,.
\end{eqnarray}
We write
\begin{equation}
    A(t,\bi z,\bi v_{h},s)\approx A(t,\bi x,\bi v_{\x},s) \,.
\end{equation}
Thus, (\ref{eq:psf2}) becomes
\begin{equation}\label{eq:psf_rho1}
    L^{\bi v_{h}}_{\bi v_{h}}(\bi z,\bi x)
    =\int \rme^{\rmi  t (\bi z-\bi x)\cdot \bXi_{\bi v_{h}}(s,\bi z)}
    Q_{\bi v_h}(\bi z,t,s)A(t,\bi z,\bi v_{h},s)dt\,d s
\end{equation}
where
\begin{equation}\label{eq:Xi}
    \bXi_{\bi v_{h}}(s,\bi z)=2\pi \nabla_{\bi z}f_d(s,\bi z,\bi v_{h})
    \,.
\end{equation}
For each $\bi z$, we make the following change of variables:
\begin{equation}\label{eq:xi}
    (t,s)\rightarrow \bxi= t\bXi_{\bi v_{h}}(s,\bi z)
\end{equation}
and write (\ref{eq:psf_rho1}) as follows:
\begin{equation}\label{eq:psf_rho2}
    L^{\bi v_{h}}_{\bi v_{h}}(\bi z,\bi x)
    =\int_{\Omega_{\bi v_{h},\bi z}} \rme^{\rmi  (\bi z-\bi x)\cdot \bxi}
    Q_{\bi v_h}(\bi z,\bxi)A(\bi z,\bi v_{h}, \bxi)\eta(\bi z,\bi v_{h},\bxi)d\bxi
\end{equation}
where
\begin{eqnarray}
    Q_{\bi v_h}(\bi z,\bxi)=Q_{\bi v_h}(\bi z,t(\bxi),s(\bxi))\,,
    \\
    A(\bi z,\bi v_{h}, \bxi)=A(t(\bxi),\bi z,\bi v_{h},s(\bxi))\,
\end{eqnarray}
and
\begin{equation}\label{eq:det}
    \eta(\bi z,\bi v_{h},\bxi)=\frac{\partial(t,s)}{\partial \bxi}=|t|^{-1}
    \left|
    \det
    \left [
    \begin{array}{c}
    \bXi_{\bi v_{h}}(s,\bi z)
    \\
    \partial_{s}\bXi_{\bi v_{h}}(s,\bi z)
    \end{array}
    \right ]
    \right|^{-1}
\end{equation}
is the determinant of the Jacobian that comes from the change of variables given in (\ref{eq:xi}).

The domain of integration in (\ref{eq:psf_rho2}) is given by
\begin{equation}\label{eq:Omega}
    \Omega_{\bi v_{h},\bi z}=\{\bxi= t \bXi_{\bi v_{h}}(s,\bi z)\,|\,A(t,\bi z,\bi v_{h},s)\neq 0,\quad t,s \in \mathbb{R}\}.
\end{equation}
We refer to $\Omega_{\bi v_{h},\bi z}$ as the \emph{data collection manifold} at $\bi z$ for $\bi v_h = \bi v_{\x}$. 
This set determines many of the properties of the reconstructed reflectivity image when $\bi v_h=\bi v_{\x}$.

Using (\ref{eq:Xi}) and (\ref{eq:fd}), we obtain
\begin{eqnarray}
    \fl \bXi_{\bi v_{h}}(s,\bi z)
    &= & -\frac{2\pi f_0}{c_0} \left \{
    [D+D^2 s]\cdot
    \left[
    \frac{(\dot{\bgamma}_{T}(s)-\v_{h})_\perp}{|\bgamma_T(s)-(\z+\v_{h} s)|}
    +
    \frac{(\dot{\bgamma}_{R}(s)-\v_{h})_\perp}{|\bgamma_R(s)-(\z+\v_{h} s)|}\right]\right.
    \nonumber
    \\
    \fl && \hspace{2.25cm}\left. +D^2 \cdot [\widehat{(\bgamma_T(s)-(\z+\v_{h} s))}+\widehat{(\bgamma_R(s)-(\z+\v_{h} s))}] \right\}
    \label{eq:bXi2}
\end{eqnarray}
where
\begin{equation}
    D=\left[
    \begin{array}{ccc}
    1&0&\partial \psi(\bi z)/\partial z_1\\
    0&1&\partial \psi(\bi z)/\partial z_2
    \end{array}
    \right ],
    \label{eq:D}
\end{equation}
\begin{equation}
    D^2=\left[
    \begin{array}{ccc}
    0&0&\frac{\partial^2 \psi(\bi z)}{\partial^2 z_1}v_{h1}+\frac{\partial^2 \psi(\bi z)}{\partial z_2\partial z_1}v_{h2}\\
    0&0&\frac{\partial^2 \psi(\bi z)}{\partial z_1 \partial z_2}v_{h1}+\frac{\partial^2 \psi(\bi z)}{\partial^2 z_2}v_{h2}
    \end{array}
    \right ]
    \label{eq:D2}
\end{equation}
and
$(\dot{\bgamma}_{T,R}-\v_{h})_\perp$ is the projection of $\dot{\bgamma}_{T,R}-\v_{h}$ onto the plane whose normal is $\widehat{\bgamma_{T,R}(s)-(\z+\v_{h} s)}$ as defined by (\ref{eq:gamaperp}).
Note that $\v_{h}=[\bi v_h,\nabla_{\bi z}\psi(\bi z) \cdot \bi v_h]$, $\bi v_h=[v_{h1},v_{h2}]$.
For the derivation of (\ref{eq:bXi2}), see \ref{app:bXi}.

To approximate the point spread function $L^{\bi v_{h}}_{\bi v_{h}}(\bi z,\bi x)$ in (\ref{eq:psf_rho2}) with the Dirac-delta function, we choose the filter as follows:
\begin{equation}\label{eq:Q_{v}}
    Q_{\bi v_h}(\bi z,\bxi)=\frac{\chi_{\Omega_{\bi v_h,\bi z}}}{\eta(\bi z,\bi v_h,\bxi)}\frac{A^*(\bi z,\bi v_h, \bxi)}{|A(\bi z,\bi v_h, \bxi)|^2}
\end{equation}
where $\chi_{\Omega_{\bi v_h,\bi z}}$ is a smooth cut-off function that prevents division by zero in (\ref{eq:Q_{v}}).

With this choice of filter, the resulting FBP operator can recover not only the correct position and orientation of a scatterer, but also the correct reflectivity at $\x$ whenever $\bi v_h=\bi v_{\x}$ in the $\bi v_h$-reflectivity image.

\subsection{Determination of the Velocity Field}\label{sec:imageContrast}
The filtered-backprojection of data results in a set of reflectivity images $q_{\bi v_h}$ in the two-dimensional position space corresponding to a range of velocity values that is suitably chosen for ground moving targets. When the hypothesized velocity $\bi v_{h}$ is equal to the correct velocity $\bi v_{\x}$, the corresponding $\bi v_{h}$-reflectivity image is focused at $\bi x$. We measure the degree to which the reflectivity images are focused with the image contrast measure \cite{Martorella2005_2,Jakowatz98}  and generate a contrast-image as follows:
\textbf{\begin{equation}\label{eq:con_I}
    I(\bi v_h)= \frac{\mathcal{M}[|q_{\bi v_h}-\mathcal{M}[q_{\bi v_h}]|^2]}{|\mathcal{M}[q_{\bi v_h}]|^2}
\end{equation}}
\noindent where $\bi v_h=[v_{h1},v_{h2}]$ is the index of the contrast-image  
and $\mathcal{M}[\cdot]$ denotes the sample mean 
over the spatial coordinates. Note that the image contrast can be viewed as the ratio of the standard deviation to the mean of the $\bi v_{h}$-reflectivity image. This figure-of-merit was previously used in \cite{Martorella2005_2,Jakowatz98} to determine target velocities from a stack of images for the conventional SAR moving target imaging. 

If there are multiple moving targets with different velocities in the scene, the contrast-image could have several peaks each one corresponding to the velocity of a different moving target. We accordingly detect the moving targets and determine their velocities by detecting the local maxima in the contrast-image $I(\bi v_h)$. A threshold can be used in the detection, which may be determined using the Constant False Alarm Rate (CFAR) criterion \cite{skolnik}.

In practice, the discretized and estimated velocity may deviate from the true velocity. In the following two sections, we analyze the velocity resolution and the error in the reflectivity image reconstruction due to error in the estimated velocity.

\section{Resolution Analysis}
\label{sec:ResoAna}
In this section, we analyze the resolution of reconstructed reflectivity images and the velocity resolution available in the collected data.
Our resolution analysis results are consistent with the Doppler ambiguity theory of ultra-narrowband CW signals \cite{radarsignal}. 


\subsection{Resolution of Reflectivity Images at the Correct Target Velocity}\label{sec:ResoAnaRef}

To determine the resolution of the reconstructed reflectivity images, we analyze the bandwidth of the PSF associated with the image fidelity operator $\mathcal{K}_{\bi v_{h}}\mathcal{F}$ at the correct target velocity.

Substituting (\ref{eq:Q_{v}}) into (\ref{eq:psf_rho2}) and the result back into (\ref{eq:rho_fbp_psf}), we obtain
\begin{eqnarray}\label{eq:image1}
    q_{\bi v_{h}}(\bi z)&:=\mathcal{K}_{\bi v_{h}}\mathcal{F}[q](\bi z)
    \nonumber
    \\
    &=\int_{\Omega_{\bi v_{h},\bi z}} {\rme}^{{\rmi}(\bi z-\bi x) \cdot \bxi} q(\bi x, \bi v_{\x})d\bi x d\bxi.
\end{eqnarray}
(\ref{eq:image1}) shows that the image $q_{\bi v_{h}}(\bi z)$ is a band-limited version of $q$ whose bandwidth is determined by the data collection manifold $\Omega_{\bi v_{h},\bi z}$ whenever the hypothesized velocity is equal to the true velocity. The larger the data collection manifold, the better the resolution of the reconstructed reflectivity image becomes.
Furthermore, as indicated by (\ref{eq:Omega}) and (\ref{eq:bXi2}), the band-width contribution of $\bxi$ to the reflectivity image at $\bi z$ is given by
{\begin{eqnarray}\label{eq:band_bxi}
    &&\hspace{-1cm}
    \frac{2\pi f_0}{c_0}L_{\phi}\left |[D+D^2 s]\cdot
    \left[
    \frac{(\dot{\bgamma}_{T}(s)-\v_{h})_\perp}{|\bgamma_T(s)-(\z+\v_{h} s)|}
    +
    \frac{(\dot{\bgamma}_{R}(s)-\v_{h})_\perp}{|\bgamma_R(s)-(\z+\v_{h} s)|}\right]
    \right.
    \nonumber
    \\
    && \hspace{3cm}\left.
    + 2\cos \frac{\theta_{TR}(\z, \v_h, s)}{2} D^2 \cdot \hat{b}_{TR}(\z, \v_h, s)\right|
\end{eqnarray}
where $L_{\phi}$ denotes the length of the support of $\phi(t)$, $\hat{b}_{TR}(\z, \v_h, s)$ denotes the unit vector in the direction of $[\widehat{(\bgamma_T(s)-(\z+\v_{h} s))}+\widehat{(\bgamma_R(s)-(\z+\v_{h} s))}]$ and $\theta_{TR}(\z, \v_h, s)$ denotes the bistatic angle formed by the transmitter and receiver with respect to the target located at $(\z +\v_{h} s)$ at time $s$. $D$ and $D^2$ are as described in (\ref{eq:D}) and (\ref{eq:D2}).

(\ref{eq:band_bxi}) shows that
as the carrier frequency of the transmitted signal $f_0$ becomes higher, the magnitude of
$\bxi$ gets larger, which results in higher resolution reflectivity image of the moving target.
Furthermore, (\ref{eq:band_bxi}) shows that the resolution depends on the range of the antenna to the moving target via the terms $|\bgamma_T(s)-(\z+\v_{h} s)|$ and $|\bgamma_R(s)-(\z+\v_{h} s)|$; and the relative speed between the transmitter (receiver) and the moving target via the terms $(\dot{\bgamma}_{T}(s)-\v_{h})_\perp$ and $(\dot{\bgamma}_{R}(s)-\v_{h})_\perp$. As the antennas move away from the target, or the relative speed decreases in certain directions, the magnitude of $\bxi$ decreases, which results in reduced resolution.
Additionally, larger number of processing windows, i.e., $s$ samples, used for imaging leads to a larger data collection manifold, and hence better resolution.
As indicated by the second line of (\ref{eq:band_bxi}), the resolution of the reflectivity image also depends on the bistatic angle $\theta_{TR}(\z, \v_h, s)$.
Larger the $\theta_{TR}(\z, \v_h, s)$, lower the resolution becomes.


We emphasize again that this analysis holds only for those reconstructed scatterers at $\bi x$ whose velocity $\bi v_{\x}$ is equal to the hypothesized velocity $\bi v_h$.

We summarize the parameters that affect the resolution of the reconstructed moving target image in Table II.
\addtocounter{table}{0}
\begin{table}[t]
\linespread{1.1}
\centering
\caption{Parameters that affect the resolution of the $\bi v_{h}$-reflectivity image}
{\small
  \begin{tabular}{l p{0.9in} p{1in}}
  \hline\hline
  Parameter & Increase($\uparrow$)  &  Resolution \\ 
  \hline 
  Carrier frequency: $f_0$ & $\uparrow$ & $\uparrow$\\
  Length of the windows $L_{\phi}$ & $\uparrow$ & $\uparrow$\\
  Distance $|\bgamma_T(s)-(\z+\v_{h} s)|$, $|\bgamma_T(s)-(\z+\v_{h} s)|$& $\uparrow$ & $\downarrow$\\
  Relative velocity $\dot{\bgamma}_{T}-\v_{h}$ or $\dot{\bgamma}_{R}-\v_{h}$ & $\uparrow$ & $\uparrow$\\
  Bistatic angle $\theta_{TR}$ & $\uparrow$ & $\downarrow$\\
  Ground topography variations & $\uparrow$ & $\uparrow$\\
  Number of $s$ samples  & $\uparrow$ & $\uparrow$\\
  \hline\hline 
  \end{tabular}}
  \\
  \hspace{-9cm} Higher ($\uparrow$) or Lower ($\downarrow$)
\end{table}

\subsection{Velocity Resolution}

Our imaging method discretizes the range of the velocity and forms a reflectivity image corresponding to each velocity sample. Therefore, the velocity resolution depends on how finely the range of the velocity can be sampled, which, in turn, depends on the ``velocity bandwidth'' available in the data. We show that, for a point target located at a fixed position, the data can be interpreted as the bandlimited Fourier transform of the phase-space reflectivity function with respect to the velocity variable and analyze the bandwidth of the data in terms of the imaging geometry, parameters of the ultra-narrowband CW signals, and other data collection parameters.


We assume that the scene consists of a moving point target located at $\bi x_0$ at $t=0$ moving with velocity $\bi v_{\x_0}$, i.e.,
\begin{equation}\label{eq:ptt_v}
    q(\bi x,\bi v)=\delta(\bi x-\bi x_0)q(\bi x_0,\bi v).
\end{equation}

Without loss of generality, we assume that $\bi v_{\x_0}=0$. Performing Taylor series expansion in the phase of the forward model given by (\ref{eq:phi}) around $\bi v=0$, we get
\begin{equation}\label{eq:phi2}
    \phi(t,\bi x,\bi v,s,\mu)\approx
    \phi(t,\bi x,0,s,\mu)+\nabla_{\bi v}\phi(t,\bi x,\bi v,s,\mu)|_{\bi v=0}\cdot \bi v\,.
\end{equation}

Substituting (\ref{eq:ptt_v}) and (\ref{eq:phi2}) into (\ref{eq:d_ori2}), we obtain
\begin{eqnarray}
    \mathcal{F}[q](s,\mu) \approx & \int
    \rme^{-\rmi \phi(t,\bi x_0,0,s,\mu)}
    \rme^{-\rmi \bi \nabla_{\bi v}\phi(t,\bi x_0,\bi v,s,\mu)|_{\bi v=0}}
    \nonumber
    \\
    & \times q(\bi x_0,\bi v) A(t,\bi x_0,\bi v,s,\mu)d\bi v dt
    \label{eq:d_ori_v}
\end{eqnarray}
where
\begin{eqnarray}\label{eq:delta_v1}
    \phi(t,\bi x_0,\bi v,s,\mu)&=&2\pi t[(\mu-1)f_0+f_d(s,\bi x_0,0)]
\end{eqnarray}
and
\begin{equation}\label{eq:delta_v2}
    \nabla_{\bi v}\phi(t,\bi x_0,\bi v,s,\mu)|_{\bi v=0}
    =2\pi t \nabla_{\bi v}f_d(s,\bi x_0,\bi v)|_{\bi v=0}\,.
\end{equation}
Note that $f_d(s,\bi x_0,0)$ in (\ref{eq:delta_v1}) represents the Doppler frequency induced by the movement of the transmitter and receiver. It does not depend on the target velocity.

Substituting (\ref{eq:delta_v1}) and (\ref{eq:delta_v2}) into (\ref{eq:d_ori_v}), we obtain
\begin{eqnarray}
    \mathcal{F}[q](s,\mu)&\approx& \int
    \left(\rme^{-\rmi 2 \pi t \nabla_{\bi v}f_d(s, \bi x_0,\bi v)|_{\bi v=0} \cdot \bi v}
    q(\bi x_0,\bi v) A(t,\bi x_0,\bi v,s,\mu)d\bi v \right)
    \nonumber
    \\
    &&\times
    \rme^{-\rmi 2 \pi t [(\mu-1)f_0+f_d(s,\bi x_0,0)]}
     dt\,.
\end{eqnarray}

Let
\begin{equation}\label{eq:symb}
    t \nabla_{\bi v}f_d(s, \bi x_0,\bi v)|_{\bi v=0}=\boldsymbol{\varsigma}\,.
\end{equation}
We see that $\boldsymbol{\varsigma}$ is the Fourier vector associated with $\bi v$. Therefore, the length and direction of $\boldsymbol{\varsigma}$ determine the velocity resolution available in the data, $d(s,\mu)$.
The bandwidth contribution of $\boldsymbol{\varsigma}$ is given as follows:
\begin{eqnarray}
    |\boldsymbol{\varsigma}| &=& |t \nabla_{\bi v}f_d(s, \bi x_0,\bi v)|_{\bi v=0}|
    \nonumber
    \\
    &= & L_{\phi} \left| D \cdot
    \left[2\cos\frac{\theta_{TR}(\x_0, \v, s)}{2}\hat{b}_{TR}(\bi x_0, \v, s)\right.\right.
    \nonumber
    \\
    & & \hspace{1cm}\left. \left.+\frac{(\dot{\bgamma}_T(s)-\v)_{\perp}\,s}{|\bgamma_T(s)-(\x_0+\v s)|}    +\frac{(\dot{\bgamma}_R(s)-\v)_{\perp}\,s}{|\bgamma_R(s)-(\x_0+\v s)|}
    \right]
    \right|
    \label{eq:band_bxi_v}
\end{eqnarray}
where $L_{\phi}, \theta_{TR}(\x_0, \v, s), \hat{b}_{TR}(\bi x_0, \v, s)$ are as defined in (\ref{eq:band_bxi}).
Note that $\x_0=[\bi x_0,\psi(\bi x_0)]$, $\v=[\bi v, \nabla_{\bi x_0}\psi(\bi x_0) \cdot \bi v]$ and $D$ is given by (\ref{eq:D}) with $\bi z$ replaced with $\bi x_0$.

Comparing (\ref{eq:band_bxi_v}) with (\ref{eq:band_bxi}), we see that similar to the reflectivity image formation,
the larger the carrier frequency $f_0$ and the support of $\phi(t)$, the higher the velocity resolution is. Furthermore, the velocity resolution also depends on the range of the antennas to the moving target via the terms $|\bgamma_T(s)-(\x_0+\v s)|$ and $|\bgamma_R(s)-(\x_0+\v s)|$; and the relative speed between the transmitter (receiver) and the target via the terms $(\dot{\bgamma}_T(s)-\v)_{\perp}$ and $(\dot{\bgamma}_R(s)-\v)_{\perp}$. 
The increase in the number of $s$ samples used for imaging also results in a larger data collection manifold and hence better resolution.
Additionally, the larger the bistatic angle $\theta_{TR}(\x_0, \v, s)$ is, the lower the velocity resolution becomes. Note that the bistatic angle $\theta_{TR}(\x_0, \v, s)$ has a larger impact on the velocity resolution than on the position resolution due to the dependence on $D$ instead of $D^2$.

Note that the parameters that affect the resolution of reflectivity images and velocity resolution identified in our analysis are consistent with the Doppler ambiguity theory of ultra-narrowband CW signals \cite{radarsignal}. 
\section{Analysis of Position Error in Reflectivity Images due to Incorrect Velocity Field}
\label{sec:PosError}
In the previous section, we show that the image fidelity operator $\mathcal{K}_{\bi v_h}\mathcal{F}$ reconstructs the singularities at the intersection of the bi-static iso-Doppler and iso-Doppler-rate manifolds defined by the following two equations:
\begin{eqnarray}
    f_d(s,\bi z,\bi v_{h})-f_d(s,\bi x,\bi v_{\x})&=0
    \label{eq:f_d}
    \\
    \dot{f}_d(s,\bi z,\bi v_{h})-\dot{f}_d(s,\bi x,\bi v_{\x})&=0
    \label{eq:dotf_d}
\end{eqnarray}
where $\dot{f}_d=\partial_s f_d$. When $\bi v_{h} = \bi v_{\x}$, one of the solutions of (\ref{eq:dotf_d}) is $\bi z = \bi x$ which shows that $\mathcal{K}_{\bi v_h}\mathcal{F}$ reconstructs a singularity that coincides with the visible singularity of the scene, $q(\bi x, \bi v_{\x})$. We shall refer to such singularities of $\mathcal{K}_{\bi v_h}\mathcal{F}[q]$ as the useful singularities \footnote[8]{Note that in addition to useful singularities, $\mathcal{K}_{\bi v_h}\mathcal{F}$ may reconstruct additional artifact singularities that are of the same strength as the useful singularities. The location of these singularities are given by the solution of (\ref{eq:dotf_d}) when $\bi v_{h} = \bi v_{\x}$.}.
If, on the other hand, $\bi v_{h} \neq \bi v_{\x}$, the useful singularities of $\mathcal{K}_{\bi v_h}\mathcal{F}[q]$ no longer coincide with the visible singularities of the scene reflectivity. In this section, we analyze the shift in the location of the useful singularities $\bi v_h$-reflectivity image due to errors in the hypothesized velocity field $\bi v_{h}$. The analysis provides the positioning error between the correct and reconstructed targets due to error in their hypothesized velocities. Additionally, it shows the geometry and degree of smearing in the reconstructed reflectivity images due to incorrect velocity information given the imaging geometry. For simplicity, we assume that the ground topography is flat for the rest of our analysis.

Suppose for the target located at $\bi x$ at $t=0$ moving with velocity $\bi v_\x$, we use an erroneous hypothesized velocity
\begin{equation}
    \bi v_{h} = \bi v_{\z}+\epsilon \triangle \bi v_{\z}
\end{equation}
in the backprojection, where $\bi v_{\z} = \bi v_{\x}$ and $\epsilon \triangle \bi v_{\z}$, $\epsilon \in \mathbb{R}$, is the error in the velocity $\bi v_{h}$. Then, the target at position $\bi x=\bi z$ is reconstructed at $\bi z_{\epsilon}=\bi z+\triangle \bi z$ and we have
\begin{eqnarray}
    f_d(s,\bi z+\triangle \bi z,\bi v_{\z}+\epsilon \triangle \bi v_{\z})-f_d(s,\bi x,\bi v_{\x})&=0
    \label{eq:f_d2}
    \\
    \dot{f}_d(s,\bi z+\triangle \bi z,\bi v_{\z}+\epsilon \triangle \bi v_{\z})-\dot{f}_d(s,\bi x,\bi v_{\x})&=0\,.
    \label{eq:dotf_d2}
\end{eqnarray}
(\ref{eq:f_d2}) and (\ref{eq:dotf_d2}) show that the visible singularity at $\bi x$ in the scene is mapped to a singularity at $\bi z_\epsilon = \bi z+\triangle \bi z$ in the reconstructed image.

We want to determine the first order approximation to the shift $\triangle \bi z$ due to the velocity error $\epsilon \triangle \bi v_{\z}$. In order to determine $\triangle \bi z$, we assume that $\epsilon\rightarrow 0$ is small and expand (\ref{eq:f_d2}) and (\ref{eq:dotf_d2}) in Taylor series around $\epsilon=0$ and keep the first-order terms in $\epsilon$.
Then, using (\ref{eq:f_d})-(\ref{eq:dotf_d}) and (\ref{eq:f_d2})-(\ref{eq:dotf_d2}) in the Taylor series expansion, we obtain 
\begin{eqnarray}
    \epsilon \partial_{\epsilon}f_d(s,\bi z,\bi v_{\z}+\epsilon \triangle \bi v_{\z})|_{\epsilon=0}
    +\nabla_{\bi z}f_d(s,\bi z,\bi v_{\z})\cdot \triangle \bi z=0
    \label{eq:partial_fd2}
    \\
    \epsilon \partial_{\epsilon} \dot{f}_d(s,\bi z,\bi v_{\z}+\epsilon \triangle \bi v_{\z})|_{\epsilon=0}
    +\nabla_{\bi z}\dot{f}_d(s,\bi z,\bi v_{\z})\cdot \triangle \bi z=0\,.
    \label{eq:partial_dotfd2}
\end{eqnarray}


Evaluating (\ref{eq:partial_fd2}) and (\ref{eq:partial_dotfd2}) for the bi-static Doppler frequency of moving targets, (\ref{eq:partial_fd2}) simplifies to
\begin{eqnarray}
    -\epsilon s \triangle \v_{\z}^{\perp,T} \cdot
    \frac{(\dot{\bgamma}_T(s)-\v_{\z})}{|\bgamma_T(s)-(\z+\v_{\z} s)|} +
    \nonumber
    \\
    -\epsilon s \triangle \v_{\z}^{\perp,R} \cdot
    \frac{(\dot{\bgamma}_R(s)-\v_{\z})}{|\bgamma_R(s)-(\z+\v_{\z} s)|}+
    \nonumber
    \\
    \hspace{0.55cm}
    -\epsilon
    \triangle \v_{\z} \cdot  \left[(\widehat{\bgamma_T(s)-(\z+\v_{\z} s)} + (\widehat{\bgamma_R(s)-(\z+\v_{\z} s)}\right]
    \nonumber
    \\
    \hspace{-0cm} =
    \triangle \bi z \cdot
    \bXi_{\bi v_\z}(s,\bi z)\frac{c_0}{2\pi f_0}
    \label{eq:partial_fd3}
\end{eqnarray}
where for flat topography,
\begin{equation}\label{eq:bXi_flat}
    \hspace{-1cm} \bXi_{\bi v_\z}(s,\bi z)\frac{c_0}{2\pi f_0}=- D \cdot
    \left[\frac{(\dot{\bgamma}_T(s)-\v_{\z})_{\perp} }{|\bgamma_T(s)-(\z+\v_{\z} s)|}+
    \frac{(\dot{\bgamma}_R(s)-\v_{\z})_{\perp}}{|\bgamma_R(s)-(\z+\v_{\z} s)|}\right].
\end{equation}
$\triangle \v_{\z}^{\perp,T,R}$ and $(\dot{\bgamma}_{T,R}(s)-\v_{\z})_{\perp}$ are the projections of $\triangle \v_{\z}$ and $\dot{\bgamma}_{T,R}(s)-\v_{\z}$ onto the plane whose normal direction is $(\widehat{\bgamma_{T,R}(s)-(\z+\v_{\z} s)}$.
Note that in (\ref{eq:partial_fd3}) and (\ref{eq:bXi_flat}), $\v_{\z} = [\bi v_{\z},\ 0]$, $\z = [\bi z, 0]$, and $\bi v_{\z} = \bi v_{\x}$ and $\z = \x$. In other words, $\z$ and $\v_{\z}$ are the correct position and velocity of the target in the image domain that is located at $\x$ moving with velocity $\v_{\x}$ in the scene.

Similarly, (\ref{eq:partial_dotfd2}) simplifies to
\begin{eqnarray}
    \hspace{-1cm}- \epsilon \triangle \v_\z^{\perp,T} \cdot  \left [
    \frac{s\,\ddot{\bgamma}_T(s)}{|\bgamma_T(s)-(\z+\v_\z s)|}
    +\frac {2 (\dot{\bgamma}_T(s)-\v_\z)_\perp }{|\bgamma_T(s)-(\z+\v_\z s)|}
    C_{T}(\z,\v_{\z},s)
    \right ]
    \nonumber
    \\
    \hspace{-1cm} + \epsilon s
    \triangle \v_\z \cdot \widehat{\bgamma_{T}(s)-(\z+\v_\z s)}
    \frac{|(\dot{\bgamma}_{T}(s)-\v_\z)_\perp|^2}
    {|\bgamma_T(s)-(\z+\v_\z s)|^2}
    \nonumber
    \\
    \hspace{-1cm} -\epsilon \triangle \v_\z^{\perp,R} \cdot  \left [
    \frac{s\,\ddot{\bgamma}_R(s)}{|\bgamma_R(s)-(\z+\v_\z s)|}
    +\frac {2 (\dot{\bgamma}_R(s)-\v_\z)_\perp }{|\bgamma_R(s)-(\z+\v_\z s)|}
    C_{R}(\z,\v_{\z},s)
    \right ]
    \nonumber
    \\
    \hspace{-1cm} +\epsilon s
    \triangle \v_\z \cdot \widehat{\bgamma_{R}(s)-(\z+\v_\z s)}
    \frac{|(\dot{\bgamma}_{R}(s)-\v_\z)_\perp|^2}
    {|\bgamma_R(s)-(\z+\v_\z s)|^2}
    \nonumber
    \\
    \hspace{-1cm}=
    -\triangle \bi z \cdot \dot{\bXi}_{\bi v_{\z}}(s,\bi z) \frac{c_0}{2\pi f_0}
    \label{eq:partial_Dotfd4}
\end{eqnarray}
where $\dot{\bXi}_{\bi v_{\z}}(s,\bi z) = \partial_s\bXi_{\bi v_{\z}}(s,\bi z)$ is the derivative of ${\bXi}_{\bi v_{\z}}(s,\bi z)$ with respect to $s$ for flat topography. For the explicit form of $C_{T,R}(\z,\v_{\z},s)$ and $\dot{\bXi}_{\bi v_{\z}}(s,\bi z)$, and the derivation of (\ref{eq:partial_fd3}) and (\ref{eq:partial_Dotfd4}), see \ref{app:Equation1} and \ref{app:Equation2}.
Note that in (\ref{eq:partial_Dotfd4}), $\ddot{\bgamma}^\perp_{T,R}$ is the projection of the acceleration, $\ddot{\bgamma}_{T,R}$, of the transmitting/receiving antenna onto the plane whose normal direction is $(\widehat{\bgamma_{T,R}(s)-(\z+\v_{\z} s)})$.

The shift $\triangle \bi z$ in the useful singularity lies at the intersection of the solution of (\ref{eq:partial_fd3}) and (\ref{eq:partial_Dotfd4}). (\ref{eq:partial_fd3}) and (\ref{eq:partial_Dotfd4}) show that when the error in the velocity $\bi v_{\z}$ is in the order of $\epsilon$, the shift in the reconstructed useful singularities is also in the order of $\epsilon$, which means that the reconstructed reflectivity images would vary smoothly with respect to the change in the velocity around the correct value.

(\ref{eq:partial_fd3}) and (\ref{eq:partial_Dotfd4}) show that for a given aperture location $'s'$, the shift in position, $\triangle \bi z $, depends on the components of the velocity error $\triangle \v_\z$ in the look directions of the transmitting/receiving antennas, $(\widehat{\bgamma_{T,R}(s)-(\z+\v_{\z} s)})$, and its projections onto the planes perpendicular to the antenna look directions. Clearly, the shift in position depends on the antenna flight trajectories. The $'s'$ dependency of the shift explains the smearing observed in the final backprojected data.

Clearly, (\ref{eq:partial_fd3}) and (\ref{eq:partial_Dotfd4}) can be used to predict the positioning errors caused by moving targets in reflectivity images reconstructed under the stationary scene assumption, in which case $\triangle \v_\z=-\v_\x$.

\vspace{0.1in}
\noindent\emph{Example - Monostatic SAR traversing a linear flight trajectory}

To understand the implications of (\ref{eq:partial_fd3}) and (\ref{eq:partial_Dotfd4}) and to illustrate the shift in position, we consider a relatively simple scenario where the transmitting and receiving antennas are colocated, traversing a linear trajectory forming a relatively short synthetic aperture.

Let $\bgamma=\bgamma_T=\bgamma_R$ denote the flight trajectory. Then (\ref{eq:partial_fd3}) and (\ref{eq:partial_Dotfd4}) become
\begin{eqnarray}
    \fl
    -\epsilon s \triangle \v_{\z}^{\perp} \cdot
    \frac{(\dot{\bgamma}(s)-\v_{\z})}{|\bgamma(s)-(\z+\v_{\z} s)|}
    - \epsilon
    \triangle \v_{\z} \cdot (\widehat{\bgamma(s)-(\z+\v_{\z} s)})
    =-
    \triangle \z \cdot
    \frac{(\dot{\bgamma}(s)-\v_{\z})_{\perp} }{|\bgamma(s)-(\z+\v_{\z} s)|}
    \nonumber
    \\
    \label{eq:partial_fd3_mono}
\end{eqnarray}
and
\begin{eqnarray}
    -\epsilon s \triangle \v_{\z}^{\perp} \cdot
    \frac{\ddot{\bgamma}(s)}{|\bgamma(s)-(\z+\v_{\z} s)|}
    \nonumber
    \\
    -2\epsilon \triangle \v_{\z}^{\perp} \cdot
    \frac{(\dot{\bgamma}(s)-\v_\z)_\perp}{|\bgamma(s)-(\z+\v_{\z} s)|}
    \nonumber
    \\
    +2\epsilon \triangle \v_{\z}^{\perp} \cdot
    \frac{(\dot{\bgamma}(s)-\v_\z)_\perp [(\dot{ \bgamma}(s) - \v_{\z}) \cdot \widehat{\bgamma(s)-(\z+\v_{\z} s)}]s}
    {|\bgamma(s)-(\z+\v_{\z} s)|^2}
    \nonumber
    \\
    +\epsilon s \triangle \v_{\z} \cdot(\widehat{\bgamma(s)-(\z+\v_{\z} s)})
    \frac{|(\dot{\bgamma}(s) - \v_{\z})_{\perp}|^2}
    {|\bgamma(s)-(\z+\v_{\z} s)|^2}
    \nonumber
    \\
    \hspace{-0.35cm}=
    -\triangle \z \cdot 
    \widehat{\bgamma(s)-(\z+\v_{\z} s)}
    \frac{|(\dot{\bgamma}(s)-\v_{\z})_\perp|^2}{|\bgamma(s)-(\z+\v_{\z} s)|^2}
    \nonumber
    \\
    -\triangle \z \cdot (\dot{\bgamma}(s)-\v_{\z})_\perp
    \frac{2(\dot{\bgamma}(s)-\v_{\z}) \cdot \widehat{\bgamma(s)-(\z+\v_{\z} s)}}
    {|\bgamma(s)-(\z+\v_{\z} s)|^2}
    \nonumber
    \\
    +\triangle \z \cdot \frac{\ddot{\bgamma}^\perp(s)}{|\bgamma(s)-(\z+\v_{\z} s)|}.
    \label{eq:partial_Dotfd4_monostatic}
\end{eqnarray}

We assume that the radar flies along a linear straight trajectory with a constant velocity and observes a region of interest in the far-field and in the boresight direction of the antenna. Since the speed of the target is usually much smaller than the speed of the antenna, we assume that the relative velocity vector, $\dot{\bgamma}-\v_\z$, is perpendicular to the radar line of sight (RLOS), $\widehat{\bgamma(s)-(\z+\v_{\z} s)}$ throughout a short synthetic aperture. Under these assumptions, $\ddot{\bgamma}=0$,
$(\dot{\bgamma}(s)-\v_{\z}) \cdot \widehat{\bgamma(s)-(\z+\v_{\z} s)}=0$ and
$(\dot{\bgamma}-\v_{\z})_\perp \approx \dot{\bgamma}-\v_{\z}$. Thus, (\ref{eq:partial_fd3_mono}) and (\ref{eq:partial_Dotfd4_monostatic}) reduce to
\begin{equation}\label{eq:partial_fd3_simple}
    |\triangle \z^\perp|=\epsilon s |\triangle \v_\z^\perp|\cos \theta+\epsilon |\triangle \v_\z^r|\frac{|\bgamma(s)-(\z+\v_{\z} s)|}{|\dot{\bgamma}-\v_\z|}
\end{equation}
and
\begin{equation}\label{eq:partial_Dotfd4_simple}
    |\triangle \z^r|=-\epsilon s |\triangle \v_\z^r|+2\epsilon |\triangle \v_\z^\perp|\cos \theta
    \frac{|\bgamma(s)-(\z+\v_{\z} s)|}{|\dot{\bgamma}-\v_\z|}
\end{equation}
where $\theta$ denote the angle between $\triangle \v_\z^\perp$ and $\dot{\bgamma}-\v_{\z}$ on the plane normal to the RLOS, $|\triangle \z^\perp|=\triangle \z \cdot \widehat{\dot{\bgamma}-\v_{\z}}$, denotes the position shift along the direction of the vector $\widehat{\dot{\bgamma}-\v_{\z}}$, i.e, perpendicular to the RLOS,
and $|\triangle \z^r|=\triangle \z \cdot \widehat{(\bgamma(s)-(\z+\v_{\z} s))}$, denotes the position shift along the RLOS. We refer to $|\triangle \z^\perp|$ and $|\triangle \z^r|$ as the \emph{tangential position error} and the \emph{radial position error}, respectively. Similarly, we refer to $|\triangle \v_\z^\perp|$ and $|\triangle \v_\z^r|$ as the \emph{tangential velocity error} and the \emph{radial velocity error}, respectively.

From (\ref{eq:partial_fd3_simple}) and (\ref{eq:partial_Dotfd4_simple}), we see that for a fixed time (aperture point) $s$, the tangential position error, $|\triangle \z^\perp|$, mainly depends on the radial component of the velocity error, $|\triangle \v_\z^r|$, due to the range term, $|\bgamma(s)-(\z+\v_{\z} s)|$. Similarly, the radial position error, $|\triangle \z^r|$, mainly depends on the tangential component of the velocity error, $|\triangle \v_\z^\perp|$. Under the far-field assumption and for a short synthetic aperture, we note that the RLOS vector,  $\widehat{(\bgamma(s)-(\z+\v_{\z} s)}$, and the second terms in (\ref{eq:partial_fd3_simple}) and (\ref{eq:partial_Dotfd4_simple}) are approximately $s$ independent. In the following section, we elaborate on this example and show the shift and smearing for a point target moving perpendicular to the antenna flight trajectory. 

\section{Numerical Simulations}\label{sec:simulation}
We performed two sets of numerical simulations to demonstrate the performance of our imaging method and to validate the theoretical results. 
In the first set of simulations, we numerically studied the reflectivity (or position) reconstruction performance and the velocity estimation performance of our method for a single point moving target. We also demonstrated the theoretical velocity error analysis described in Section \ref{sec:PosError} using the experimental results of the first set of simulations. In the second set of simulations, we demonstrated the performance of our imaging method for multiple moving targets. Different transmitter and receiver trajectories were used in the two sets of simulations. In the first set of simulations, we considered a monostatic antenna traversing a straight linear trajectory. In the second set of simulations, we considered a bistatic setup where both the transmitter and receiver are traversing a circular trajectory.

For all the numerical experiments, we assumed that a single-frequency continuous waveform operating at $f_0=\omega_0/2\pi=800\mathrm{MHz}$ is being transmitted. We used (\ref{eq:rec_4}) and (\ref{eq:d_ori}) to generate the data.
We used (\ref{eq:rec_5}) to generate the received signal and (\ref{eq:d_ori}) to generate the data used for imaging and chose the windowing function $\phi$ in (\ref{eq:d_ori}) to be a Hanning function.

\subsection{Simulations for a Point Moving Target}\label{sec:sim_point}
We considered a scene of size $256\times 256\,\mathrm{m}^2$ with flat topography centered at $[11,11,0]\mathrm{km}$. The scene was discretized into $128\times128$ pixels, where $[0,0,0]\mathrm{m}$ and $[256,256,0]\mathrm{m}$ correspond to the pixels $(1,1)$ and $(128,128)$, respectively. We assumed that a point moving target with unit reflectivity was located at the center of the scene at time $t=0$ moving with velocity $[0,6.2,0]\mathrm{m/s}$. Note that this position corresponds to the $(65.65)\mathrm{th}$ pixel in the reconstructed scene.

We considered a monostatic antenna traversing a straight linear trajectory, $\gamma_L(s)=(s,0,6.5)\,\mathrm{km}$, at a constant speed. Hence, $s=vt$ where $v=261\mathrm{m/s}$ is the radar velocity. \figref{fig:setup_1} shows the 2D view of the scene with the target and antenna trajectories.
The aperture length used for the image was $5.5\mathrm{e}3\mathrm{m}$, as indicated by the red line.

\begin{figure}
	\centering
    \includegraphics[width=2.5in]{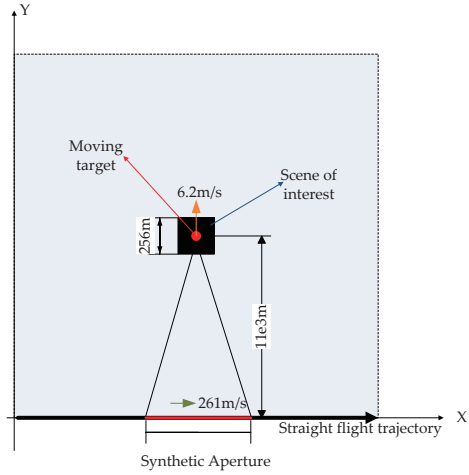}
	\caption{2D illustration of the simulation setup for a single moving target using a monostatic antenna. The dark region shows the scene considered. The red dot shows the position of the point target with the arrow indicating the direction of the target velocity. The radar platform traverses a straight linear trajectory, as shown by the black line. The aperture used for the image is shown by the red line.}
	\label{fig:setup_1}
\end{figure}

We assumed that the velocity of the target is in the range of $[-10,10]\times[-10,10]\mathrm{m/s}$ and implemented the velocity estimation in two stages, each one using a different discretized step: We first discretized the entire velocity space into a $21\times 21$ grid with a step size of $1\mathrm{m/s}$, from which we obtain an initial estimate of the target velocity, $\tilde{\bi v}_{\x,0}$. Then, we discretized a small region of size $[-1,1]\times[-1,1]\mathrm{m/s}$ around the initial velocity estimate 
into a $21\times 21$ grid with a step size of $0.1\mathrm {m/s}$ to refine our velocity estimate obtained in the first stage.

We reconstructed $\tilde{\rho}_{\bi v_h}(\bi x)$ images via the FBP method as described in Section \ref{sec:ReflectFBP} with $f_0=0.8\mathrm{e}9\mathrm{GHz}$, $L_\phi=42.67\mathrm{ms}$ and the aperture sampling frequency, $f_s=97.1869\mathrm{Hz}$.

We formed the contrast images for each velocity estimation stage as described in Section \ref{sec:imageContrast}. The results are shown in \figref{fig:contrastImage1_Linear} and \figref{fig:contrastImage2_Linear}. The red circle shows the velocity estimation. The initial velocity estimate of $\tilde{\bi v}_{\x,0}=[0,7]\mathrm{m/s}$, is shown in \figref{fig:contrastImage1_Linear}, which is close to the true target velocity, $\bi v_\x=[0,6.2]\mathrm{m/s}$. The bright region around the peak in the contrast image shown in \figref{fig:contrastImage1_Linear} indicates that the image contrast varies smoothly with the hypothesized velocity. 

The contrast image obtained using a finer dscretization step is shown in \figref{fig:contrastImage2_Linear}. This contrast image results in a velocity estimate of $[-0.4,6]\mathrm{m/s}$ which is shown by the red circle. The estimate deviates slightly from the true value shown by a black circle. Looking at \figref{fig:contrastImage2_Linear}, we see that the refinement of velocity estimation is not as good as expected.  This may be explained by the velocity resolution provided by the linear flight trajectory and the short aperture as well as waveform parameters. 

\begin{figure}
    \centering
    \subfigure[]{
    \label{fig:contrastImage1_Linear}
    \includegraphics[width=2.6in]{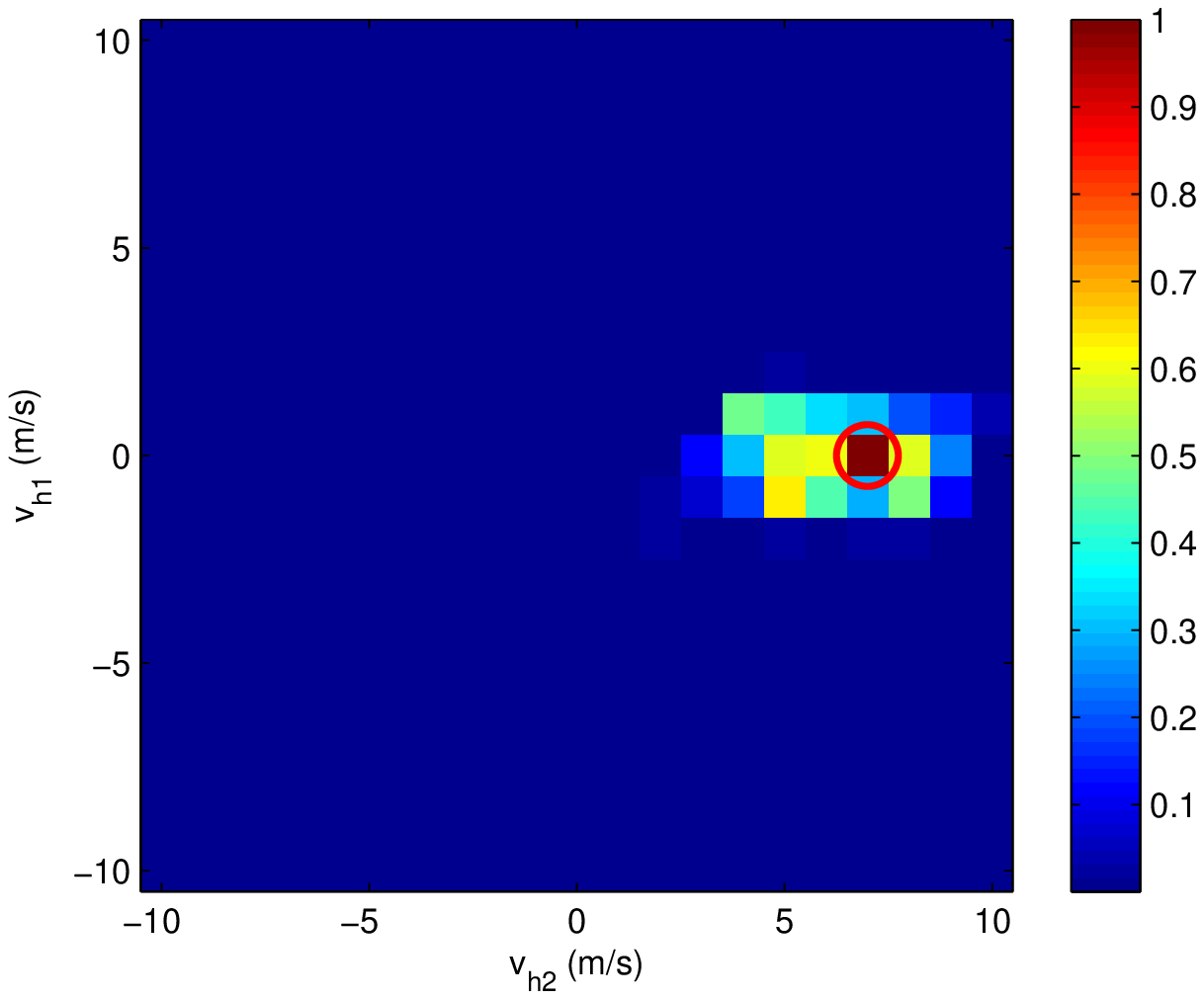}
    }
    \hspace{-0.35in}
    \subfigure[]{
    \label{fig:contrastImage2_Linear}
    \includegraphics[width=2.6in]{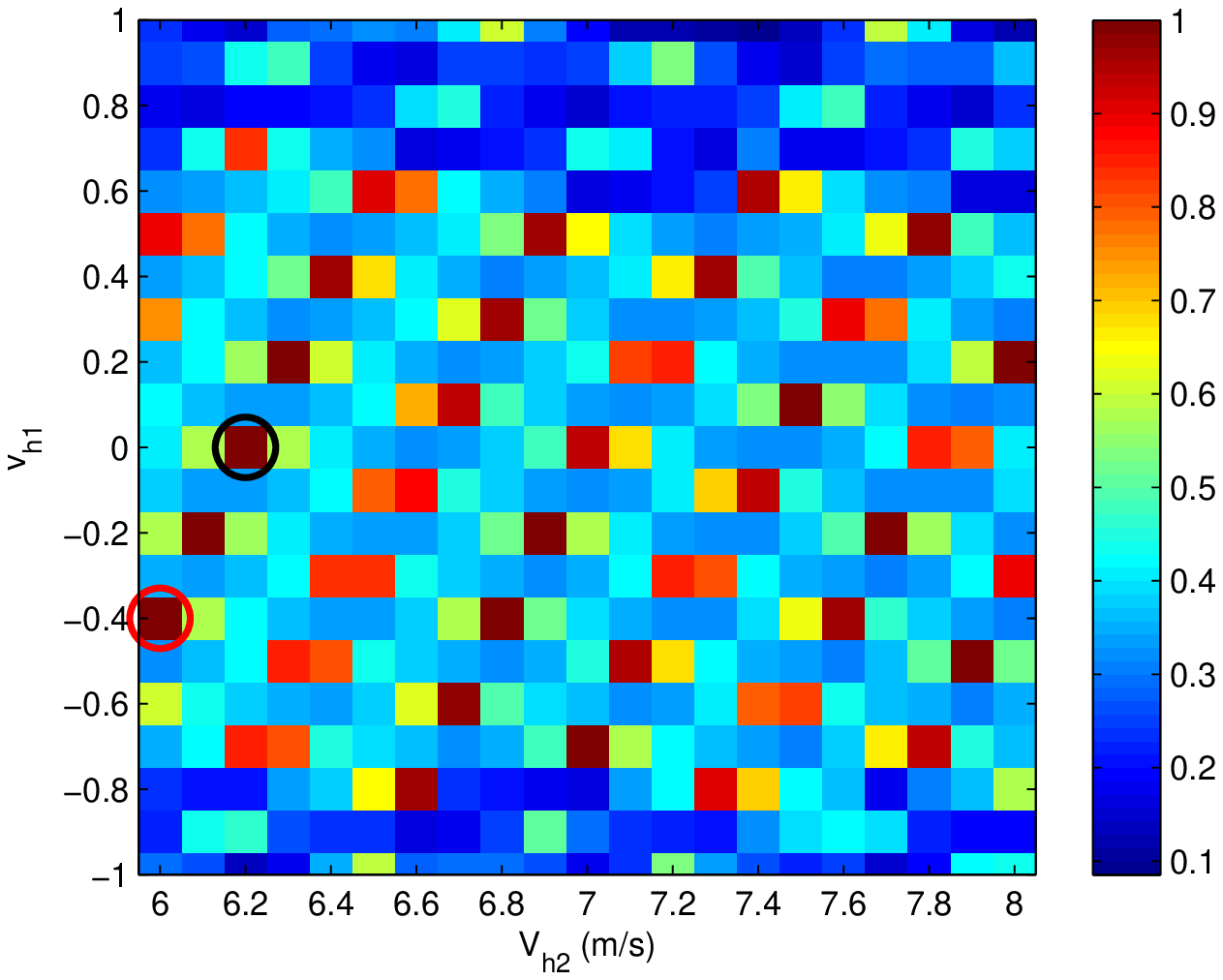}
    }
    \caption{The contrast images formed in the two-stage velocity estimation.(a) Contrast image formed for the entire velocity space discretized using a step size of $1\mathrm{m/s}$. The estimated velocity is $\tilde{\bi v}_{\x,0}=[0,7]\mathrm{m/s}$ as indicated by the red circle. (b) Contrast image formed for a small region of size $[-1,1]\times[-1,1]\mathrm{m/s}$ around $\tilde{\bi v}_{\x,0}$ using a step size of $0.1\mathrm{m/s}$. This corresponds to the region $[-1,1]\times[6,8]\mathrm{m/s}$. The estimated velocity at this stage is $\tilde{\bi v}_\x=[-0.4,6]\mathrm{m/s}$, as shown by the red circle. The black circle shows the true target velocity.}
\end{figure}

\figref{fig:PSF_estimated} shows the reconstructed reflectivity image of the moving target when $\bi v_h=\tilde{\bi v}_\x=[-0.4,6]\mathrm{m/s}$. Note that the black circle shows the true target location. \figref{fig:PSF_CorrectV} shows the reconstructed reflectivity image of the target when the hypothesized velocity is equal to the true target velocity, i.e., $\bi v_h=\bi v_\x=[0,6.2]\mathrm{m/s}$. We see that the moving target in \figref{fig:PSF_estimated} is reconstructed  almost as good as the one in \figref{fig:PSF_CorrectV} with the exception of slight energy spread and a position error due to error in the estimated velocity. 
In the following subsection, we present a quantitative numerical study to demonstrate the results of the analysis described in Section \ref{sec:PosError}.

\begin{figure}[t]
    \centering
    \subfigure[]{
    \label{fig:PSF_estimated}
    \includegraphics[width=2.65in]{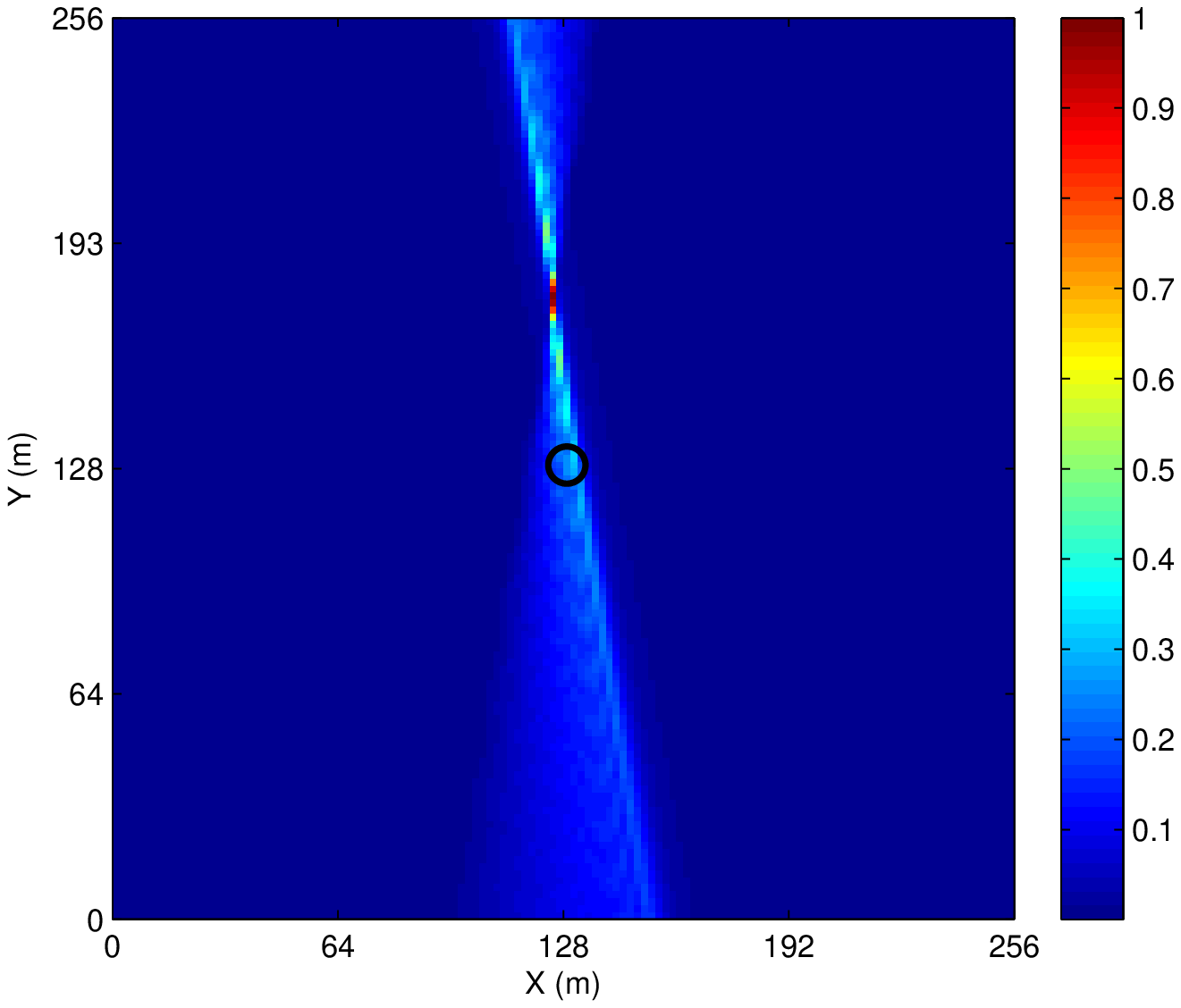}
    }
    \hspace{-0.35in}
    \subfigure[]{
     \label{fig:PSF_CorrectV}
    \includegraphics[width=2.65in]{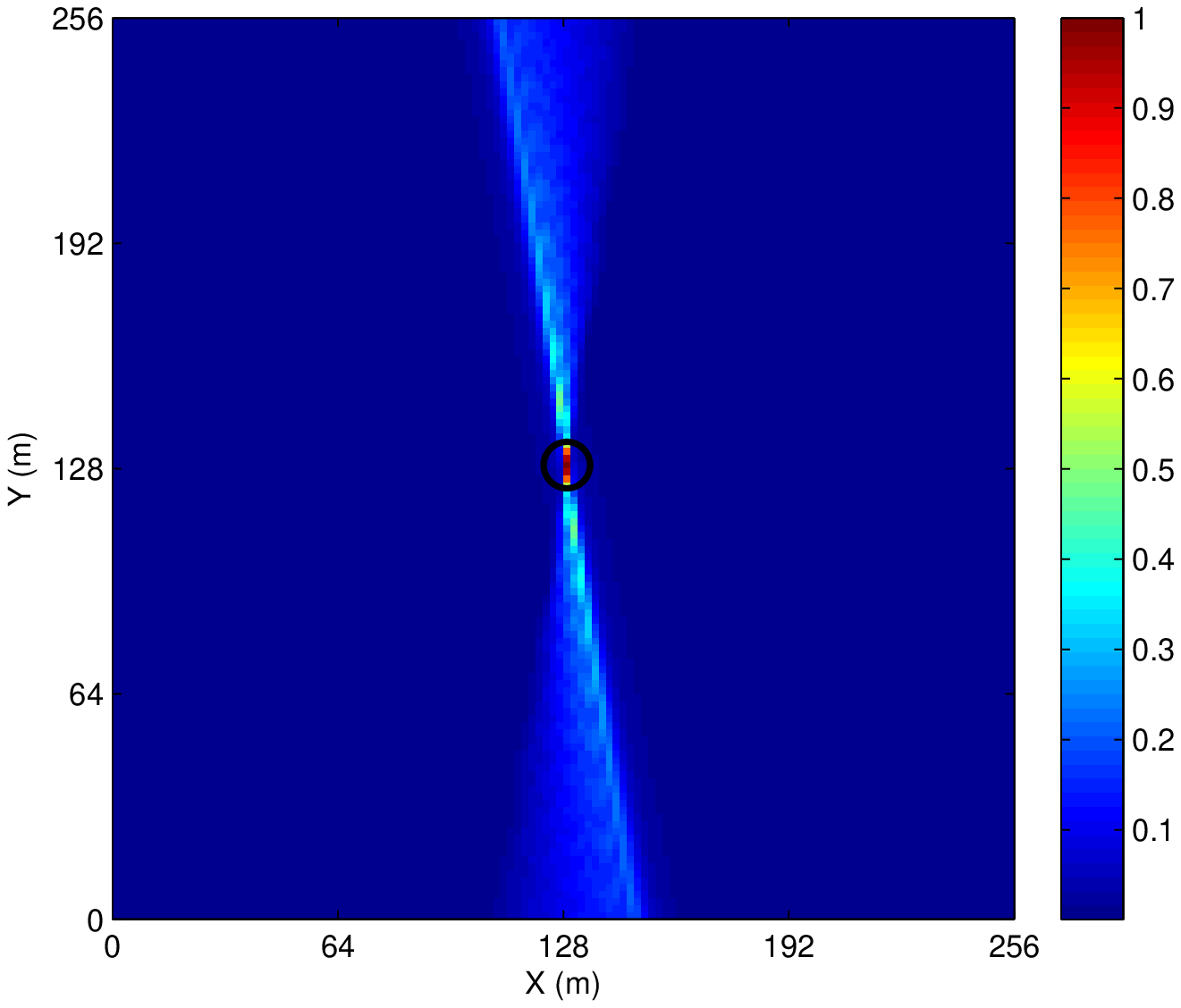}
    }
    \caption{Reconstructed reflectivity images (a) with the estimated velocity, $\tilde{\bi v}_\x=[-0.4,6]\mathrm{m/s}$; and (b) with the correct target velocity, i.e., $\bi v_\x=[0,6.2]\mathrm{m/s}$. The black circle indicates the true target position at time $t=0$.}
\end{figure}

\subsection{Numerical Analysis of the Position Error due to Velocity Error}
We use the simulation results obtained in the previous subsection to demonstrate the theoretical analysis presented in Section \ref{sec:PosError}. Note that the geometry considered in the simulation is consistent with the example given in subsection \ref{sec:PosError}.

\figref{fig:PSF_VerrorRadial} and \figref{fig:PSF_VerrorPerp} show the reflectivity images reconstructed using $\bi v_h=[0,6.7]\mathrm{m/s}$ and $\bi v_h=[0.5,6.2]\mathrm{m/s}$, respectively. Note that the former has a radial velocity error of $|\triangle \v_\z^r|=0.5\mathrm{m/s}$ and the latter has a tangential velocity error of $|\triangle \v_\z^\perp|=0.5\mathrm{m/s}$. The true position of the moving target is shown by a red circle in \figref{fig:PSF_VerrorRadial} and \figref{fig:PSF_VerrorPerp}.

As compared to the image reconstructed using the correct velocity shown in \figref{fig:PSF_CorrectV}, we see that in \figref{fig:PSF_VerrorRadial}, there is an obvious horizontal (or tangential) position shift, while in \figref{fig:PSF_VerrorPerp}, there is an even larger vertical (or radial) position shift. This is predicted by (\ref{eq:partial_fd3_simple}) and (\ref{eq:partial_Dotfd4_simple}) in Section \ref{sec:PosError}, which state that the velocity error in the tangential direction would lead to roughly twice the radial position error that would result from the same magnitude of velocity error in the radial direction. 
Table III compares the position shift errors that are measured from the reconstructed images and the ones predicted by (\ref{eq:partial_fd3_simple}) and (\ref{eq:partial_Dotfd4_simple}), as well as the estimated target positions (in pixel indices) and the corresponding reflectivity values. 
We also note the smearing in \figref{fig:PSF_VerrorRadial} and \figref{fig:PSF_VerrorPerp} due to the velocity error. This can be seen more clearly by comparing the X and Y profiles of the reconstructed images, as shown in \figref{fig:PSF_VerrorRadial_XP}, \figref{fig:PSF_VerrorRadial_YP} and \figref{fig:PSF_VerrorPerp_XP} and \figref{fig:PSF_VerrorPerp_YP}. Note that the X and Y profiles were
shifted to the center for ease of comparison with the results
obtained using the correct velocity.
\addtocounter{table}{0}
\begin{table}[t]
\linespread{1.1}
\centering
\caption{Analysis of the reflectivity images reconstructed using erroneous velocities}
{\small
  \begin{tabular}{l p{0.9in} p{1in} p{0.6in} p{0.6in}}
  \hline\hline
  Velocity error & Measured position shift ($\mathrm{m}$) & Analytic position shift ($\mathrm{m}$) & Estimated target location & Target reflectivity \\ 
  \hline 
  No (Correct velocity) & 0 & 0 & $(65,65)^\mathrm{th}$ & 1\\
  $|\triangle \v_\z^r|=0.5\mathrm{m/s}$ & 22 ($|\triangle \z^\perp|$) & 21.07 ($|\triangle \z^\perp|$) & $(76,59)^\mathrm{th}$ & 0.6145\\
  $|\triangle \v_\z^\perp|=0.5\mathrm{m/s}$ & 52 ($|\triangle \z^r|$) & 42.14 ($|\triangle \z^r|$) & $(62,39)^\mathrm{th}$ & 0.7237\\
  \hline\hline 
  \end{tabular}}
\end{table}

\begin{figure}
    \centering
    \subfigure[]{
    \label{fig:PSF_VerrorRadial}
    \includegraphics[width=3in]{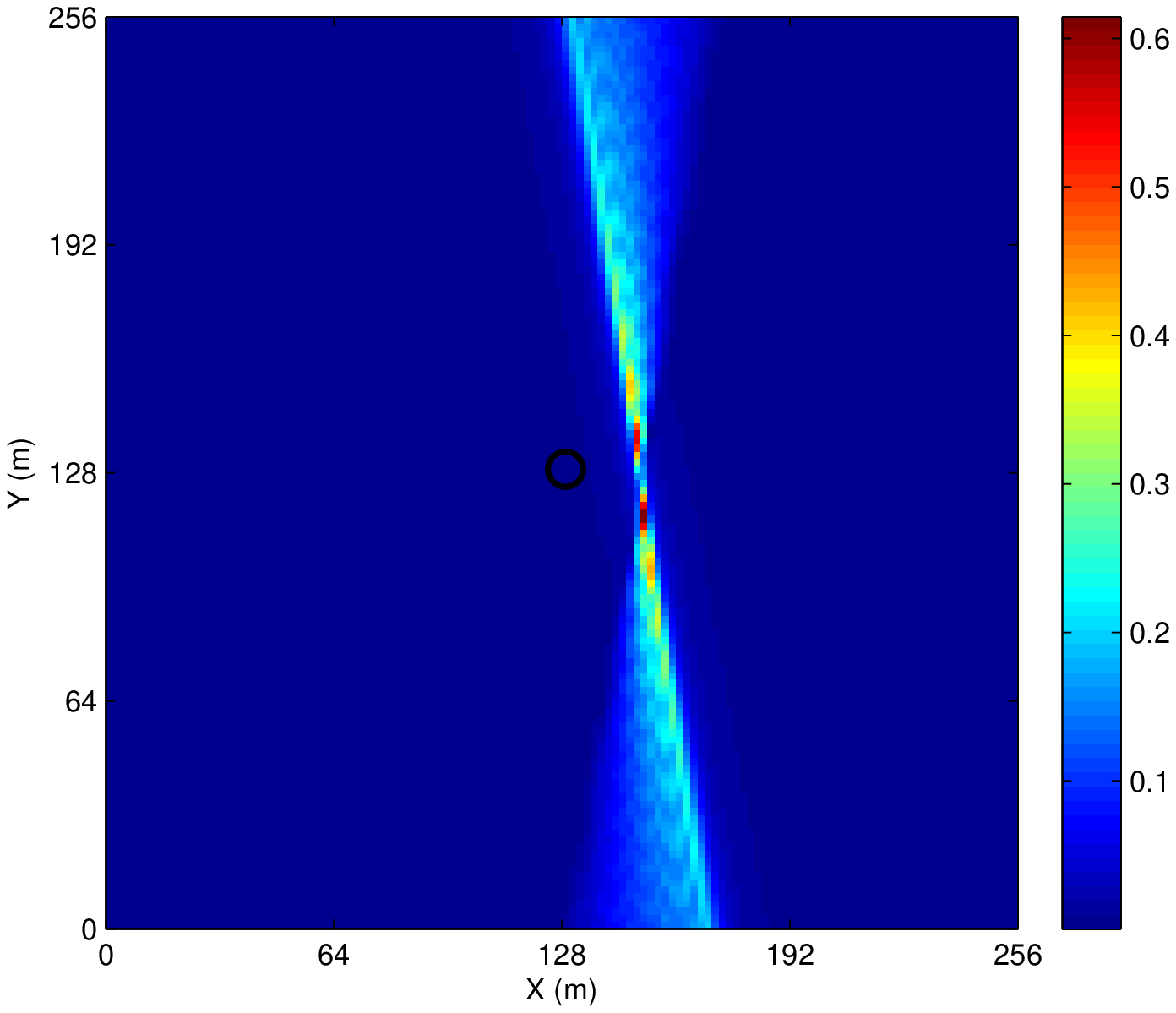}
    }
    \\
    \subfigure[]{
    \label{fig:PSF_VerrorRadial_XP}
    \includegraphics[width=2.4in]{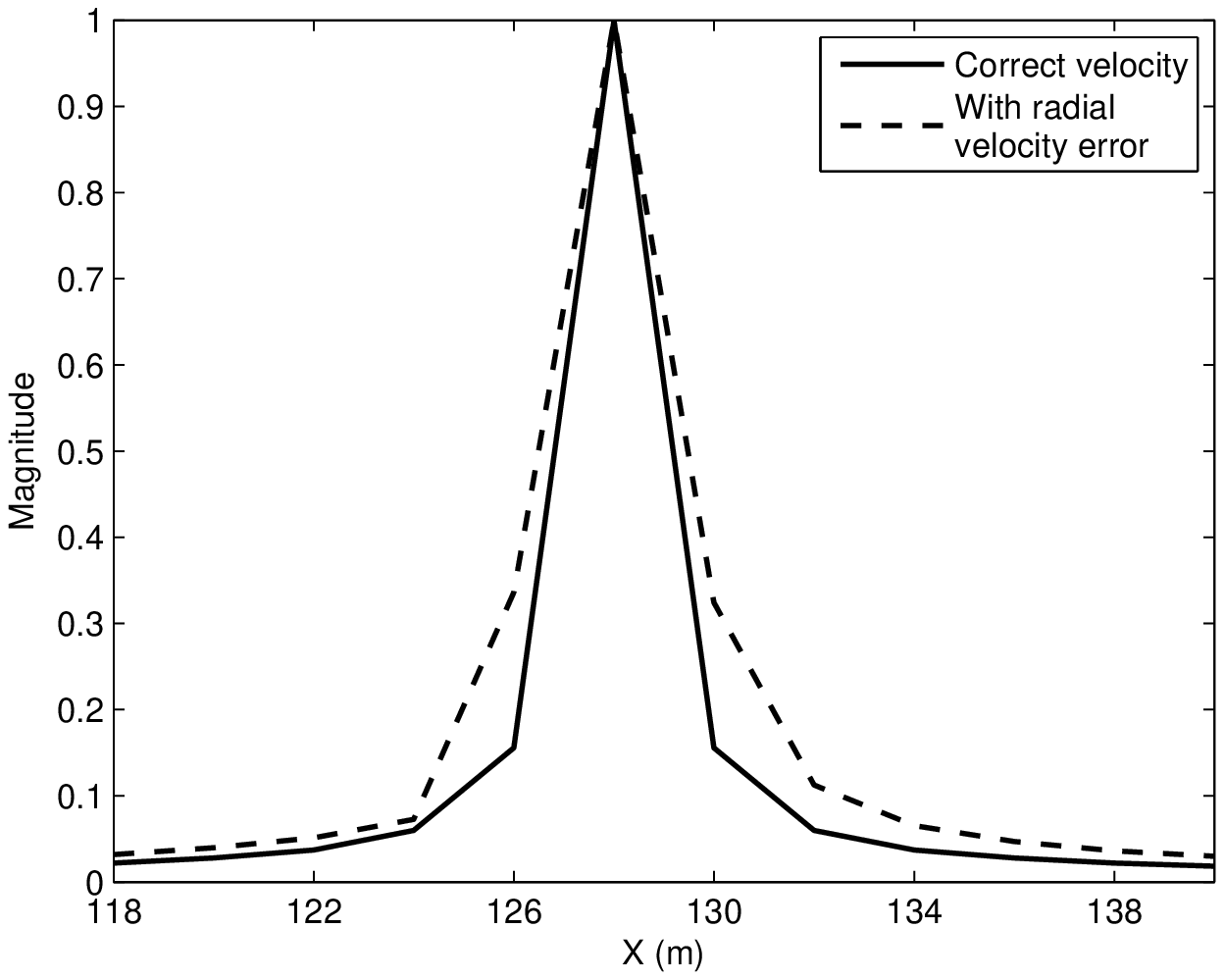}
    }
    \subfigure[]{
    \label{fig:PSF_VerrorRadial_YP}
    \includegraphics[width=2.4in]{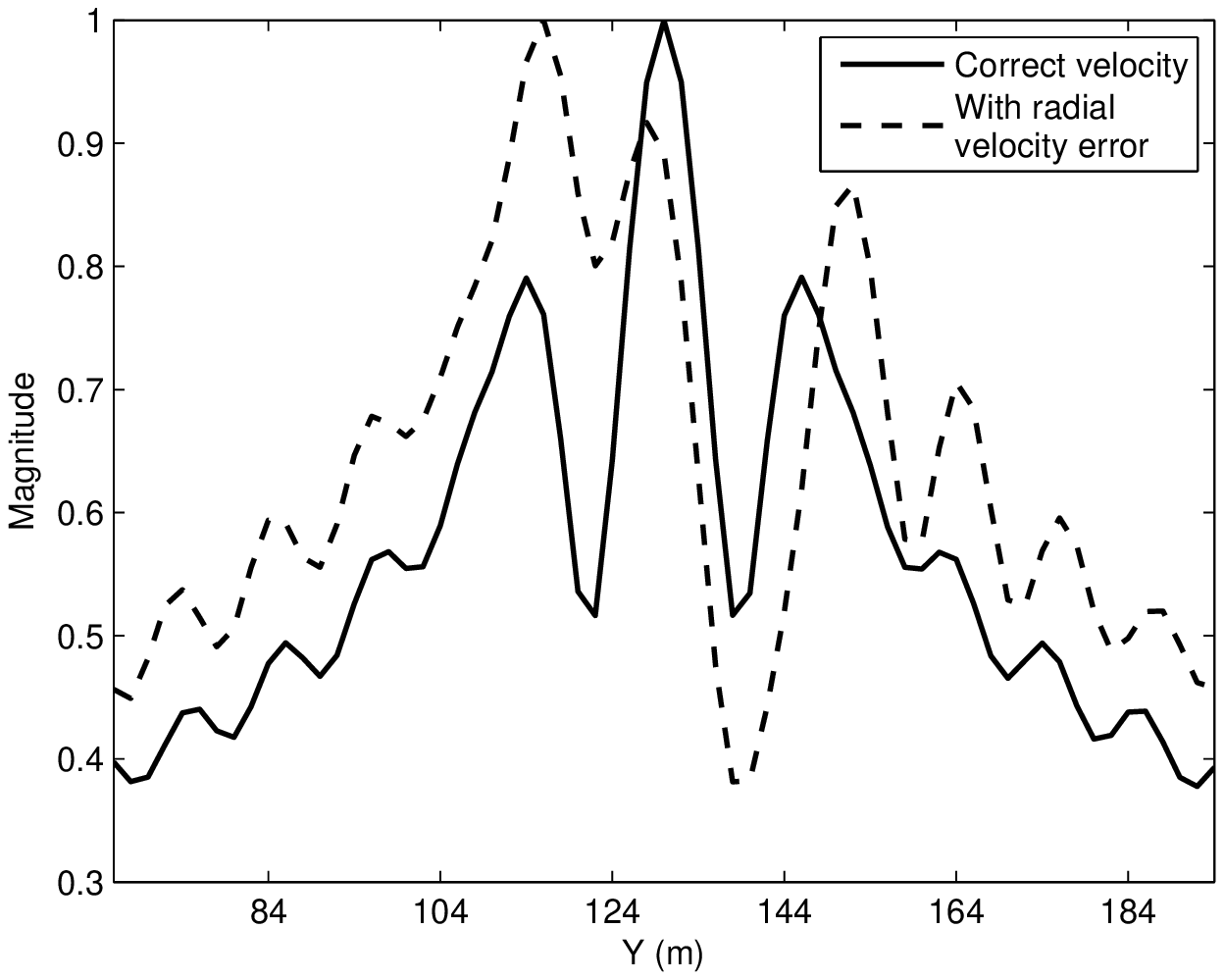}
    }
    \caption{(a) Reconstructed reflectivity image when $\bi v_h=[0,6.7]\mathrm{m/s}$ with a radial velocity error, $\triangle \v_\z^r=0.5\mathrm{m/s}$. The black circle indicates the true target position at time $t=0$. (b) X profiles, and (c) Y profiles. Solid lines show the X and Y profiles of the image reconstructed using the correct velocity shown in \figref{fig:PSF_CorrectV}.}
    \label{fig:PSF_VerrorRadial_Total}
\end{figure}

\begin{figure}
    \centering
    \subfigure[]{
    \label{fig:PSF_VerrorPerp}
    \includegraphics[width=3in]{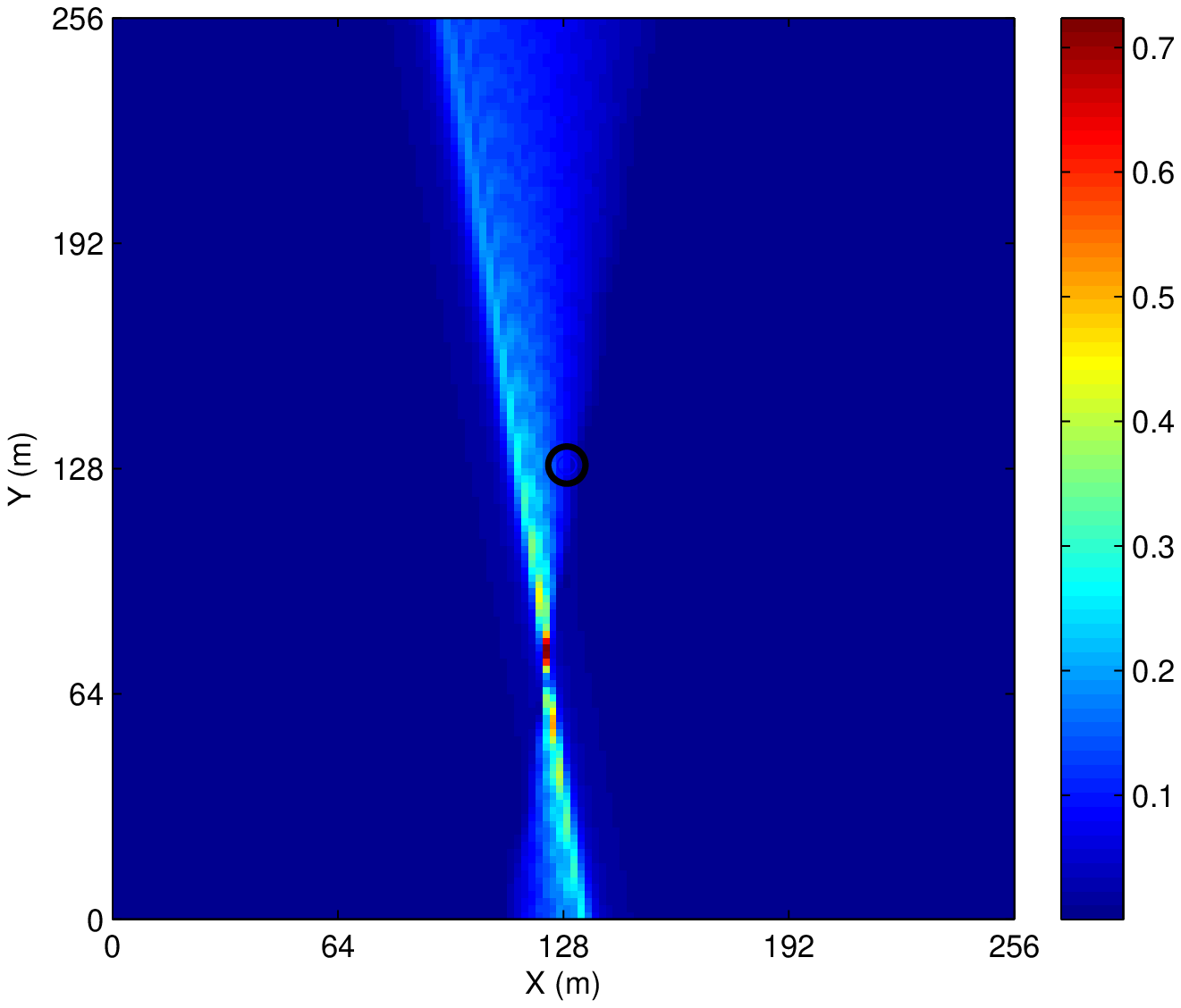}
    }
    \\
    \subfigure[]{
    \label{fig:PSF_VerrorPerp_XP}
    \includegraphics[width=2.4in]{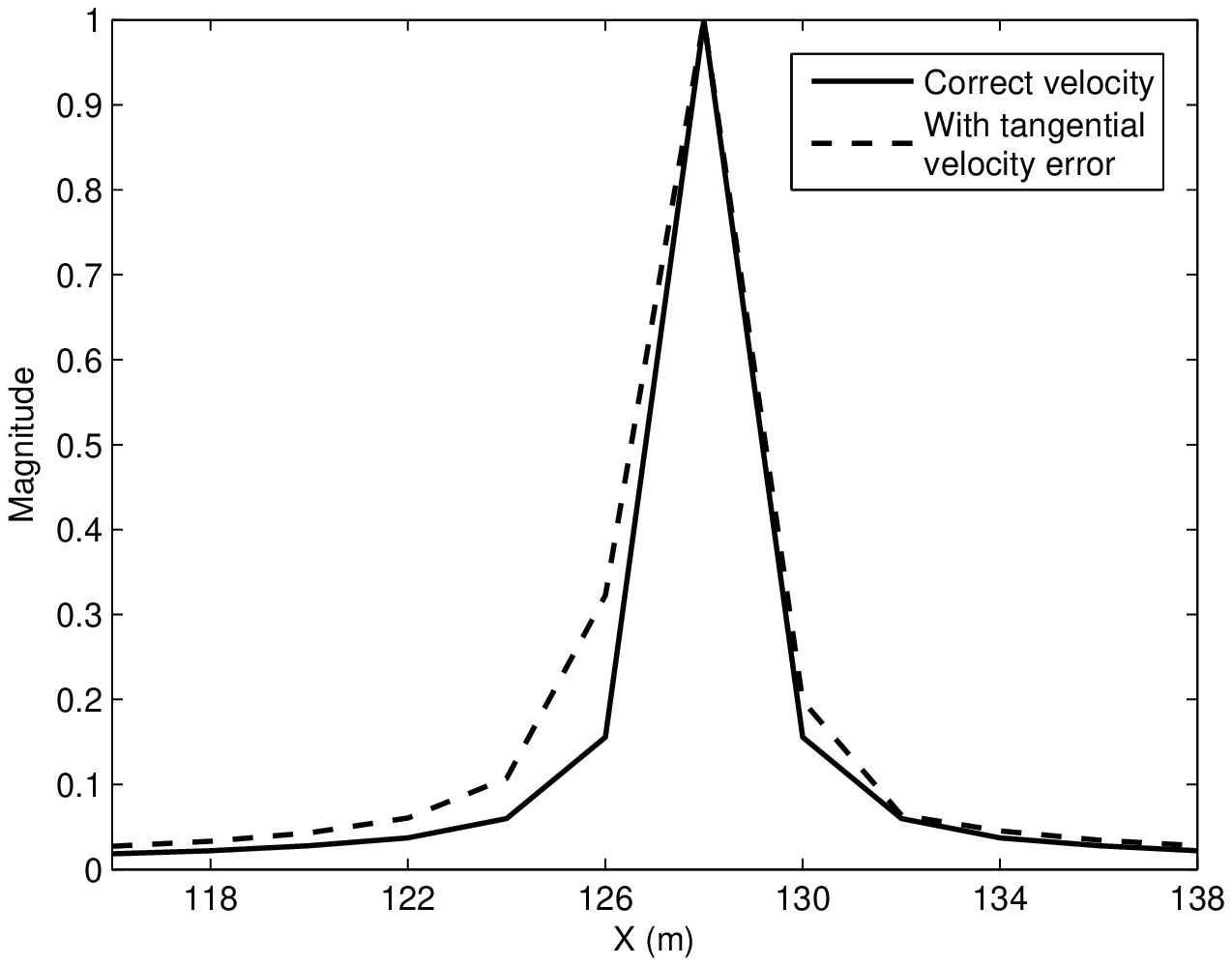}
    }
    \subfigure[]{
    \label{fig:PSF_VerrorPerp_YP}
    \includegraphics[width=2.4in]{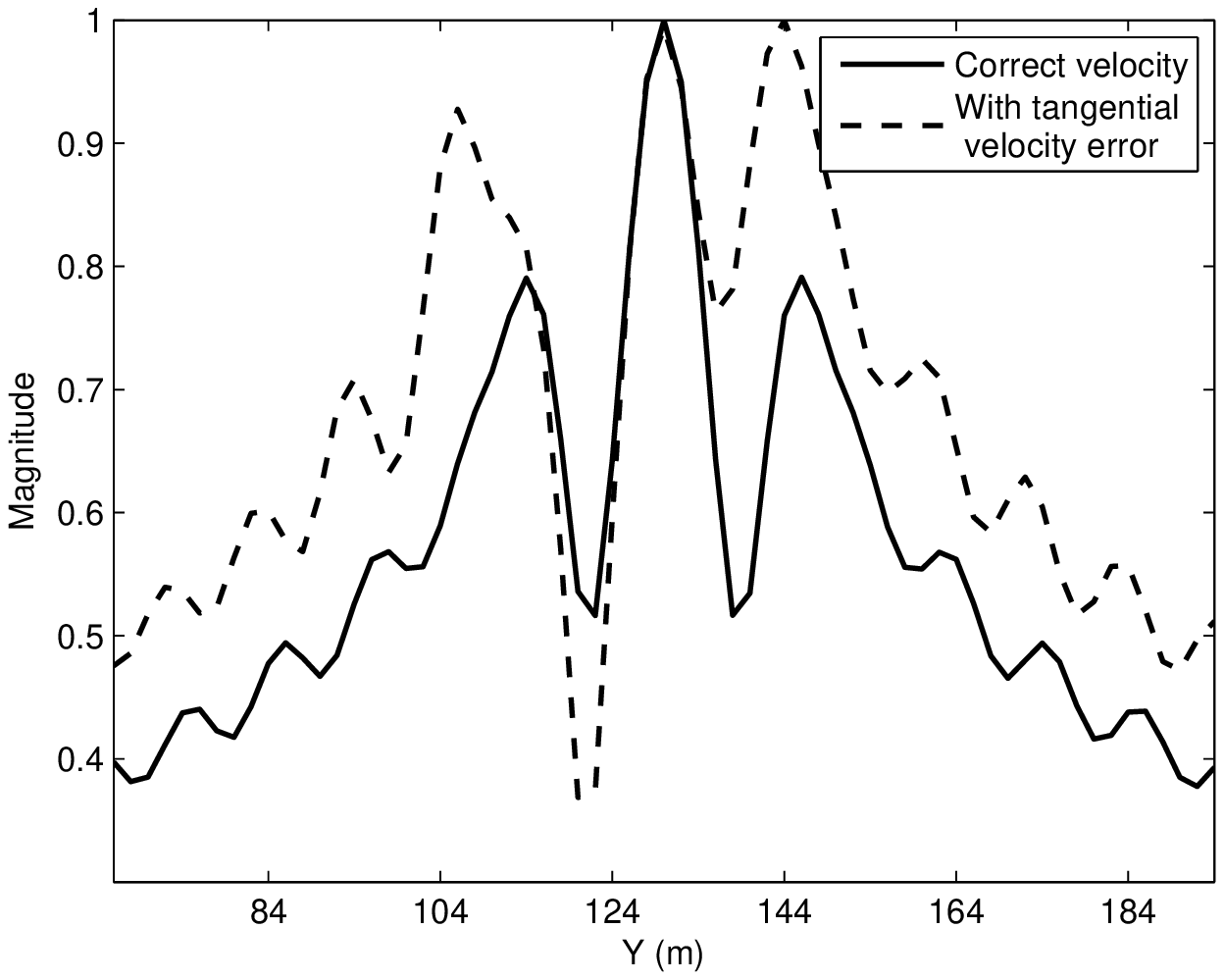}
    }
    \caption{(a) Reconstructed reflectivity image when $\bi v_h=[0.5,6.2]\mathrm{m/s}$ with a tangential velocity error, $\triangle \v_\z^\perp=0.5\mathrm{m/s}$. The black circle indicates the true target position at time $t=0$. (b) X profiles and (c) Y profiles with a comparison with the profiles of \figref{fig:PSF_CorrectV}.}
    \label{fig:PSF_VerrorPerp_Total}
\end{figure}

\subsection{Simulations for Multiple Moving Targets}
In this subsection, we perform simulations for a scene containing multiple moving targets to demonstrate the performance of our method in detecting and estimating the location and velocity of multiple moving targets.

We considered a scene of size $1100\times 1100\,\mathrm{m}^2$ with flat topography centered at $[11,11,0]\mathrm{km}$. The scene was discretized into $128\times128$ pixels, where $[0,0,0]\mathrm{m}$ and $[1100,1100,0]\mathrm{m}$ correspond to the pixels $(1,1)$ and $(128,128)$, respectively. \figref{fig:setup} shows the scene with a static extended target and multiple moving targets along with their corresponding velocities.

\begin{figure}
	\centering
    \includegraphics[width=2.5in]{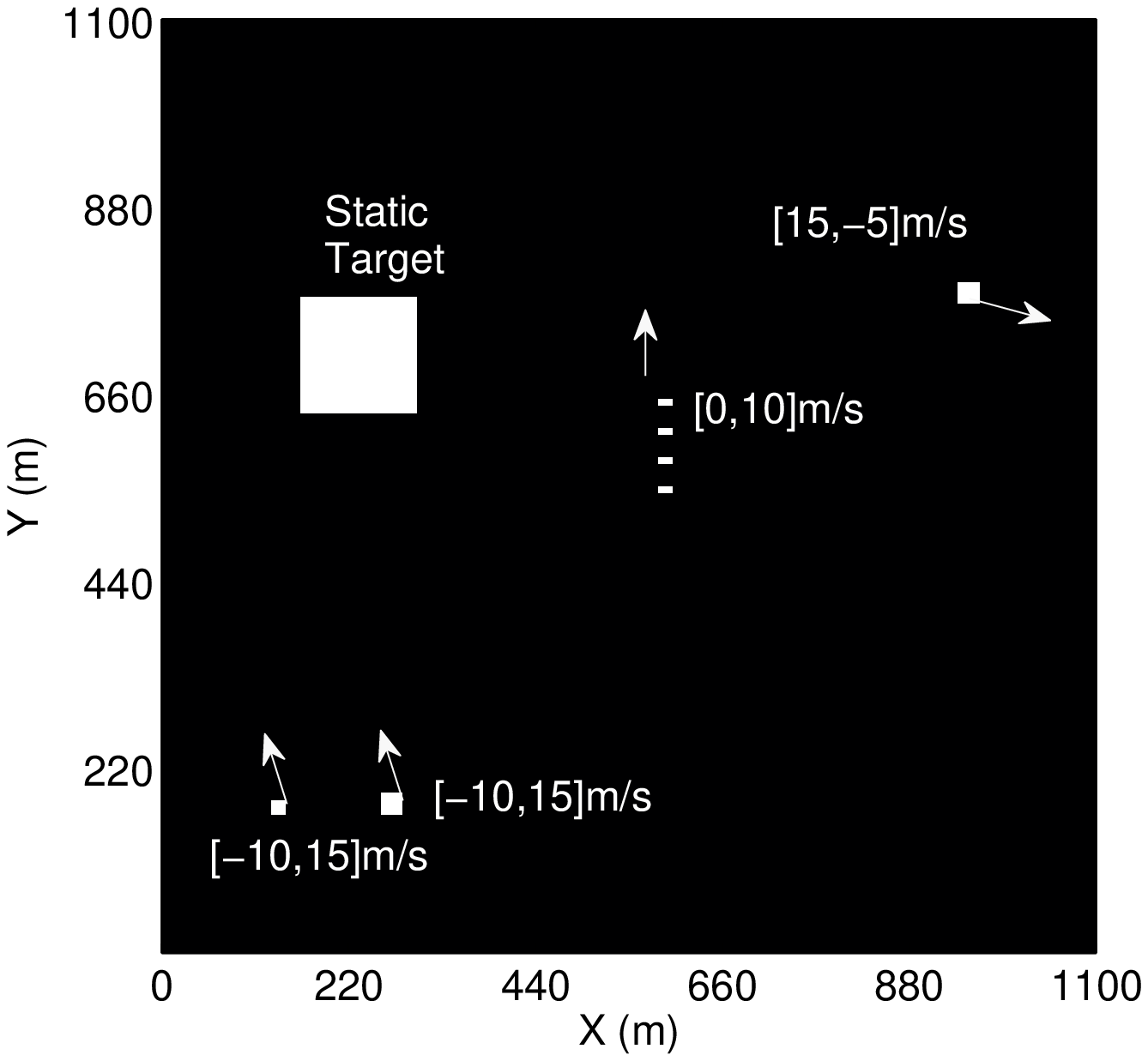}
	\caption{The moving target scene considered in the numerical simulations.}
	\label{fig:setup}
\end{figure}

We assumed that the transmitter and receiver were traversing a circular trajectory given by
    $\gamma_C(s)=(11+11\cos(s),11+11\sin(s),6.5)\,\mathrm{km}$.
Let $\bgamma_T(s)$ and $\bgamma_R(s)$ denote the trajectories of the transmitter and receiver. We set $\bgamma_T(s)=\gamma_C(s)$ and $\bgamma_R(s)=\gamma_C(s-\frac{\pi}{4})$. Note that the variable $s$ in $\gamma_C$ is equal to $\frac{V}{R}t$ where $V$ is the speed of the receiver or the transmitter, and $R$ is the radius of the circular trajectory. We set the speed of the transmitter and receiver to $261\,\mathrm{m/s}$.
\figref{sim} shows the 3D view of the transmitter and receiver trajectories and the scene.
\begin{figure}[]
    \centering
    \includegraphics[width=3in]{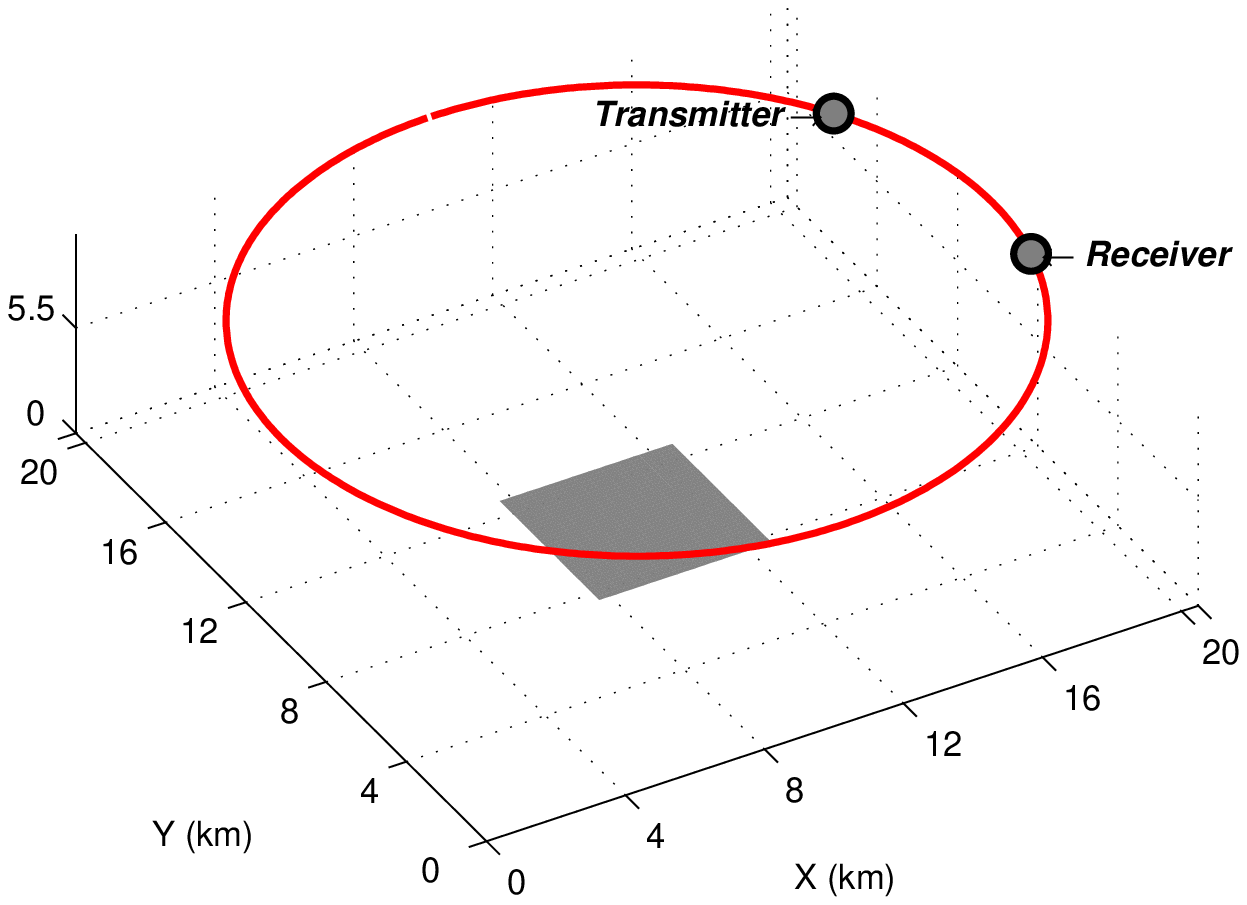}
    \caption{A 3-D illustration of the simulation setup. The dark region denotes the scene (moving targets are not displayed) considered in the simulations. The transmitter/receiver antennas traverse a continuum of positions along the circular trajectory as shown by the red line. At a certain time instant, the transmitter and receiver are located at the positions indicated by the solid dots.}
    \label{sim}
\end{figure}

The length of the signal was set to $L_\phi=0.1707\mathrm{s}$. The circular trajectory was uniformly sampled into 2048 points, corresponding to $f_s=7.7339$Hz. 

We assumed that the velocity of the targets is in the range of $[-20,20]\times[-20,20]\mathrm{m/s}$ and discretized the target velocity space into a $41\times 41$ grid with the discretization step equal to $1\mathrm {m/s}$. Thus, the velocity estimation precision is $1 \mathrm{m/s}$ in our simulations. We reconstructed $\tilde{\rho}_{\bi v_h}(\bi x)$ images via the FBP method as described in Section \ref{sec:ReflectFBP}.

We form the contrast image as described in Section \ref{sec:imageContrast}. The result is shown in \figref{fig:contrastImage}.
We see from \figref{fig:contrastImage} that there are four dominant peaks marked with red circles. This indicates that there are four different velocities associated with the moving target scene. The velocities
where the peaks are located are $[-10,15,0],[0,10,0],[15,-5,0]\mathrm{m/s}$ and $[0,0,0]\mathrm{m/s}$. 
The estimated velocities are equal to the true target velocities used in the simulations. 

\figref{fig:MovTarImages} presents the reconstructed reflectivity images corresponding to the estimated velocities, i.e., $[-10,15,0],[0,10,0],[15,-5,0]\mathrm{m/s}$ and $[0,0,0]\mathrm{m/s}$. We see that the targets are well-focused in the images formed using the correct velocity associated with each target. 
Note that \figref{fig:staticTarget} is the image reconstructed with $\v_h=[0,0,0]\mathrm{m/s}$. In this case, the moving target imaging method described here is equivalent to the static target imaging method that we introduced  
in \cite{LB2010}. As expected only the static target is reconstructed in \figref{fig:staticTarget}.

\begin{figure}
	\centering
    \includegraphics[width=3in]{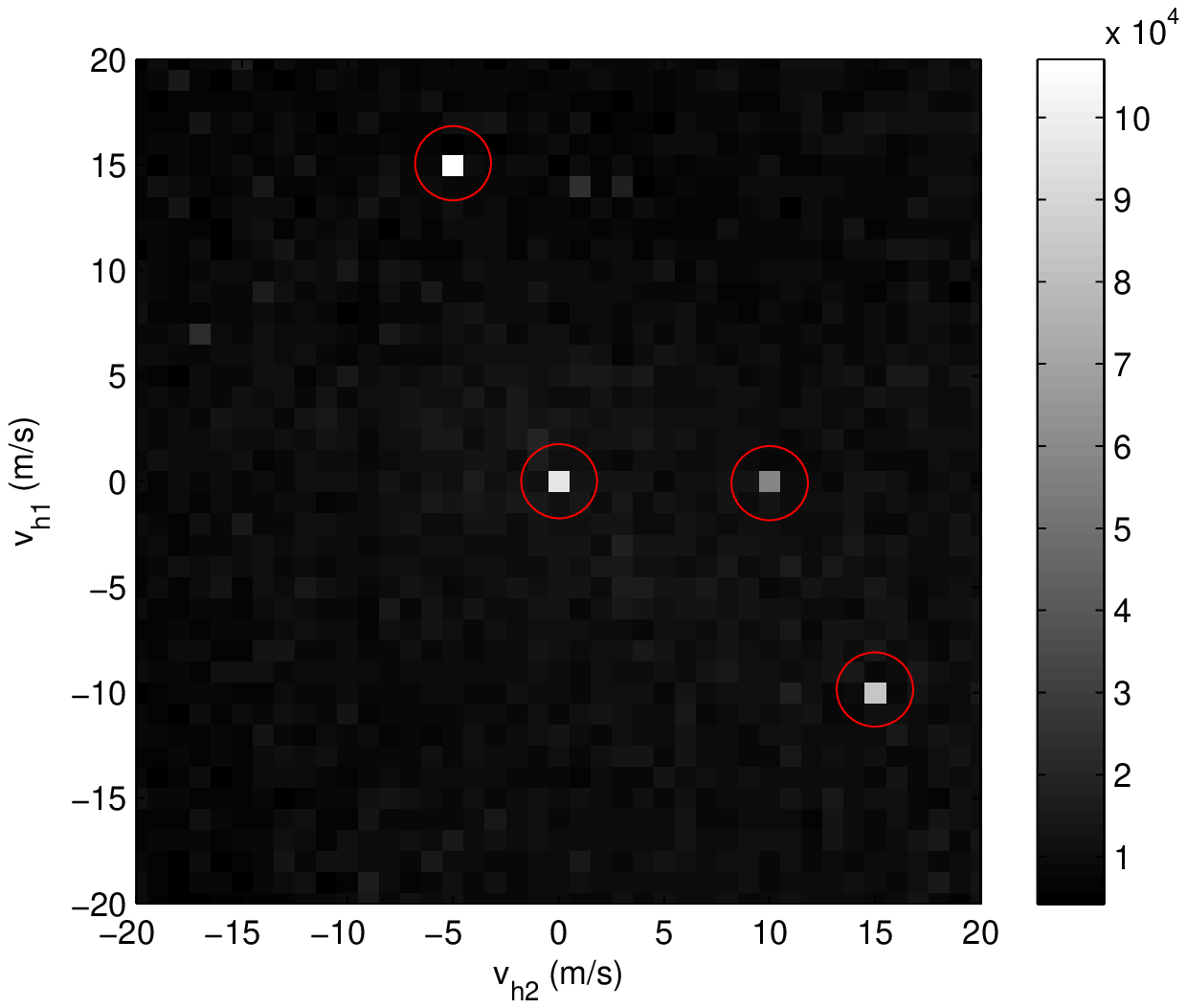}
	\caption{The contrast-image obtained from the reflectivity images reconstructed using a range of hypothesized velocities.}
	\label{fig:contrastImage}
\end{figure}

\begin{figure}
  \centering
  \subfigure[] {
    \label{fig:staticTarget}
    \includegraphics[width=2.5in]{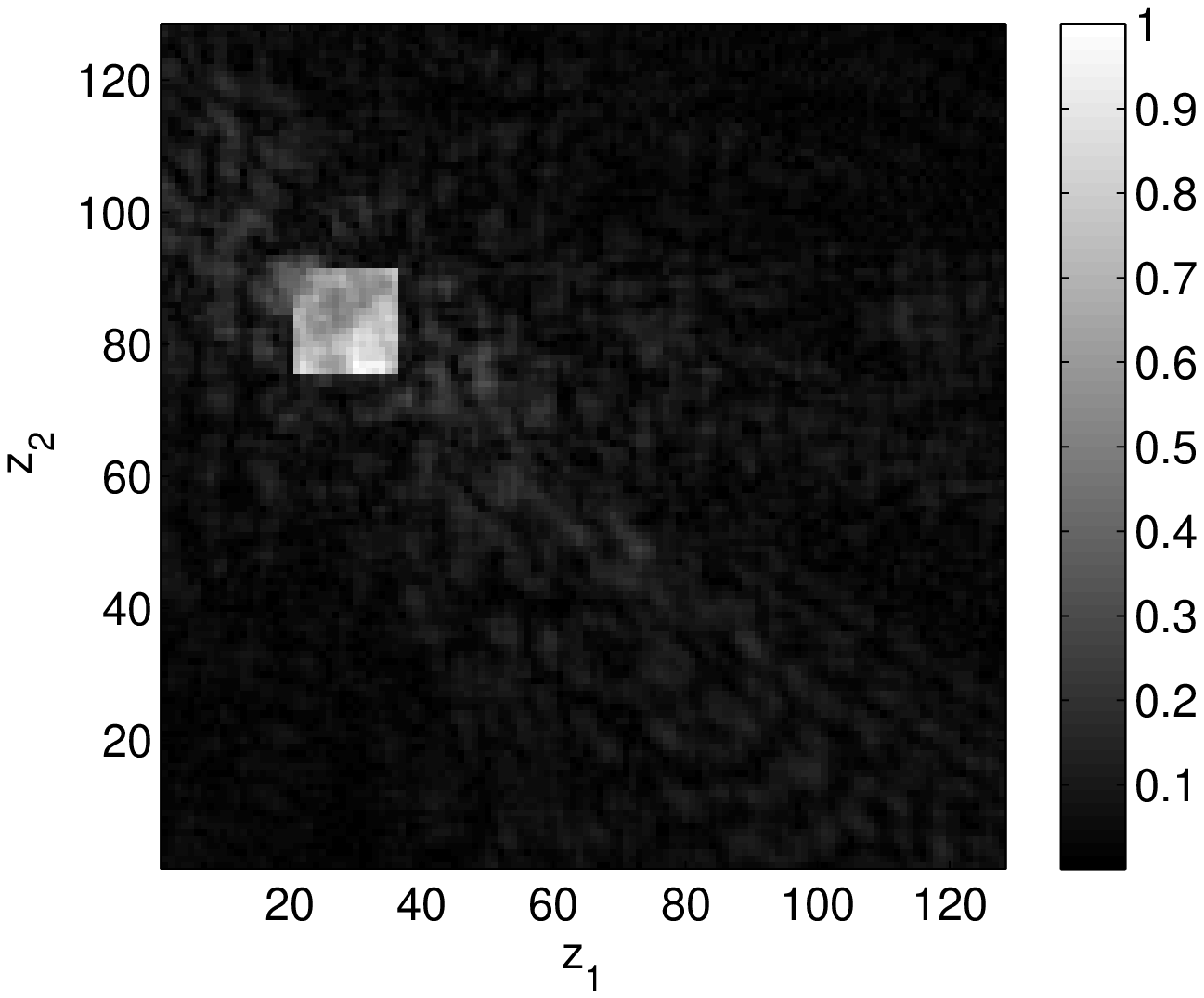}
  }
  \hspace{-0.35cm}
  \subfigure[] {
    \label{fig:MovTar1and2}
    \includegraphics[width=2.5in]{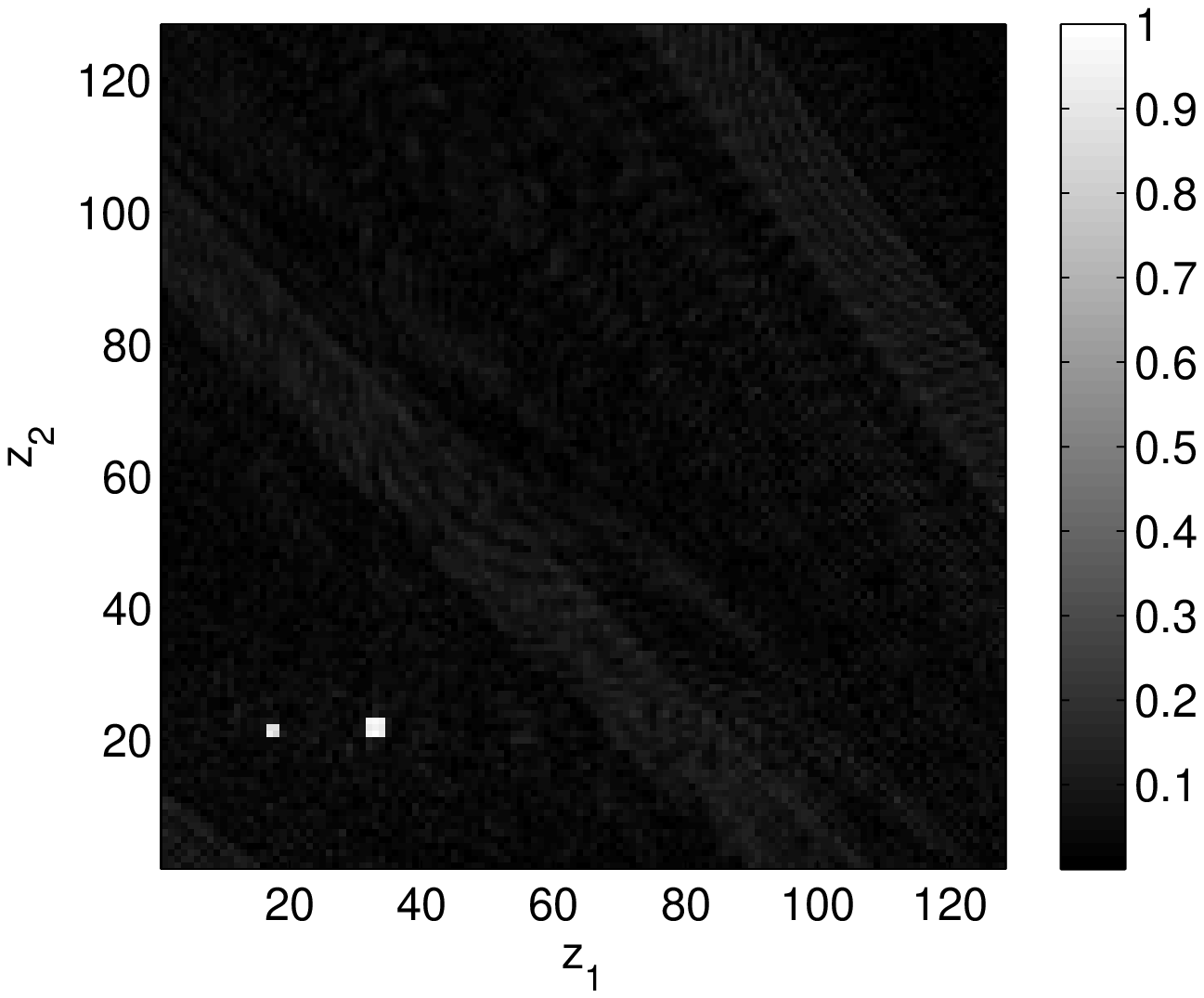}
  }
  \\
    \subfigure[] {
    \label{fig:MovTar3to6}
    \includegraphics[width=2.5in]{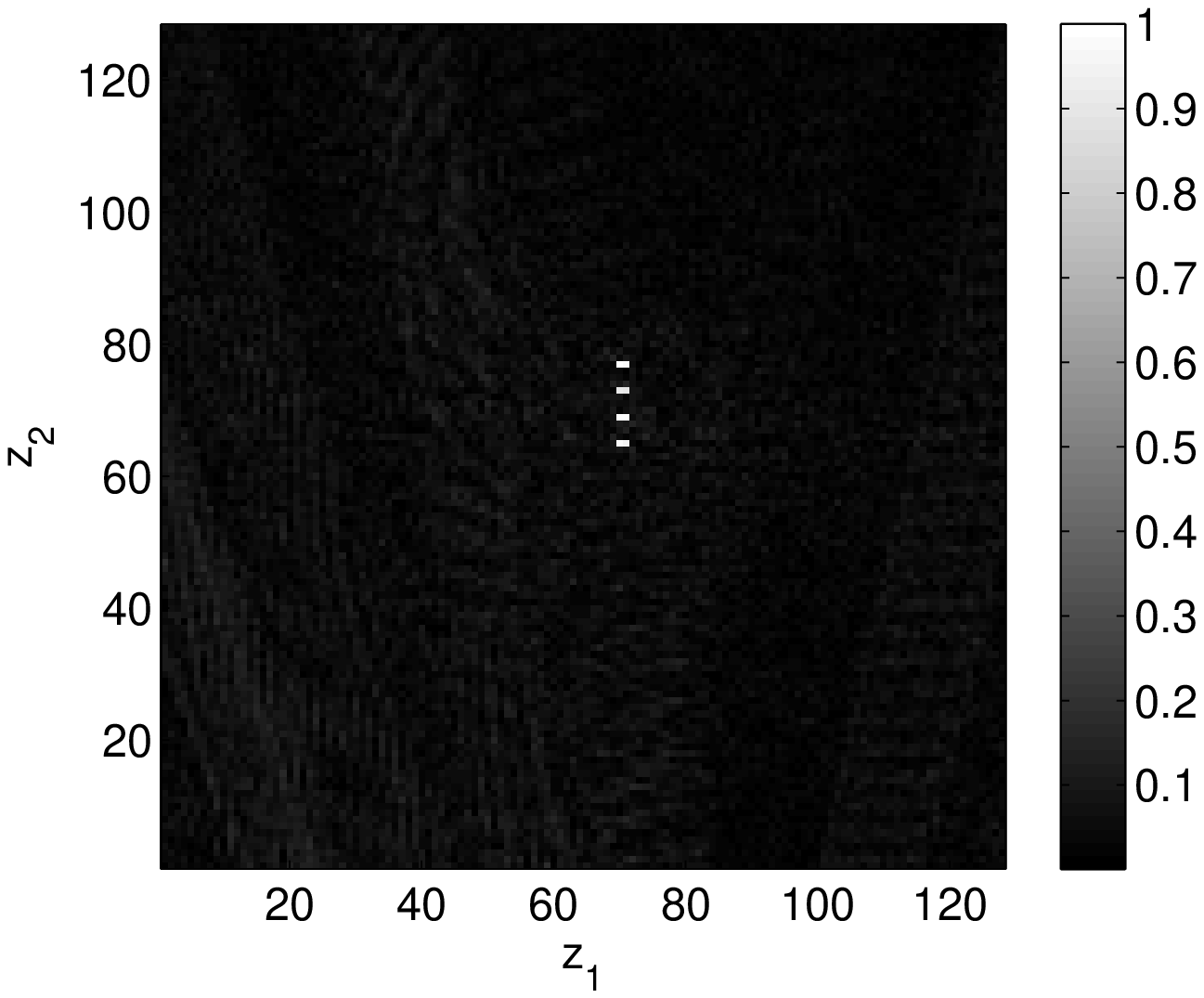}
  }
  \hspace{-0.35cm}
  \subfigure[] {
    \label{fig:MovTar7}
    \includegraphics[width=2.5in]{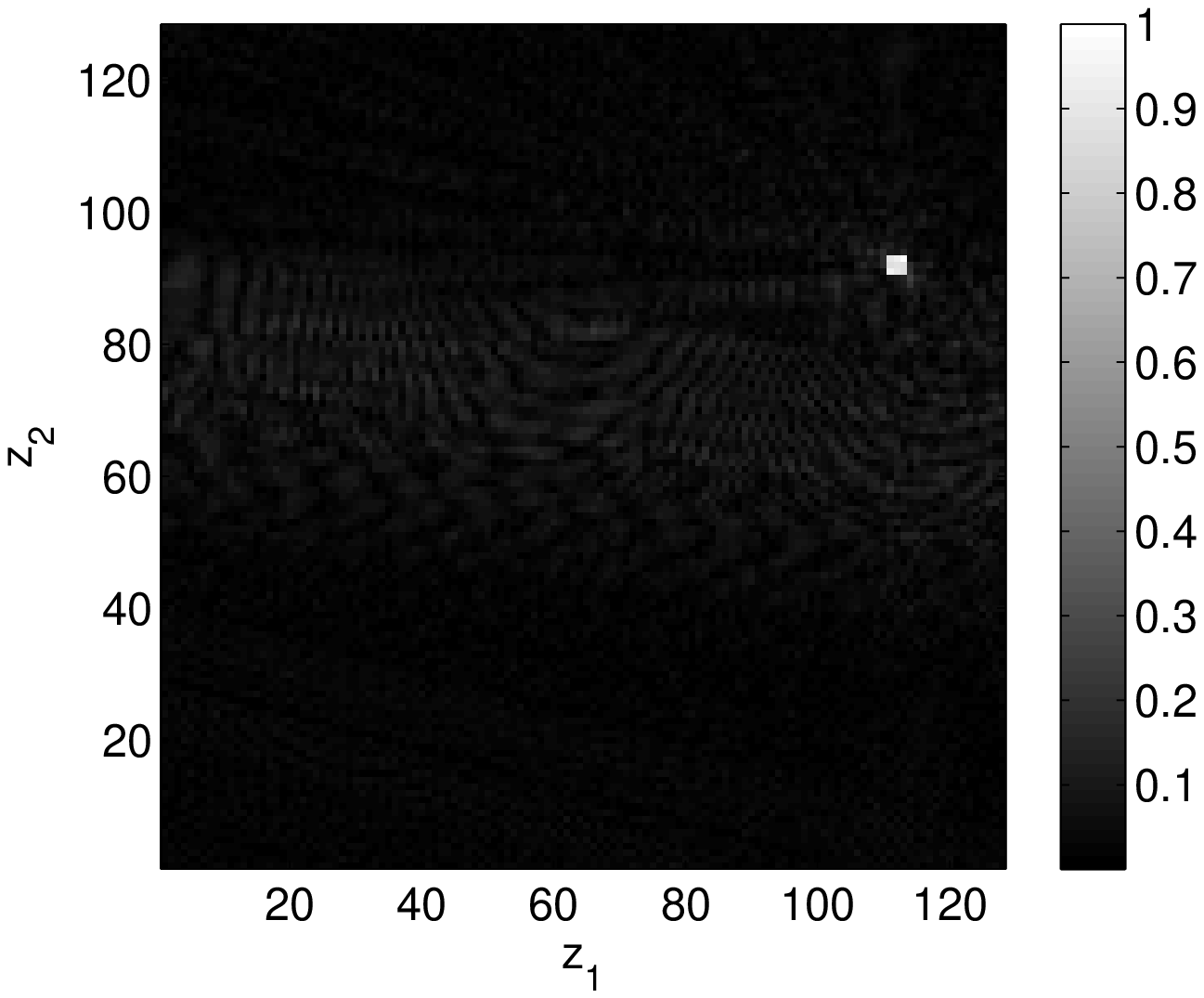}
  }
  \caption{Reconstructed images corresponding to the velocities: (a) $\v_h=[0,0,0]\mathrm{m/s}$; (b) $\v_h=[-10,15, 0]\mathrm{m/s}$; (c) $\v_h=[0,10,0]\mathrm{m/s}$; (d) $\v_h=[15,-5,0]\mathrm{m/s}$.\label{fig:MovTarImages}}
\end{figure}

\section{Conclusion}\label{sec:conclusion}
We have introduced a novel method for the synthetic aperture imaging of moving targets using ultra-narrowband transmitted waveforms. Starting from the first principle, we developed a novel forward model by correlating the received signal with a scaled version of the transmitted signal over a finite time interval. Unlike the conventional wideband SAR forward model, which is based on the start-stop approximation and high resolution delay measurements, this model does not use start-stop approximation and is based on the temporal Doppler induced by the movement of antennas and moving targets. 
The analysis of the forward model shows that the data used for reconstruction is the projections of the phase-space reflectivity onto the four-dimensional bistatic iso-Doppler manifolds. 
We next developed a FBP-type image reconstruction method to reconstruct the reflectivity of the scene and used a contrast optimization method to estimate the velocity of moving targets. The reflectivity reconstruction involves backprojecting the correlated signal onto the two-dimensional cross-sections of the four-dimensional iso-Doppler manifolds which we referred to as the position-space iso-Doppler contours for a range of hypothesized velocities. We showed that when the hypothesized velocity is equal to the true velocity of a scatterer, the singularity is reconstructed at the correct position and orientation. The PSF analysis shows that the visible singularities reconstructed are those that are at the intersection of position-space bistatic iso-Doppler curves and bistatic iso-Doppler-rate curves corresponding to the correct target velocity. We designed the filter so that the strength of the singularities are preserved at the correct velocity. The resulting filter depends not only on the antenna beam patterns, but also on the hypothesized velocity of targets. Using the image contrast optimization, we estimated the velocity of moving targets from a stack of reflectivity images.

We have analyzed the resolution of the reconstructed reflectivity images and the velocity resolution available in the data by analyzing the PSF of the imaging operator and the temporal Doppler bandwidth of the correlated data. Our analysis shows that both the reflectivity and velocity resolutions are determined by the temporal duration and the carrier frequency of the transmitted waveforms. These findings are consistent with the
Doppler ambiguity theory of CW waveforms. Additionally, our analysis has identified various
other factors, such as the relative velocity between the antennas and the moving targets, the range of the antennas to the moving targets, the bistatic angle and the variation in the ground topography, etc. that affect the reflectivity and velocity resolution.

We have analyzed the error in reconstructed reflectivity images
due to error in target velocity. Our analysis leads to several important
results in moving target imaging. First, it shows that the position error
primarily depends on the component of the velocity error in the antenna
look direction and the projection of the velocity error onto the planes
perpendicular to the look direction and the trajectories of the
antennas. Secondly, our analysis explains the artifacts expected due to
moving targets when the image is reconstructed under a stationary scene
assumption. Finally, it shows that the position error in the backprojected
data is small when the error in the estimated velocity is small. Additionally,
our error analysis method can be easily applied to understand and analyze the
positioning errors due to errors in antenna positions.

We presented extensive numerical simulations to verify our theoretical analysis and to illustrate the performance of our imaging method.

We considered the bistatic scenario where the transmitting and receiving antennas are sufficiently far apart. The results for the monostatic case can be deduced by simply setting the two antenna trajectories to be equal.

Our moving target imaging method can be easily extended to incorporate imaging of airborne-targets and complex target motion models. While in the current paper, we assumed that the velocity of each target remains constant throughout the synthetic aperture, the forward model and the image formation method can be extended to include higher order kinetic parameters. We leave the investigation of this topic as a future research.

Our imaging method can be implemented efficiently by using fast backprojection algorithms \cite{Ni,UHS} or fast Fourier integral operator computation methods \cite{Demanet07,Demanet10}, and by utilizing parallel processing on graphics processing units \cite{CCL_2012}.

Although our imaging scheme was developed in a deterministic setting, it is also applicable when the measurements are corrupted by additive white Gaussian noise \cite{Katie}. When a priori information for the scene to be reconstructed is available and additive noise is colored, FBP-type inversion method presented in this paper can be extended as described in \cite{ycy06}.

Finally, while our primarily interest is in radar imaging, our method is also applicable to other similar imaging problems such as those that may arise in acoustics.

\section*{Acknowledgement}
This work was supported by the Air Force Office of
Scientific Research (AFOSR) under the agreements FA9550-09-1-0013 and FA9550-12-1-0415,
and by the National Science Foundation (NSF) under Grant No. CCF-08030672 and CCF-1218805.

\appendix
\section{}\label{app:Appendix1}
Let $\bgamma_{T(R)}=(\gamma_{T(R)}^1,\gamma_{T(R)}^2,\gamma_{T(R)}^3)^T$, $\dot{\bgamma}_{T(R)}=(\dot{\gamma}_{T(R)}^1,\dot{\gamma}_{T(R)}^2,\dot{\gamma}_{T(R)}^3)^T$
and $\v=(v_1,v_2,v_3)^T$.
We write
\begin{eqnarray}
    \hspace{-2cm}\widehat{\bgamma_T(s)-(\x+\v s)} \cdot (\dot{\bgamma}_T(s)-\v)
    &=\frac{1}{|\bgamma_T(s)-(\x+\v s)|}
    [(\gamma_T^1(s)-(x_1+v_1 s))(\dot{\gamma}_T^1(s)-v_1)
    \nonumber
    \\
    \hspace{-3cm}& +(\gamma_T^2(s)-(x_2+v_2 s))(\dot{\gamma}_T^2(s)-v_2)
    \nonumber
    \\
    \hspace{-3cm}& +(\gamma_T^3(s)-(x_3+v_3 s))(\dot{\gamma}_T^3(s)-v_3)]\,.
    \label{eq:fd_T}
\end{eqnarray}
Note that $\x=(\bi x,\psi(\bi x)), \bi x=(x_1,x_2).$

Thus,
\begin{eqnarray}
    &\fl \hspace{1.5cm}\frac{\partial \widehat{\bgamma_T(s)-(\x+\v s)} \cdot (\dot{\bgamma}_T(s)-\v) }{\partial s}
    \nonumber
    \\
    &\fl\hspace{1cm}=\left[
    \frac{-1}{|\bgamma_T(s)-(\x+\v s)|^2}
    \frac{\bgamma_T(s)-(\x+\v s)}{|\bgamma_T(s)-(\x+\v s)|}
    \cdot (\dot{\bgamma}_T(s)-\v)
    \right]
    \nonumber
    \\
    &\fl \hspace{1.5cm}\times[(\bgamma_T(s)-(\x+\v s)) \cdot (\dot{\bgamma}_T(s)-\v)]
    \nonumber
    \\
    & \fl \hspace{1.5cm}+
    \frac{1}{|\bgamma_T(s)-(\x+\v s)|}
    [(\dot{\gamma}_T^1(s)-v_1)^2+(\gamma_T^1(s)-(x_1+v_1 s))\ddot{\gamma}_T^1(s)
    \nonumber
    \\
    &\fl \hspace{1.5cm}+
    (\dot{\gamma}_T^2(s)-v_2)^2+(\gamma_T^2(s)-(x_2+v_2 s))\ddot{\gamma}_T^2(s)
    \nonumber
    \\
    &\fl \hspace{1.5cm}
    +(\dot{\gamma}_T^3(s)-v_3)^2+(\bgamma_T^3(s)-(x_3+v_3 s))\ddot{\gamma}_T^3(s)]
    \nonumber
    \\
    & \fl\hspace{1cm} =\frac{-1}{|\bgamma_T(s)-(\x+\v s)|}
    [\widehat{(\bgamma_T(s)-(\x+\v s))}\cdot (\dot{\bgamma}_T(s)-\v)]^2
    \nonumber
    \\
    &\fl \hspace{1.5cm}+
    \frac{1}{|\bgamma_T(s)-(\x+\v s)|}
    [|\dot{\bgamma}_T(s)-\v|^2+(\bgamma_T(s)-(\x+\v s))\cdot \ddot{\bgamma}_T(s)]
    \nonumber
    \\
    &\fl \hspace{1cm}=
    \frac{1}{|\bgamma_T(s)-(\x+\v s)|}[|\dot{\bgamma}_T(s)-\v|^2-
    (\widehat{(\bgamma_T(s)-(\x+\v s))}\cdot (\dot{\bgamma}_T(s)-\v))^2]
    \nonumber
    \\
    & \fl\hspace{1.5cm} +\widehat{(\bgamma_T(s)-(\x+\v s))}\cdot \ddot{\bgamma}_T(s)
    \nonumber
    \\
    &\fl \hspace{1cm} =
    \frac{1}{|\bgamma_T(s)-(\x+\v s)|}|(\dot{\bgamma}_{T}(s)-\v)_\perp|^2
    +\underbrace{\widehat{(\bgamma_T(s)-(\x+\v s))}\cdot \ddot{\bgamma}_T(s)}_{a_{T,r}(s,\x)}
    \label{eq:Doprate2_apd}
\end{eqnarray}
where
\begin{equation}\label{eq:gamaperpT_apd}
    \fl (\dot{\bgamma}_{T}(s)-\v)_\perp=(\dot{\bgamma}_T(s)-\v)-\widehat{(\bgamma_T(s)-(\x+\v s))}
    (\widehat{(\bgamma_T(s)-(\x+\v s))} \cdot \dot{(\bgamma}_T(s)-\v))
\end{equation}
denotes the projection of the relative velocity between the transmitter and the moving target
$\dot{\bgamma}_T(s)-\v$ onto the plane whose normal vector is along
$\widehat{\bgamma_T(s)-(\x+\v s)}$, $a_{T,r}(s,\x)$ denotes the
projection of the transmitter acceleration $\ddot{\bgamma}_T(s)$ along $\widehat{\bgamma_T(s)-(\x+\v s)}$.
We see that the summation of the two terms of (\ref{eq:Doprate2_apd}) is the total radial acceleration of the transmitter evaluated at $s$ with respect to the moving target located at $\x+\v s$ on the ground at $s$.

Using (\ref{eq:Doprate2_apd}) and (\ref{eq:fd}), we obtain
\begin{eqnarray}
    \fl \dot{f}_d(s,\x,\v)=& \frac{f_0}{c_0}
    \left[
    \frac{1}{|\bgamma_T(s)-(\x+\v s)|}|(\dot{\bgamma}_{T}(s)-\v)_\perp|^2
    +\widehat{(\bgamma_T(s)-(\x+\v s))}\cdot \ddot{\bgamma}_T(s)
    \right.
    \nonumber
    \\
    \fl & \left.+ \frac{1}{|\bgamma_R(s)-(\x+\v s)|}|(\dot{\bgamma}_{R}(s)-\v)_\perp|^2
    +\widehat{(\bgamma_R(s)-(\x+\v s))}\cdot \ddot{\bgamma}_R(s)\right]
\end{eqnarray}
where similar to (\ref{eq:gamaperpT_apd}),
\begin{equation}\label{eq:gamaperpR_apd}
    \fl (\dot{\bgamma}_{R}(s)-\v)_\perp=(\dot{\bgamma}_R(s)-\v)-\widehat{(\bgamma_R(s)-(\x+\v s))}
    (\widehat{(\bgamma_R(s)-(\x+\v s))} \cdot \dot{(\bgamma}_R(s)-\v))
\end{equation}
denotes the projection of the relative velocity between the receiver and the moving target
$\dot{\bgamma}_R(s)-\v$ onto the plane whose normal vector is along
$\widehat{\bgamma_R(s)-(\x+\v s)}$.

\section{}\label{app:bXi}
Using (\ref{eq:Xi}), we have
\begin{equation}\label{eq:bXi_ap1}
    \bXi_{\bi v_0}(s,\bi x)=2\pi \nabla_{\bi x} f_d(s,\bi x,\bi v_0)=\left[
    \begin{array}{c}
    \partial f_d/\partial x_1\\
    \partial f_d/\partial x_2
    \end{array}
    \right].
\end{equation}

The first-order partial differential of (\ref{eq:fd_T}) with respect to $x_1$ is given by
\begin{eqnarray}
    & \fl \hspace{0.5cm}\frac{\partial \widehat{(\bgamma_T(s)-(\x+\v s))}\cdot (\dot{\bgamma}_T(s)-\v)}{\partial x_1}
    \nonumber
    \\
    & \fl =
    \frac{-1}{|\bgamma_T(s)-(\x+\v s)|^2}
    \nonumber
    \\
    & \fl \hspace{0.25cm} \times \left[
    \frac{-(\gamma_T^1(s)-x_1)-(\frac{\partial \psi}{\partial x_1}+ \frac{\partial^2 \psi}{\partial^2 x_1}v_1 s+\frac{\partial^2 \psi}{\partial x_2 \partial x_1}v_2 s)
    (\gamma_T^3(s)-(\psi(x_1,x_2)+v_3 s)) }
    {|\bgamma_T(s)-(\x+\v s)|}\right.
    \nonumber
    \\
    & \fl \hspace{6cm} \left.\times
    ((\bgamma_T(s)-(\x+\v s))\cdot (\dot{\bgamma}_T(s)-\v))\right]
    \nonumber
    \\
    & \fl \hspace{0.25cm}
    +\frac{1}{|\bgamma_T(s)-(\x+\v s)|}
    \left [-(\dot{\gamma}^1_T(s)-v_1)
    -(\frac{\partial \psi}{\partial x_1}+ \frac{\partial^2 \psi}{\partial^2 x_1}v_1 s+\frac{\partial^2 \psi}{\partial x_2 \partial x_1}v_2 s)
    (\dot{\gamma}^3_T(s)-v_3)\right.
    \nonumber
    \\
    & \fl \hspace{3.5cm} \left.
    -(\gamma_T^3(s)-(\psi(x_1,x_2)+v_3 s))(\frac{\partial^2 \psi}{\partial^2 x_1}v_1 +\frac{\partial^2 \psi}{\partial x_2 \partial x_1}v_2) \right]\,.
    \label{eq:partial_x_1}
\end{eqnarray}
Note that $v_3=\nabla_{\bi x}\psi(\bi x)\cdot [v_1,v_2]=\frac{\partial \psi}{\partial x_1}v_1+\frac{\partial \psi}{\partial x_2}v_2$.

Similarly, the first-order partial differential of (\ref{eq:fd_T}) with respect to $z_2$ can be expressed as follows:
\begin{eqnarray}
    & \fl \hspace{0.5cm} \frac{\partial \widehat{(\bgamma_T(s)-(\x+\v s))}\cdot (\dot{\bgamma}_T(s)-\v)}{\partial x_2}
    \nonumber
    \\
    & \fl =
    \frac{-1}{|\bgamma_T(s)-(\x+\v s)|^2}
    \nonumber
    \\
    & \fl \hspace{0.25cm} \times \left[
    \frac{-(\gamma_T^2(s)-x_2)-(\frac{\partial \psi}{\partial x_2}+ \frac{\partial^2 \psi}{\partial x_1 \partial x_2}v_1 s+\frac{\partial^2 \psi}{\partial^2 x_2}v_2 s)
    (\gamma_T^3(s)-(\psi(x_1,x_2)+v_3 s)) }
    {|\bgamma_T(s)-(\x+\v s)|}\right.
    \nonumber
    \\
    & \fl \hspace{6cm} \left.\times
    ((\bgamma_T(s)-(\x+\v s))\cdot (\dot{\bgamma}_T(s)-\v))\right]
    \nonumber
    \\
    & \fl \hspace{0.25cm}
    +\frac{1}{|\bgamma_T(s)-(\x+\v s)|}
    \left [-(\dot{\gamma}^2_T(s)-v_2)
    -(\frac{\partial \psi}{\partial x_2}+ \frac{\partial^2 \psi}{\partial x_1 \partial x_2}v_1 s+\frac{\partial^2 \psi}{\partial^2 x_2}v_2 s)
    (\dot{\gamma}^3_T(s)-v_3) \right.
    \nonumber
    \\
    & \fl \hspace{3.5cm} \left.
    -(\gamma_T^3(s)-(\psi(x_1,x_2)+v_3 s))(\frac{\partial^2 \psi}{\partial x_1 \partial x_2}v_1 +\frac{\partial^2 \psi}{\partial^2 x_2 }v_2) \right]\,.
    \label{eq:partial_x_2}
\end{eqnarray}

We define
\begin{equation}
    D=\left[
    \begin{array}{ccc}
    1&0&\partial \psi(\bi x)/\partial x_1\\
    0&1&\partial \psi(\bi x)/\partial x_2
    \end{array}
    \right ]
\end{equation}
and
\begin{equation}\label{eq:D2_append}
    D^2=\left[
    \begin{array}{ccc}
    0&0&\frac{\partial^2 \psi(\bi x)}{\partial^2 x_1}v_1+\frac{\partial^2 \psi(\bi x)}{\partial x_2\partial x_1}v_2\\
    0&0&\frac{\partial^2 \psi(\bi x)}{\partial x_1 \partial x_2}v_1+\frac{\partial^2 \psi(\bi x)}{\partial^2 x_2}v_2
    \end{array}
    \right ].
\end{equation}
Hence
\begin{eqnarray}
    & \fl \hspace{0.5cm} \left [
    \begin{array}{c}
        \frac{\partial \widehat{(\bgamma_T(s)-(\x+\v s))}\cdot (\dot{\bgamma}_T(s)-\v)}{\partial x_1}
        \\
        \frac{\partial \widehat{(\bgamma_T(s)-(\x+\v s))}\cdot (\dot{\bgamma}_T(s)-\v)}{\partial x_2}
    \end{array}
    \right]
    \nonumber
    \\
    & \fl =- [D+D^2 s] \cdot
    \frac{(\dot{\bgamma}_{T}(s)-\v)_\perp}{|\bgamma_T(s)-(\x+\v s)|}
    -D^2 \cdot \widehat{(\bgamma_T(s)-(\x+\v s))}
    \label{eq:bXi_T}
\end{eqnarray}
where
\begin{equation}
    \fl (\dot{\bgamma}_{T}(s)-\v)_\perp=(\dot{\bgamma}_T(s)-\v)-\widehat{(\bgamma_T(s)-(\x+\v s))}
    \widehat{(\bgamma_T(s)-(\x+\v s))} \cdot (\dot{\bgamma}_T(s)-\v)\,.
\end{equation}

Thus, using (\ref{eq:fd}), applying the derivation in (\ref{eq:partial_x_1}), (\ref{eq:partial_x_2}) and (\ref{eq:bXi_T}) to each component of
$\partial \widehat{(\bgamma_R(s)-(\x+\v s))}\cdot (\dot{\bgamma}_R(s)-\v)/\partial \bi x$,
we obtain
\begin{eqnarray}
    \fl \bXi_{\bi v_0}(s,\bi x)
    &= & -\frac{2\pi f_0}{c_0} \left \{
    [D+D^2 s]\cdot
    \left[
    \frac{(\dot{\bgamma}_{T}(s)-\v_0)_\perp}{|\bgamma_T(s)-(\x+\v_0 s)|}
    +
    \frac{(\dot{\bgamma}_{R}(s)-\v_0)_\perp}{|\bgamma_R(s)-(\x+\v_0 s)|}\right]\right.
    \nonumber
    \\
    \fl && \hspace{1.5cm}\left. +D^2 \cdot [\widehat{(\bgamma_T(s)-(\x+\v_0 s))}+\widehat{(\bgamma_R(s)-(\x+\v_0 s))}]\right\}\,.
    \label{eq:bXi_ap2}
\end{eqnarray}
Note that $D^2$ in (\ref{eq:bXi_ap2}) is given by (\ref{eq:D2_append}) with $v_1,v_2$ replaced with $v_{0,1},v_{0,2}$ where $\v_0=[v_{0,1},v_{0,2},\nabla_{\bi x}\psi(\bi x) \cdot [v_{0,1},v_{0,2}]]$.

\section{}\label{app:Equation1}
Using (\ref{eq:fd}), we have
\begin{eqnarray}
    \hspace{-2cm}
    f_d(s,\bi z,\bi v_\z+\epsilon \triangle \v_\z)&=\frac{f_0}{c_0}
    \left[
    (\widehat{\bgamma_T(s)-(\z+\v_\z s+\epsilon \triangle \v_\z s)})\cdot (\dot{\bgamma}_T(s)-\v_\z-\epsilon \triangle \v_\z)
    \right.
    \nonumber
    \\
    \hspace{-2cm} & \left. +
    (\widehat{\bgamma_R(s)-(\z+\v_\z s+\epsilon \triangle \v_\z s)})\cdot (\dot{\bgamma}_R(s)-\v_\z-\epsilon \triangle \v_\z)
    \right]\,.
    \label{eq:f_d_deltaV}
\end{eqnarray}

Now we calculate the first derivative of (\ref{eq:f_d_deltaV}) with respect to $\epsilon$. Let us consider the derivative of the first item in the square bracket in (\ref{eq:f_d_deltaV}).
\begin{eqnarray}
    \hspace{-1cm}& \frac{d}{d \epsilon}
    \widehat{(\bgamma_T(s)-(\z+\v_\z s+\epsilon \triangle \v_\z s))}
    \cdot
    (\dot{\bgamma}_T(s)-\v_\z-\epsilon \triangle \v_\z)
    \nonumber
    \\
    \hspace{-1cm}=&
    (\widehat{\bgamma_T(s)-(\z+\v_\z s+\epsilon \triangle \v_\z s)})'|_{\epsilon=0}\cdot (\dot{\bgamma}_T(s)-\v_\z-\epsilon \triangle \v_\z)
    \nonumber
    \\
    \hspace{-1cm}
    & +
    \widehat{(\bgamma_T(s)-(\z+\v_\z s+\epsilon \triangle \v_\z s))} \cdot (-\triangle \v_\z)|_{\epsilon=0}
    \label{eq:partial}
\end{eqnarray}
where $[\,]'$ denotes the first derivative with respect to $\epsilon$,
%
\begin{eqnarray}
    \hspace{-1cm} &(\widehat{\bgamma_T(s)-(\z+\v_\z s+\epsilon \triangle \v_\z s)})'|_{\epsilon=0}
    \nonumber
    \\
    =&
    \frac{-[\triangle \bi v s-[(\widehat{\bgamma_T(s)-(\z+\v_\z s)}) \cdot \triangle \v_\z s](\widehat{\bgamma_T(s)-(\z+\v_\z s)})]}
    {|\bgamma_T(s)-(\z+\v_\z s)|}
    \nonumber
    \\
    =&
    \frac{-s \triangle \v_\z^{\perp,T}}{|\bgamma_T(s)-(\z+\v_\z s)|}\,.
    \label{eq:partial_11}
\end{eqnarray}
Note that $\triangle \v_\z^{\perp,T}$ is the projection of $\triangle \v_\z$ on the plane whose normal direction is along$(\widehat{\bgamma_T(s)-(\z+\v_\z s)})$.

Thus, substituting (\ref{eq:partial_11}) into (\ref{eq:partial}), and the result back into (\ref{eq:partial}), we have
%
\begin{eqnarray}
    & \partial_{\epsilon}f_d(s,\bi z,\bi v_\z+\epsilon \triangle \bi v_\z)|_{\epsilon=0}
    \nonumber
    \\
    =&
    \frac{f_0}{c_0}\left \{-
    \frac{s \triangle \v_\z^{\perp,T} \cdot (\dot{\bgamma}_T(s)-\v_\z)}{|\bgamma_T(s)-(\z+\v_\z s)|}
    -
    \triangle \v_\z \cdot (\widehat{\bgamma_T(s)-(\z+\v_\z s))}
    \right.
    \nonumber
    \\
    & \hspace{0.5cm} \left.
    -\frac{s \triangle \bi v_\z^{\perp,R} \cdot (\dot{\bgamma}_R(s)-\v_\z)}{|\bgamma_R(s)-(\z+\v_\z s)|}
    -
    \triangle \v_\z \cdot (\widehat{\bgamma_R(s)-(\z+\v_\z s))}
    \right \}\,.
    \label{eq:partial_e}
\end{eqnarray}

We assume flat topography. From (69) of the manuscript, we have
\begin{eqnarray}
    \hspace{-1cm} \nabla_{\bi z}f_d(s,\bi z,\bi v_\z)&=
    \frac{f_0}{c_0}
    \left\{
    D \cdot \left[
    \frac{(\dot{\bgamma}_T(s)-\v_\z)^{\perp}}{|\bgamma_T(s)-(\z+\v_\z s)|}+
    \frac{(\dot{\bgamma}_R(s)-\v_\z)^{\perp}}{|\bgamma_R(s)-(\z+\v_\z s)|}
    \right]
    \right\}
    \nonumber
    \\
    &=\frac{f_0}{c_0} \left\{-\frac{2 \pi f_0}{c_0} \bXi_{\bi v_\z}(s,\bi z)\right\}\,.
    \label{eq:partial_z}
\end{eqnarray}

Plugging (\ref{eq:partial_e}) and (\ref{eq:partial_z}) into (\ref{eq:partial_fd2}), we obtain
\begin{eqnarray}
    &-\epsilon s \left[
    \frac{\triangle \v_\z^{\perp,T} \cdot (\dot{\bgamma}_T(s)-\v_\z)}{|\bgamma_T(s)-(\z+\v_\z s)|}
    +\frac{\triangle \v_\z^{\perp,R} \cdot (\dot{\bgamma}_R(s)-\v_\z)}{|\bgamma_R(s)-(\z+\v_\z s)|}\right]
    \nonumber
    \\
    & -\epsilon
    \triangle \v_\z \cdot [\widehat{\bgamma_T(s)-(\z+\v_\z s)}
    +\widehat{\bgamma_R(s)-(\z+\v_\z s)} ]
    \nonumber
    \\
    & = \triangle \bi z \cdot \bXi_{\bi v_\z}(s,\bi z)\,.
    \label{eq:partial_fd3_app}
\end{eqnarray} 
\section{}\label{app:Equation2}
Using (\ref{eq:Doprate2}), we have
\begin{eqnarray}
    \fl \hspace{1cm}\dot{f}_d(s,\bi z,\bi v_\z+\epsilon \triangle \bi v_\z)&=\frac{f_0}{c_0}
    \left[\frac{|(\dot{\bgamma}_{T}(s)-(\v_\z+\epsilon \triangle \v_\z))_\perp|^2}{|\bgamma_T(s)-(\z+(\v_\z+\epsilon \triangle \v_\z) s)|} \right.
    \nonumber
    \\
    & \hspace{1cm} \left.+\widehat{(\bgamma_T(s)-(\z+(\v_\z+\epsilon \triangle \v_\z) s))}\cdot \ddot{\bgamma}_T(s)\right.
    \nonumber
    \\
    \fl & \hspace{1cm} \left.
    +\frac{|(\dot{\bgamma}_{R}(s)-(\v_\z+\epsilon \triangle \v_\z))_\perp|^2}{|\bgamma_R(s)-(\z+(\v_\z+\epsilon \triangle \v_\z) s)|}\right.
    \nonumber
    \\
    & \hspace{1cm} \left. +\widehat{(\bgamma_R(s)-(\z+(\v_\z+\epsilon \triangle \v_\z) s))}\cdot \ddot{\bgamma}_R(s)\right]\,.
    \label{eq:Doprate22}
\end{eqnarray}

Let
\begin{equation}\label{eq:star}
    \fl \hspace{1cm} \star=\frac{1}{|\bgamma_T(s)-(\z+(\v+\epsilon \triangle \bi v) s)|}|(\dot{\bgamma}_{T}(s)-(\v+\epsilon \triangle \bi v))_\perp|^2\,.
\end{equation}
%
Calculating the first derivative of (\ref{eq:star}) with respect to $\epsilon$, we obtain
\begin{eqnarray}
    \fl \hspace{1cm}\frac{\partial \star}{\partial \epsilon}&=\left[\frac{1}{|\bgamma_T(s)-(\z+(\v_\z+\epsilon \triangle \v_\z) s)|} \right]' |(\dot{\bgamma}_{T}(s)-(\v_\z+\epsilon \triangle \v_\z))_\perp|^2
    \nonumber
    \\
    \fl \hspace{1cm} &+\frac{1}{|\bgamma_T(s)-(\z+(\v_\z+\epsilon \triangle \v_\z) s)|}
    (|(\dot{\bgamma}_{T}(s)-(\z+(\v_\z+\epsilon \triangle \v_\z))_\perp|^2)' \,,
    \label{eq:partial_star}
\end{eqnarray}
where $[\,]'$ denotes the first derivative with respect to $\epsilon$,
\begin{eqnarray}
    \fl \hspace{1cm}\left[\frac{1}{|\bgamma_T(s)-(\z+(\v_\z+\epsilon \triangle \v_\z) s)|} \right]'
    &=\frac{\widehat{\bgamma_T(s)-(\z+\v_\z s+\epsilon \triangle \v_\z s)} \cdot \triangle \v_\z s}{|\bgamma_T(s)-(\z+\v_\z s+\epsilon \triangle \v_\z s)|^2}
    \label{eq:result_11}
\end{eqnarray}
and
\begin{eqnarray}
    \fl \hspace{1cm} (|(\dot{\bgamma}_{T}(s)-(\v_\z+\epsilon \triangle \v_\z))_\perp|^2)' =&
    2|(\dot{\bgamma}_{T}(s)-(\v_\z+\epsilon \triangle \v_\z))_\perp|\,
    \nonumber
    \\
    & \times |(\dot{\bgamma}_{T}(s)-(\v_\z+\epsilon \triangle \v_\z))_\perp|' \,.
    \label{eq:result_2}
\end{eqnarray}

\noindent In (\ref{eq:result_2}),
\begin{eqnarray}
    \fl \hspace{1cm}  |(\dot{\bgamma}_{T}(s)-(\v_\z+\epsilon \triangle \v_\z))_\perp|'
    & =\frac{1}
    {|(\dot{\bgamma}_{T}(s)-(\v_\z+\epsilon \triangle \v_\z))_\perp|}
    \nonumber
    \\
    & \hspace{-1cm} \times (\dot{\bgamma}_{T}(s)-(\v_\z+\epsilon \triangle \v_\z))_\perp \cdot
    (\dot{\bgamma}_{T}(s)-(\v_\z+\epsilon \triangle \v_\z))'_\perp \,.
    \nonumber
    \\
    \label{eq:result_3}
\end{eqnarray}
where
\begin{eqnarray}
    &\fl \hspace{1cm} (\dot{\bgamma}_{T}(s)-(\v_\z+\epsilon \triangle \v_\z))'_\perp|_{\epsilon=0}
    \nonumber
    \\
    &\fl \hspace{1cm} =-\triangle \v_\z-
    \left\{
    \frac{-s \triangle \v_\z^{\perp,T}}{|\bgamma_T(s)-(\z+\v_\z s)|}
    (\widehat{\bgamma_T(s)-(\z+\v_\z s)})\cdot (\dot{\bgamma}_T(s)-\v_\z)
    \right.
    \nonumber
    \\
    & \fl \hspace{1cm} \hspace{1.5cm} \left.
    + \widehat{\bgamma_T(s)-(\z+\v_\z s)}
    \left[
    \frac{-s \triangle \v_\z^{\perp,T}}{|\bgamma_T(s)-(\z+\v_\z s)|} \cdot (\dot{\bgamma}_T(s)-\v_\z)
    \right.
    \right.
    \nonumber
    \\
    &\fl \hspace{1cm}  \hspace{1.5cm}\left. \left. -\triangle \bi v \cdot \widehat{\bgamma_T(s)-(\z+\v s)}
    \right]
    \right\}
    \nonumber
    \\
     &\fl \hspace{1cm}= -\triangle \v_\z^{\perp,T} 
    \left[1- \frac{s\,\widehat{\bgamma_T(s)-(\z+\v_\z s)})\cdot (\dot{\bgamma}_T(s)-\v_\z)}
    {|\bgamma_T(s)-(\z+\v_\z s)|} \right]+
    \nonumber
    \\
    &\fl \hspace{1.5cm} \widehat{\bgamma_T(s)-(\z+\v_\z s)} \frac{s\,\triangle \v_\z^{\perp,T} \cdot (\dot{\bgamma}_T(s)-\v_\z)}
    {|\bgamma_T(s)-(\z+\v_\z s)|}.
    \label{eq:result_4}
\end{eqnarray}
where 
$\triangle \v_\z^{\perp,T,R}$ is the projection of $\triangle \v_\z$ on the plane whose normal direction is along $(\widehat{\bgamma_{T,R}(s)-(\z+\v_\z s)})$.

Substituting (\ref{eq:result_4}) into (\ref{eq:result_3}) and then the result back into (\ref{eq:result_2}), we have
\begin{eqnarray}
    \fl \hspace{1cm} (|(\dot{\bgamma}_{T}(s)-(\v+\epsilon \triangle \bi v))_\perp|^2)'|_{\epsilon=0}
    \nonumber
    \\
    \fl \hspace{1cm} =  -2 \triangle \v_\z^{\perp,T} \cdot (\dot{\bgamma}_T(s)-\v_\z)_\perp
    \left[1- \frac{s\,\widehat{\bgamma_T(s)-(\z+\v_\z s)})\cdot (\dot{\bgamma}_T(s)-\v_\z)}{|\bgamma_T(s)-(\z+\v_\z s)|} \right].
    \label{eq:result_5}
\end{eqnarray}

Using (\ref{eq:result_5}), (\ref{eq:result_11}) and (\ref{eq:partial_star}), we obtain
\begin{eqnarray}
    \fl \hspace{1cm}\frac{\partial \star}{\partial \epsilon}|_{\epsilon=0}
    =\triangle \v_\z \cdot \widehat{\bgamma_{T}(s)-(\z+\v_\z s)}
    \frac{s\,|(\dot{\bgamma}_{T}(s)-\v_\z)_\perp|^2}
    {|\bgamma_T(s)-(\z+\v_\z s)|^2}
    \nonumber
    \\
    \hspace{-1.5cm} -2\frac{\triangle \v_\z^{\perp,T} \cdot (\dot{\bgamma}_T(s)-\v_\z)_\perp}
    {|\bgamma_T(s)-(\z+\v-\z s)|}
    \left[1- \frac{s\,\widehat{\bgamma_T(s)-(\z+\v_\z s)})\cdot (\dot{\bgamma}_T(s)-\v_\z)}{|\bgamma_T(s)-(\z+\v_\z s)|} \right].
    \label{eq:result_6}
\end{eqnarray}

Thus, using (\ref{eq:result_6}) and (\ref{eq:partial_11}), we have
\begin{eqnarray}
    \fl \hspace{1cm} \partial_{\epsilon}\dot{f}_d (s,\bi z,\bi v+\epsilon \triangle \bi v)|_{\epsilon=0}
    \nonumber
    \\
    \fl \hspace{1cm} =\frac{f_0}{c_0}
    \left [
    -\triangle \v_\z^{\perp,T} \cdot  \left (
    \frac{s\,\ddot{\bgamma}_T(s)}{|\bgamma_T(s)-(\z+\v_\z s)|} 
    +\frac {2 (\dot{\bgamma}_T(s)-\v_\z)_\perp }{|\bgamma_T(s)-(\z+\v_\z s)|}
    C_{T}(\z,\v_{\z},s)
    \right )
    \right.
    \nonumber
    \\
    \fl \hspace{2cm} \left.+
    \triangle \v_\z \cdot \widehat{\bgamma_{T}(s)-(\z+\v_\z s)}
    \frac{s\,|(\dot{\bgamma}_{T}(s)-\v_\z)_\perp|^2}
    {|\bgamma_T(s)-(\z+\v_\z s)|^2}
    \right.
    \nonumber
    \\
    \fl \hspace{2cm} \left. -\triangle \v_\z^{\perp,R} \cdot  \left (
    \frac{s\,\ddot{\bgamma}_R(s)}{|\bgamma_R(s)-(\z+\v_\z s)|}
    +\frac {2 (\dot{\bgamma}_R(s)-\v_\z)_\perp }{|\bgamma_R(s)-(\z+\v_\z s)|}
    C_{R}(\z,\v_{\z},s)
    \right )
    \right.
    \nonumber
    \\
    \fl \hspace{2cm} \left.+
    \triangle \v_\z \cdot \widehat{\bgamma_{R}(s)-(\z+\v_\z s)}
    \frac{s\,|(\dot{\bgamma}_{R}(s)-\v_\z)_\perp|^2}
    {|\bgamma_R(s)-(\z+\v_\z s)|^2}
    \right]
    \label{eq:e_result}
\end{eqnarray}
where 
\begin{eqnarray}
    \fl \hspace{1cm} C_{T,R}(\z,\v_{\z},s) = 
    1- \frac{s\,\widehat{\bgamma_{T,R}(s)-(\z+\v_\z s)} \cdot (\dot{\bgamma}_{T,R}(s)-\v_\z)}{|\bgamma_{T,R}(s)-(\z+\v_\z s)|}
    \label{eq:C_T_app}
\end{eqnarray}

Now let us consider $\nabla_\bi z \dot{f}_d(s,\bi z,\bi v_\z)$. We assume flat topography.
We have
\begin{eqnarray}
    \fl \hspace{1cm} \nabla_\bi z \left( \frac{|(\dot{\bgamma}_{T}(s)-\v_\z)_\perp|^2}{|\bgamma_T(s)-(\z+\v_\z s)|}
    +\widehat{\bgamma_T(s)-(\z+\v_\z s)}\cdot \ddot{\bgamma}_T(s)\right)
    \nonumber
    \\
    \fl \hspace{1cm} = \nabla_\bi z \left( \frac{|(\dot{\bgamma}_{T}(s)-\v_\z)_\perp|^2}{|\bgamma_T(s)-(\z+\v_\z s)|}\right)
    +\nabla_\bi z (\widehat{\bgamma_T(s)-(\z+\v_\z s)}\cdot \ddot{\bgamma}_T(s))
    \nonumber
    \\
    \fl \hspace{1cm} = \left(\nabla_\bi z \frac{1}{|\bgamma_T(s)-(\z+\v_\z s)|}\right) |(\dot{\bgamma}_{T}(s)-\v_\z)_\perp|^2
    \nonumber
    \\
    \fl \hspace{1.5cm} +\frac{1}{|\bgamma_T(s)-(\z+\v_\z s)|} \nabla_\bi z |(\dot{\bgamma}_{T}(s)-\v_\z)_\perp|^2
    \nonumber
    \\
    \fl \hspace{1.5cm} +(\nabla_\bi z \widehat{\bgamma_T(s)-(\z+\v_\z s)})\cdot \ddot{\bgamma}_T(s)\,.
    \label{eq:rz1}
\end{eqnarray}
In (\ref{eq:rz1}),
\begin{equation}\label{eq:rz2}
    \fl \hspace{1cm} \nabla_\bi z \frac{1}{|\bgamma_T(s)-(\z+\v_\z s)|}
    =\frac{D \cdot \widehat{\bgamma_T(s)-(\z+\v_\z s)}}{|\bgamma_T(s)-(\z+\v_\z s)|^2}\,,
\end{equation}
\begin{eqnarray}
    \fl \hspace{1cm} \nabla_\bi z \left (\widehat{\bgamma_T(s)-(\z+\v s)}\right)
    = \frac{\widehat{\bgamma_T(s)-(\z+\v_\z s)}}{|\bgamma_T(s)-(\z+\v_\z s)|}
    [D \cdot \widehat{\bgamma_T(s)-(\z+\v_\z s)}]
    \nonumber
    \\
    \fl \hspace{5cm} -\frac{1}{|\bgamma_T(s)-(\z+\v s)|}D
    \label{eq:rz4}
\end{eqnarray}
where for flat topography,
\begin{equation}
    D=\left[\begin{array}{ccc}
    1 & 0 & 0\\
    0 & 1 & 0
    \end{array}\right]\,,
\end{equation}
and
\begin{eqnarray}
    \fl \hspace{1cm} \nabla_\bi z |(\dot{\bgamma}_{T}(s)-\v_\z)_\perp|^2
    = 2 \nabla_\bi z (\dot{\bgamma}_{T}(s)-\v_\z)_\perp \cdot
    (\dot{\bgamma}_{T}(s)-\v_\z)_\perp
    \label{eq:rz22}
\end{eqnarray}
where
\begin{eqnarray}
    \fl \hspace{1cm}  \nabla_\bi z (\dot{\bgamma}_{T}(s)-\v_\z)_\perp
    \nonumber
    \\
    \fl \hspace{1cm} = -\frac{\widehat{\bgamma_T(s)-(\z+\v_\z s)}}{|\bgamma_T(s)-(\z+\v_\z s)|}
    [D \cdot \widehat{\bgamma_T(s)-(\z+\v_\z s)}]
    (\dot{\bgamma}_{T}(s)-\v_\z) \cdot
    \widehat{\bgamma_T(s)-(\z+\v_\z s)} 
    \nonumber
    \\
    \fl \hspace{1.5cm}   +\frac{(\dot{\bgamma}_{T}(s)-\v_\z) \cdot \widehat{\bgamma_T(s)-(\z+\v s)}}{|\bgamma_T(s)-(\z+\v_\z s)|}D
    \nonumber
    \\
    \fl \hspace{1.5cm}   -\frac{\widehat{\bgamma_T(s)-(\z+\v_\z s)}}{|\bgamma_T(s)-(\z+\v_\z s)|}
    [D \cdot \widehat{\bgamma_T(s)-(\z+\v_\z s)} ]
    (\dot{\bgamma}_{T}(s)-\v_\z) \cdot
    \widehat{\bgamma_T(s)-(\z+\v_\z s)}
    \nonumber
    \\
    \fl \hspace{1.5cm}
    +\frac{\widehat{\bgamma_T(s)-(\z+\v_\z s)}}{|\bgamma_T(s)-(\z+\v_\z s)|}
    (D \cdot \dot{\bgamma}_{T}(s)-\v_\z)\,.
    \label{eq:rz5}
\end{eqnarray}
Substituting (\ref{eq:rz5}) into (\ref{eq:rz22}), we obtain
\begin{eqnarray}
    \fl \hspace{1cm} \nabla_\bi z |(\dot{\bgamma}_{T}(s)-\v)_\perp|^2
    = 2\frac{(\dot{\bgamma}_{T}(s)-\v_\z) \cdot \widehat{\bgamma_T(s)-(\z+\v s)}}{|\bgamma_T(s)-(\z+\v s)|}
    D  \cdot (\dot{\bgamma}_{T}(s)-\v)_\perp
    \label{eq:rz6}
\end{eqnarray}

Using (\ref{eq:rz6}), (\ref{eq:rz2}) and (\ref{eq:rz4}), (\ref{eq:rz1}) becomes
\begin{eqnarray}
    \fl \hspace{1cm} \nabla_\bi z \left( \frac{|(\dot{\bgamma}_{T}(s)-\v_\z)_\perp|^2}{|\bgamma_T(s)-(\z+\v_\z s)|}
    +\widehat{\bgamma_T(s)-(\z+\v_\z s)}\cdot \ddot{\bgamma}_T(s)\right)
    \nonumber
    \\
    \fl \hspace{1cm} =-D \cdot \left\{-
    \frac{2(\dot{\bgamma}_{T}(s)-\v_\z) \cdot \widehat{\bgamma_T(s)-(\z+\v_\z s)}}{|\bgamma_T(s)-(\z+\v_\z s)|^2}(\dot{\bgamma}_{T}(s)-\v_\z)_\perp
    \right.
    \nonumber
    \\
    \fl \hspace{1.5cm} \left.
    -\frac{|(\dot{\bgamma}_{T}(s)-\v_\z)_\perp|^2}{|\bgamma_T(s)-(\z+\v_\z s)|^2}
    \widehat{\bgamma_T(s)-(\z+\v_\z s)}
    \right.
    \nonumber
    \\
    \fl \hspace{1.5cm} + \left.\frac{ \ddot{\bgamma}^\perp_T(s)}{|\bgamma_T(s)-(\z+\v_\z s)|}
    \right\}\,.
    \label{eq:z_result}
\end{eqnarray}

Using (\ref{eq:z_result}), (\ref{eq:e_result}), considering the fact that $\nabla_\bi z (\partial_s{f}_d)=\partial_s (\nabla_\bi z f_d)=\frac{1}{2\pi} \partial_s \bXi_{\bi v_\z}$, after rearrangement, we obtain
\begin{eqnarray}
    \hspace{-1cm}- \epsilon \triangle \v_\z^{\perp,T} \cdot  \left [
    \frac{s\,\ddot{\bgamma}_T(s)}{|\bgamma_T(s)-(\z+\v_\z s)|}
    +\frac {2 (\dot{\bgamma}_T(s)-\v_\z)_\perp }{|\bgamma_T(s)-(\z+\v_\z s)|}
    C_{T}(\z,\v_{\z},s)
    \right ]
    \nonumber
    \\
    \hspace{-1cm} + \epsilon s
    \triangle \v_\z \cdot \widehat{\bgamma_{T}(s)-(\z+\v_\z s)}
    \frac{|(\dot{\bgamma}_{T}(s)-\v_\z)_\perp|^2}
    {|\bgamma_T(s)-(\z+\v_\z s)|^2}
    \nonumber
    \\
    \hspace{-1cm} -\epsilon \triangle \v_\z^{\perp,R} \cdot  \left [
    \frac{s\,\ddot{\bgamma}_R(s)}{|\bgamma_R(s)-(\z+\v_\z s)|}
    +\frac {2 (\dot{\bgamma}_R(s)-\v_\z)_\perp }{|\bgamma_R(s)-(\z+\v_\z s)|}
    C_{R}(\z,\v_{\z},s)
    \right ]
    \nonumber
    \\
    \hspace{-1cm} +\epsilon s
    \triangle \v_\z \cdot \widehat{\bgamma_{R}(s)-(\z+\v_\z s)}
    \frac{|(\dot{\bgamma}_{R}(s)-\v_\z)_\perp|^2}
    {|\bgamma_R(s)-(\z+\v_\z s)|^2}
    \nonumber
    \\
    \hspace{-1cm}=
    -\triangle \bi z \cdot \dot{\bXi}_{\bi v_{\z}}(s,\bi z) \frac{c_0}{2\pi f_0}
    \label{eq:partial_Dotfd4_app}
\end{eqnarray}
where
\begin{eqnarray}
   \hspace{-2cm}\dot{\bXi}_{\bi v_{\z}}(s,\bi z) \frac{c_0}{2\pi f_0}
    =& D \cdot \left \{ \frac{|(\dot{\bgamma}_{T}(s)-\v_{\z})_\perp|^2}{|\bgamma_T(s)-(\z+\v_{\z} s)|^2}
    \widehat{\bgamma_T(s)-(\z+\v_{\z} s)} \right.
    \nonumber
    \\
    & \hspace{0.5cm} \left. + \frac{|(\dot{\bgamma}_{R}(s)-\v_{\z})_\perp|^2}{|\bgamma_R(s)-(\z+\v_{\z} s)|^2}
    \widehat{\bgamma_R(s)-(\z+\v_{\z} s)} \right.
    \nonumber
    \\
    & \hspace{0.5cm} \left.
    +2\frac{(\dot{\bgamma}_{T}(s)-\v_{\z}) \cdot \widehat{\bgamma_T(s)-(\z+\v_{\z} s)}}
    {|\bgamma_T(s)-(\z+\v_{\z} s)|^2}
    (\dot{\bgamma}_{T}(s)-\v_{\z})_\perp \right.
    \nonumber
    \\
    & \hspace{0.5cm} \left.+2\frac{(\dot{\bgamma}_{R}(s)-\v_{\z}) \cdot \widehat{\bgamma_R(s)-(\z+\v_{\z} s)}}{|\bgamma_R(s)-(\z+\v_{\z} s)|^2}
    (\dot{\bgamma}_{R}(s)-\v_{\z})_\perp \right.
    \nonumber
    \\
    & \hspace{0.5cm} \left.-\frac{\ddot{\bgamma}^\perp_T(s)}{|\bgamma_T(s)-(\z+\v_{\z} s)|}
     -\frac{\ddot{\bgamma}^\perp_R(s)}{|\bgamma_R(s)-(\z+\v_{\z} s)|} \right \}.
    \label{eq:DotOmega_app}
\end{eqnarray}

\section*{References}

\bibliography{microlocal}

\end{document}